\documentclass[fleqn,usenatbib]{mnras}
\usepackage{newtxtext,newtxmath}
\usepackage[T1]{fontenc}
\usepackage{graphicx}
\usepackage{dcolumn}% Align table columns on decimal point

\usepackage[normalem]{ulem}
\usepackage{color}
\usepackage[table]{xcolor}

\newcommand{\arp}[1]{ \textcolor{magenta}{\textbf{ARP: #1}}}

\newcommand{\VEV}[1]{\left\langle #1 \right\rangle}

\DeclareRobustCommand{\VAN}[3]{#2}
\let\VANthebibliography\thebibliography
\def\thebibliography{\DeclareRobustCommand{\VAN}[3]{##3}\VANthebibliography}

\usepackage{subcaption}
\usepackage{amsmath}	% Advanced maths commands
\newcommand{\be}{\begin{equation}}
\newcommand{\ee}{\end{equation}}
\newcommand{\p}{\mathcal{P}}
\newcommand{\pf}{\tilde{\p}}
\newcommand{\If}{\tilde{I}}
\newcommand{\Ndet}{N_{\rm{det}}}
\newcommand{\im}{\rm{i}}
\newcommand{\R}{\mathcal{R}}

\title[EXCLAIM Science]{Extragalactic Science with the Experiment for Cryogenic Large-aperture Intensity Mapping}

\author[A.~R.~Pullen et al.]{Anthony R.~Pullen,$^{1,2}$\thanks{E-mail: anthony.pullen@nyu.edu}
Patrick C.~Breysse,$^{1}$
Trevor Oxholm,$^{3}$
Eric R.~Switzer,$^{4}$\newauthor
Christopher J.~Anderson,$^{4}$
Emily Barrentine,$^{4}$
Alberto D.~Bolatto,$^{5}$
Giuseppe Cataldo,$^{4}$\newauthor
Thomas Essinger-Hileman,$^{4}$
Abhishek Maniyar,$^{1}$
Thomas Stevenson,$^{4}$
Rachel S.~Somerville,$^{2}$\newauthor
Carrie Volpert,$^{5}$
Edward Wollack,$^{4}$
Shengqi Yang,$^{6,1}$
L.~Y.~Aaron Yung,$^{4}$
and Zilu Zhou$^{1}$
\\
$^{1}$Department of Physics, New York University, 726 Broadway, New York, NY, 10003, U.S.A.\\
$^{2}$Center for Computational Astrophysics, Flatiron Institute, New York, NY 10010, U.S.A.\\
$^{3}$Department of Physics, University of Wisconsin-Madison, 1150 University Ave., Madison, WI, 53706, U.S.A.\\
$^{4}$NASA Goddard Space Flight Center, Greenbelt, MD, 20771, U.S.A.\\
$^5$Department of Astronomy and Joint Space-Science Institute, University of Maryland, College Park, MD 20742, U.S.A.\\
$^6$Carnegie Observatories, 813 Santa Barbara St., Pasadena, CA, 91101, U.S.A.
}

\date{Accepted XXX. Received YYY; in original form ZZZ}

\pubyear{2022}

\begin{document}
\label{firstpage}
\pagerange{\pageref{firstpage}--\pageref{lastpage}}
\maketitle

\begin{abstract}
The EXperiment for Cryogenic Large-Aperture Intensity Mapping (EXCLAIM) is a balloon-borne cryogenic telescope that will survey the spectrum of diffuse emission from both the Milky Way and the cosmic web to probe star formation, the interstellar medium, and galaxy evolution across cosmic time.  EXCLAIM's primary extragalactic science survey maps 305 deg$^2$ along the celestial equator with an $R=512$ spectrometer over the frequency range $\nu=420-540$\,GHz, targeting emission of the [CII] line over redshifts $2.5<z<3.5$ and several CO lines for $z<1$. Cross-correlation with galaxy redshift catalogs isolates line emission from the large-scale structure at target redshifts. In this paper, we forecast the sensitivity for both the two-point and conditional one-point cross-correlation. We predict that EXCLAIM will detect both the [CII]-QSO cross-power spectrum and the conditional voxel intensity distribution (CVID) at various redshifts under a broad range of [CII] intensity models, allowing it to differentiate among these models in the literature. These forecasts for the power spectra include the effects of line interlopers and continuum foreground contamination. We then convert the joint [CII] constraints from both the cross-power spectrum and the CVID into constraints on the [CII] halo luminosity-mass relation $L_\mathrm{[CII]}(M)$ model parameters and the star formation rate density (SFRD) from [CII] emission.  
%Depending on the true [CII] model, we predict an SNR from EXCLAIM as high as $\sim4$ at $z=2.7$ for the SFRD in the context of current models of star formation over cosmic time. 
We also develop sensitivity estimates for CO, showing the ability to differentiate between models.
\end{abstract}

\begin{keywords}
ISM: molecules -- galaxies: high-redshift -- large-scale structure of the universe -- submillimeter: ISM -- diffuse radiation
\end{keywords}

%%%%%%%%%%%%%%%%%%%%%%%%%%%%%%%%%%%%%%%%%%%%%%%%%%

%%%%%%%%%%%%%%%%% BODY OF PAPER %%%%%%%%%%%%%%%%%%

\section{Introduction}

Galaxies form the building blocks for the large-scale structure of our Universe, and a fundamental question in astronomy is how galaxies form and evolve.  The star formation history is a significant tracer of galaxy evolution, often quantified in terms of the star formation rate \citep{2014ARA&A..52..415M}.  Young galaxies with a large amount of cold gas tend to produce stars rapidly, while the rate decreases when this fuel is exhausted.  In addition, stellar winds, active galactic nuclei (AGN), and other astrophysical processes tend to disrupt star formation in the later stages of galactic evolution, causing the star formation rate to decrease by a factor of approximately 20 from redshift $z=2$ to the present \citep{2020ApJ...902..111W}.  Measurements of the gas content and chemistry of galaxies over cosmic time are necessary to understand galactic processes in a detailed manner. Traditional galaxy surveys construct large catalogues of the spectra of star-forming galaxies; however, these samples are biased in that they only detect the brightest galaxies, missing the more numerous population of fainter star-forming galaxies.  In addition, these surveys tend to have areas less than a square degree, leading to field-to-field scattering that limits the characterization of the global properties of galaxies. Selection and field size effects complicate the reconciliation of existing observations and models; see e.g. \citet{2019ApJ...882..137P}.

The EXperiment for Cryogenic Large-Aperture Intensity Mapping (EXCLAIM) is a dedicated balloon mission designed to map the emission of the [CII] line ($\nu_{\rm [CII]}=158.7\,\mu{\rm m}$) over redshifts $2.5<z<3.5$ and several CO rotational lines ($\nu_{{\rm CO},J}=115\,J$ GHz) for $J=4-7$ for redshifts $z<1$ with a spectral resolution $R=512$ and covering $420-540$\,GHz ($555-714\,\mu{\rm m}$). In addition, EXCLAIM will measure [CI], CO(4-3), and O$_2$ line emission in the galactic plane, tracing star formation and molecular gas. Previous papers have described the EXCLAIM instrument and survey parameters \citep{2021JATIS...7d4004S, 2020JLTP..199.1027A, 2020SPIE11445E..24C}, the optical design \citep{2020SPIE11453E..0HE}, and the $\mu$-spec spectrometer \citep{2020SPIE11453E..0MM, 2022arXiv220802786V}. This paper will focus on the expected science from the extragalactic survey. We will describe the statistical properties of these maps, which will allow us to measure the intensities of these lines at various redshifts and probe the properties of galaxies and how they evolve.  

EXCLAIM employs the novel technique of \emph{line intensity mapping} (LIM) \citep{2010JCAP...11..016V,2011JCAP...08..010V,2017arXiv170909066K,2019BAAS...51c.101K, 2022arXiv220615377B}.  Traditional galaxy surveys have probed star formation and galaxy evolution through direct multi-color imaging, such as through the Cosmic Assembly Near-infrared Deep Extragalactic Legacy Survey (CANDELS) \citep{2011ApJS..197...35G}, the Hubble Ultra-Deep Field (HUDF) \citep{2013ApJ...763L...7E,2013ApJS..209....3K}, the UKIRT Deep Sky Survey \citep{2009MNRAS.394....3D}, and the VIMOS-VLT Deep Survey \citep{2012A&A...539A..31C}.  Complementary work by spectroscopic surveys has been performed by the ALMA SPECtroscopic Survey (ASPECS) in the Hubble Ultra-Deep Field \citep{2016ApJ...833...67W} and GOODS-N survey \citep{2006ApJ...653.1004R}.  

Designing EXCLAIM as a LIM survey has several advantages. First, line intensity mapping integrates the light from all galaxies, making it highly complementary to surveys of individual objects, which are subject to survey-specific selection limits and effects. By trading spectra of individual galaxies for spectra of pixels on the sky, LIM surveys can map a much larger sky area than a traditional counts survey in equal time, reducing sample variance \citep{Keating2020}. Intensity mapping measures surface brightness rather than flux, relaxing requirements on aperture size, making the approach ideal for balloon and space platforms. Cosmologically, intensity mapping measures the large-scale clustering of emitting sources, so it is also sensitive to the halo and cosmological context of star formation processes in galaxies.  Several tentative detections have been reported in the literature for the 21 cm line \citep{2009MNRAS.394L...6P,2010Natur.466..463C,2013MNRAS.434L..46S, 2013ApJ...763L..20M, 2018MNRAS.476.3382A, 2022MNRAS.510.3495W, 2022arXiv220201242C, 2022arXiv220601579C}, the CO lines \citep{2016ApJ...830...34K, Keating2020, 2022ApJ...927..161K}, and the [CII] line \citep{2019MNRAS.489L..53Y}. \citet{2019BAAS...51c.101K} provide a recent list of LIM surveys that are planned or operational. The EXCLAIM measurement, which will map [CII] at a high-frequency resolution, will be an improvement over the previous [CII]-quasar cross-power constraints \citep{2018MNRAS.478.1911P,2019MNRAS.489L..53Y} by isolating the cosmic infrared background (CIB) to low-$k_\parallel$ modes \citep{2019ApJ...872...82S}.

A major challenge for performing a LIM survey is foreground contamination. Foregrounds for LIM surveys come primarily from two astrophysical sources.  The first is \emph{continuum emission}, which can either be separate from, e.g.~Milky Way emission, or correlated with, e.g.~the CIB, large-scale structure (LSS).  It has been shown that a LIM survey with high spectral resolution is superior to broadband mapping for mitigating continuum foregrounds, because the Fourier-transformed signal tends to isolate continuum emission at large scales along the line of sight due to its spectral smoothness \citep{2019ApJ...872...82S}.  A second source is line interlopers, when emission lines from either higher or lower than the target redshift contaminate the signal in the observation plane.  The contribution from line interlopers tends to additively bias the auto-power spectrum; however, it was shown in \citet{2016ApJ...825..143L} that its contribution can be modeled in a likelihood analysis and marginalized over. For the intensity-galaxy cross-power spectrum we discuss below, interlopers do not bias the signal but add to its variance. The upper atmosphere also has a forest of emission lines \citep{2021JATIS...7d4004S} uncorrelated with the astronomical signal. Its influence on the noise can be mitigated through optimal weighting of the frequency channels.

In order to reject uncorrelated and contaminating variance from foregrounds and instrumental effects, we will cross-correlate EXCLAIM maps with galaxy and quasar redshift catalogs from multiple galaxy surveys that are available in Stripe 82 on the celestial equator \citep{2009ApJS..182..543A}. Surveys of interest include the Sloan Digital Sky Survey (SDSS) \citep{2015ApJS..219...12A,2020ApJS..249....3A}, Hyper-Suprime CAM (HSC) \citep{2019PASJ...71..114A}, and the Hobby-Eberly Telescope Dark Energy Experiment (HETDEX) \citep{2008ASPC..399..115H}. Here, we will focus on the SDSS Baryon Oscillation Spectroscopic Survey (BOSS) using two statistical measures of cross-correlation. First, we will describe the two-point CO-galaxy and [CII]-quasar cross-power spectra using EXCLAIM and galaxy and quasar data cubes.  Second, we will construct a one-point cross-correlation between the [CII] and quasar maps using the \emph{conditional voxel intensity distribution} (CVID) statistic \citep{Breysse2017,Breysse2019}.  Both cross-correlation and the CVID robustly reject variance from uncorrelated foregrounds.

The EXCLAIM survey consists of several ${\sim} 100\,{\rm deg}^2$ Galactic plane (GP) regions and a 305\,deg$^2$ extragalactic (EG) survey outside the GP. In this paper, we will focus on the EG survey; the GP survey will be presented in later work. The EG survey will have several applications. For one, it has been shown through both measurements \citep{DeLooze2014, Herrera-Camus2015, Herrera-Camus2018, 2016ApJ...829L..11P, 2016ApJ...833...71A} and simulations \citep{2017ApJ...834...36H,2018A&A...609A.130L} that [CII] luminosities are strongly correlated with the star formation rate (SFR) of a galaxy.  EXCLAIM's measurements of [CII] over the redshift range $2.5 < z < 3.5$ will trace the star formation rate density (SFRD) during the cosmic ``high noon,'' after which the SFRD falls by a factor of $20$ \citep{2014ARA&A..52..415M}. 

The intermediate $J$ CO transitions targeted by EXCLAIM, on the other hand, trace a mixture of H$_2$ mass and star formation activity (since their excitation depends on the temperature of the gas, usually determined by star formation) \citep{2013ARA&A..51..207B}. Using them as mass tracers necessitates including corrections that depend on the excitation of the gas. This limitation is shared by most studies of molecular gas at high redshifts \citep[e.g.][]{2020ApJ...902..111W}. Thus, in principle measurements of CO line intensities from the EG survey could probe the SFRD over the redshift range $z<1$ (subject to the instrument sensitivity) Finally, it is also possible that with a sufficiently sensitive measurement of [CII] or CO, spectral features in the power spectrum, such as baryon acoustic oscillations, may allow us to constrain the Hubble rate \citep{2019PhRvL.123y1301B}.

In this paper, we present the scientific capabilities of the EXCLAIM mission.  After reviewing the technical specifications of the EXCLAIM survey and co-areal galaxy surveys, we present models for both the cross-power spectra and the CVID.  We then show forecasts for the sensitivity with which EXCLAIM can measure these statistics.  Forecasts include the effect of contamination and variance from the Milky Way, the cosmic infrared background, extragalactic interlopers, and atmospheric emission. We defer consideration of the auto-power, which is not EXCLAIM's primary science, to future work.  We then determine how these measurements will inform star formation models. These methods are developed in the context of the EXCLAIM survey but are applicable more generally.

The plan for the paper is as follows. In Sec.~\ref{S:survey}, we describe the EXCLAIM instrument and survey, and we describe the galaxy surveys for cross-correlations in Sec.~\ref{S:galaxy}.  Sec.~\ref{S:model} defines theoretical and measurement models of line intensities, the power spectra, and the CVID, while we present forecasts, including the effects of interlopers and continuum foregrounds in Sec.~\ref{S:stats}.  We discuss in Sec.~\ref{S:science} the constraints on star formation we will derive from EXCLAIM data, and we conclude in Sec.~\ref{S:conclude}.

\section{EXCLAIM Survey} \label{S:survey}

\subsection{Instrument}
EXCLAIM is a cryogenic balloon-borne telescope designed to perform a large-area survey over a single night during a conventional balloon flight from North America. To achieve low-background observations between narrow atmospheric lines from the stratosphere at altitudes exceeding 30~km, EXCLAIM will employ a completely cryogenic telescope housed in a liquid-helium dewar that couples light onto spectrometers integrated on microfabricated silicon with a spectral resolution $R = 512$ covering a band of 420--540~GHz. EXCLAIM employs cryogenic, flight electronics, and flight software heritage from the ARCADE 2 and PIPER missions,~\citep{ARCADE2_Instrument_Singal_2011, ARCADE2_Fixsen_Results_2011, PIPER_Lazear_2014, PIPER_Flight_Pawlyk_2018, doi:10.1063/5.0048800} which flew similar open bucket dewar instruments.

At float altitudes, the ambient pressure of less than 10~Torr pumps on the helium bath, lowering its temperature to ${\sim}1.7$~K, below its superfluid transition. Dedicated superfluid helium pumps, in combination with helium boil-off gas, keep the temperature of the entire optical chain below 5~K. Cryogenic optics reduce the instrumental background below the expected sky loading in the darkest channels of ${\sim}0.1$~fW. 

The EXCLAIM optics consist of a two-mirror off-axis Gregorian telescope with a projected aperture of 75~cm. A folding flat between the primary and secondary mirrors allows the telescope to fit within the volume of the dewar. The secondary mirror redirects and collimates light as it enters a vertical superfluid-tight receiver ``submarine'' that houses infrared-blocking and band-defining filters, a cold Lyot stop, a final silicon lens, and the focal plane. An adiabatic demagnetization refrigerator (ADR) provides a base temperature of 100~mK for the spectrometers. EXCLAIM will fly six identical $\mu$-Spec integrated spectrometer chips~\citep{uSpec_Cataldo_2014, uSpec_Noroozian_2015, uSpec_Barrentine_2016, uSpec_Second_Gen_Cataldo_2018, 2022arXiv220802786V}. Optical coupling to the spectrometers is via hyper-hemispherical silicon lenslets that focus light onto dipole slot antennas.

The $\mu$-Spec integrated spectrometers combine all elements of a traditional diffraction-grating spectrometer in a compact package using planar transmission lines on a silicon chip. The grating is replaced by a niobium microstrip delay line network that launches signals into a 2D parallel-plate waveguide region with emitting and receiving feeds arranged in a Rowland configuration \citep{2015AcAau.114...54C}. EXCLAIM operates at the second grating order selected by an on-chip order-sorting filter and free-space band-defining filters. A set of 355 microwave kinetic inductance detectors (MKIDs) made from 20~nm-thick Al half-wave resonators detect the signal \citep{2022arXiv220802786V}. The MKIDs are read out using a microwave frequency-domain multiplexing scheme.

\begin{table}
\centering
\begin{tabular}{ |l|l|c| } 
\hline
\rowcolor{lightgray} Parameter& Symbol [unit]& Value \\ 
\hline
Observed frequency range&$\nu_{\mathrm{obs}}$ [GHz] & 420--540\\ 
\hline
Observed wavelength range&$\lambda_{\mathrm{obs}}$ [$\mu{\rm m}$] & 555--714 \\ 
\hline
Spectral resolution &$R=\nu/\Delta\nu$ & 512\\ 
\hline
Beam FWHM at 480\,GHz & $\theta_{\rm FWHM}$ [arcmin] & 4.33 \\ 
\hline
Sky Area (Extragalactic, GP)& $\Omega$ [deg$^2$]& 305, 100 \\ 
\hline
%System temperature & $T_{\mathrm{sys}}$ [K]& 3 \\
%\hline
Number of spectrometers&$N_{\mathrm{specs}}$& 6 \\ 
\hline
Survey duration & $ t_{\rm surv}$ [hr] & 10.5 \\ 
\hline
%On-sky time / map pixel&$\tau_{\mathrm{\rm pix}}$ [s]& 0.24\\ 
%\hline
\end{tabular}
\caption{Properties of the EXCLAIM survey, including instrument parameters.}
\label{table:1}
\end{table}

\subsection{Sky Coverage}

The EXCLAIM survey will consist of a 305\,deg$^2$ extragalactic survey and at least two 100\,deg$^2$ Galactic plane surveys, accessible from a Ft. Sumner, NM flight.  The extragalactic survey is centered on the Stripe 82 (S82) field that has been mapped by multiple surveys, most notably SDSS. We can also cover parts of the HETDEX and HSC fields from this location.

The scan strategy for the surveys is constrained by the conditions to: 1) cover the declination range for S82, 2) sample the point spread function (PSF) in pixels 1/3 of the FWHM in sky drift and scan directions, and 3) work within the physical bounds of the attitude control system.  The EXCLAIM telescope is fixed to an altitude of $45^\circ$, while the azimuth is scanned through the attitude control system.  A sinusoidal scan strategy in azimuth with a peak-to-peak throw of $7^\circ$ and a period of 14\,seconds satisfies these conditions, with a declination range of $\pm 1.3^\circ$ around the celestial equator, covering S82. Throughout, the PSF width will be specified through its full width at half-max (FWHM).

\begin{figure}
    \centering
    \includegraphics[width=0.45\textwidth]{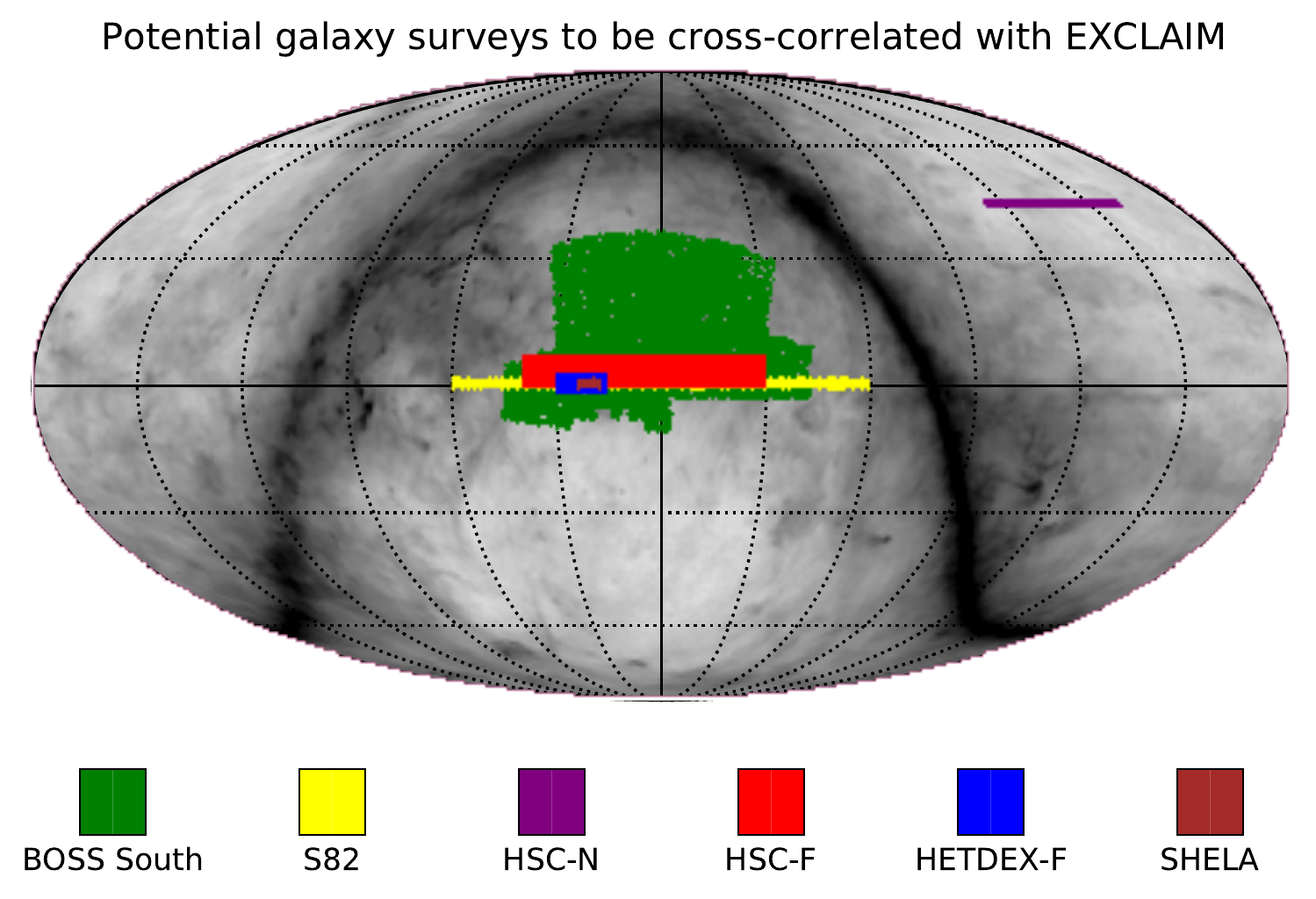}

    \caption{Potential galaxy survey regions that can be cross-correlated with the EXCLAIM survey, along with known dust map measured at 256\,GHz by the Planck satellite \citep{2013A&A...553A..96D}.  Our fiducial galaxy survey to cross-correlate with EXCLAIM is the Stripe 82 survey. This figure shows which surveys could be scanned over an EXCLAIM flight, but not all regions will be surveyed.}
    \label{fig:survey}
\end{figure}

Fig.~\ref{fig:survey} shows the nominal EXCLAIM survey footprint along with footprints for S82, HETDEX, and HSC in relation to the Milky Way foreground. The HSC-North region is visible for the first three hours of the survey, filling ${\sim} 86\%$ of the total region, whereas S82 and HSC-Fall are visible for most of the rest of the night. Furthermore, much of the S82 and HSC-Fall regions are observable twice, as they rise and then set through the EXCLAIM horizon.  Not all regions will be mapped on one flight, but there is broad flexibility for which a combination of regions can  be mapped per flight.  Note that September 1 marks the beginning of the NASA Columbia Scientific Balloon Facility campaign season for Ft. Sumner, lasting until October 20. The entire S82 region (including HSC-Fall and HETDEX-F) will be visible for any launch date during the campaign. In contrast the observable portion of HSC-N decreases with later dates until October 16, after which it is not visible at any point in the night.

We will define our fiducial survey as just the Stripe 82 region when constructing science forecasts in this analysis.  This region comprises 260\,deg$^2$.  We will also briefly investigate how these forecasts will change when altering the survey to include regions from HSC and HETDEX.

\section{Galaxy and Quasar Surveys} \label{S:galaxy}

The EXCLAIM collaboration will cross-correlate maps from the extragalactic survey with galaxy and quasar maps constructed using catalogs from the BOSS, HSC, and HETDEX surveys, as summarized in Table \ref{tab:gal_surveys}. S82, our fiducial survey, covers the celestial equator over 21:00:00 < RA < 4:00:00 (hms) and -1.25 < dec < 1.25 deg. Within S82, the Spitzer-HETDEX Exploratory Large-Area Survey (SHELA) \citep{2016ApJS..224...28P} covers 0:58:00 < RA < 1:46:00 (hms) and -1.25 < dec < 1.25 deg.

\begin{table*}
    \centering
    \begin{tabular}{|l|l|l|l|l|}
         \hline \rowcolor{lightgray} Survey
         &$\bar n$ ($h$/Mpc)$^{3}$&$A$ (deg$^2$)& $z$ range&$\sigma_z$\\
         \hline
         
         BOSS-S82 ([CII] / CO(6-5) / CO(5-4) / CO(4-3))&$10^{-6}$ / $3\times 10^{-4}$ / $10^{-2}$ / $2\times 10^{-2}$&273&2.5-3.5 / 0--0.6&0 (effectively)\\
         %QSO / CMASS / LOWZ / MAIN
         \hline
         
         HETDEX-F ([CII] / CO(6-5) / CO(5-4) / CO(4-3))&$7\times10^{-4}$ / $2.7\times 10^{-3}$ / $10^{-2}$ / $10^{-2}$&350&1.9--3.5 / 0--0.5&(6 / 2 / 2 / 2)$\times10^{-3}$\\
         \hline
         
         HETDEX-SHELA ([CII] / CO(6-5) / CO(5-4) / CO(4-3))&$7\times10^{-4}$ / $2.7\times 10^{-3}$ / $10^{-2}$/ $10^{-2}$&34&1.9--3.5 / 0--0.5&(6 / 2 / 2)$\times10^{-3}$\\
         \hline
         
         HSC-N ([CII] / CO(6-5) / CO(5-4) / CO(4-3))&$3.6\times10^{-3}$ / 0.11 / 0.38 / 0.38&125&2.5--3.5 / 0--0.6&0.03 / 0.03 / 0.1 / 0.1\\
         \hline
    \end{tabular}
    \caption{Properties of large-scale structure surveys, where $\bar n$ is the mean number density of galaxies, $A$ is the overlapping sky area, and $\sigma_z$ is the redshift precision of the galaxies.  For the $z$-range column, the left interval is for [CII] or QSO, and the right interval is for the CO lines or galaxies.  For the other columns, the values relate to the corresponding surveys in the first column. }
    \label{tab:gal_surveys}
\end{table*}

Given the BOSS redshift coverage, for S82 we will cross-correlate EXCLAIM maps with the QSO survey for $2.5<z<3.5$ \citep{Eftekharzadeh2015}, MAIN for $0<z<0.2$ \citep{guo2015}, LOWZ for $0.2<z<0.4$ \citep{manera2015}, and CMASS for $0.4<z<0.6$ \citep{reid2016}. The latter three surveys overlap with the targeted CO lines, and the QSO survey will overlap with the targeted [CII] line. The redshift ranges of these samples overlap with the relevant ranges for the emission lines of our survey, leading to mean number densities of $10^{-6}$, $3\times 10^{-4}$, $10^{-2}$, and $2\times 10^{-2}$ ($h$/Mpc)$^{3}$ for the redshift ranges of [CII], CO(6-5), CO(5-4), and CO(4-3), respectively. Redshift errors in these surveys [$\Delta z\sim(0.5-2)\times 10^{-3}$] are less than the $\Delta z\sim(2-7)\times 10^{-3}$ of EXCLAIM's channels, where the lower and upper ends of the ranges correspond to the CO and [CII] surveys, respectively.

The HETDEX Fall and SHELA surveys will also cover regions within S82. HETDEX will obtain redshifts via Ly$\alpha$ emission over the range $1.9<z<3.5$ and [OII] over the range $z<0.5$ \citep{Hill2021}. The HETDEX Visible Integral-Field Replicable Unit Spectrograph (VIRUS) instrument will measure these spectra through $R\sim 800$ spectroscopy, resulting in small redshift errors we model as $\sigma_z = (1+z) / R$. We consider two HETDEX fields: the fall survey (HETDEX-F) and SHELA. We model the HETDEX-F mean galaxy number density as $7\times 10^{-4}$, $2.7\times 10^{-3}$, $10^{-2}$, and $10^{-2}$ ($h$/Mpc)$^{3}$ for the redshift ranges of [CII], CO(6-5), CO(5-4), and CO(4-3), respectively.

Two regions within the HSC-Wide survey overlap with EXCLAIM's potential footprint: Fall (HSC-F) and North (HSC-N). HSC-F covers much of the S82 region, whereas HSC-N may be visible during the beginning of the EXCLAIM survey campaign. Photometric catalogs within Data Release 2 feature mean galaxy number densities of $3.6\times 10^{-3}$, 0.11, 0.38 and 0.38 ($h$/Mpc)$^{3}$, for CII, CO(6-5), CO(5-4), and CO(4-3), respectively \citep{Nishizawa2020}. We take a photometric redshift uncertainty of $\sigma_z\sim 0.03$ for $0<z<0.6$ and 0.10 for $2.5<z<3.5$.

\section{Intensity Statistics Models} \label{S:model}

In this section, we present models for the statistics of the CO and [CII] intensity fields that EXCLAIM will measure.  Specifically, we model the intensity power spectra for CO and [CII], including the auto-power and the cross-power with quasars and galaxies.  We also review the formalism for the CVID, which probes the full luminosity function for a line.  These are the models we will consider as observables in the forecast section of this paper.

\subsection{Intensity Models}

Here, we define the intensity models for CO and [CII].  The intensity for line $X$ is given as
\begin{eqnarray}
I_\nu^X=\frac{1}{4\pi\nu_X}\frac{c}{H(z_X)}\int dM\,n(M,z) f_{\rm duty}(M, z)L_X(M,z)\, ,
\end{eqnarray}
where $\nu_X$ is the rest frequency for line $X$, $z_X=\nu_X/\nu-1$ is the emission redshift of the line, $H(z)$ is the Hubble rate of expansion at redshift $z$, $n(M,z)$ is the halo mass function, $f_{\rm duty}(M,z)$ is the duty cycle giving the fraction of halos at a given $M$ and $z$ that form stars, and $L_X(M,z)$ is the luminosity-mass relation at redshift $z$.  The quantity $n(M,z)dM$ gives the number density of halos within a halo mass bin $[M,M+dM]$.  In our forecasts, we assume a fixed cosmology, a Sheth-Tormen halo mass function \citep{2002MNRAS.329...61S}, and a fixed $f_{\rm duty}$, which means that the variation in the intensity for each line will come from  $L_X(M,z)$.

We calculate the CO luminosity-mass relation following the formalism of \citet{Li2016} and \citet{Keating2020}.  While there are many CO line models in the literature (e.g. see Table 3 from \citet{Li2016} and \citet{2022ApJ...929..140Y} for many examples of CO models), predicted CO luminosities can differ by as much as 2 orders of magnitude. In this work, we choose the \citet{Li2016} CO model because it is consistent with the recent CO LIM measurement from \citet{Keating2020}. We begin with the catalog of relations between halo mass and SFR from $0<z<8$ as described in \citet{Behroozi2013}. The IR luminosity is calculated as $L_{\rm IR} = {\rm SFR} / (\delta_{MF}\times 10^{-10})$ where $\delta_{MF}$ is the initial mass function taken to be $\delta_{MF}=1.0$ \citep{1998ApJ...498..541K, carilli2013}. The luminosity functions for CO lines are calculated as
\begin{equation}
    \log L_{\rm IR} = \alpha_{\rm CO}(J) L_{\rm CO}'(J) + \beta_{\rm CO}(J)
    \label{eqn:L_IR_Lp_CO}
\end{equation}
where $L_{\rm CO}'$ is the CO line luminosity in units of K km s$^{-1}$ pc$^2$ and $\alpha_{\rm CO}$ and $\beta_{\rm CO}$ are calculated for specific J lines through Table 4 in \citet{Kamenetzky2016}.  Based on their model, we set $f_{\rm duty}=1$. Note that the final step relating $L_{\rm IR}$ to $L_{\rm CO}'$ was only included for $J{=}1$ in \citet{Li2016}; the generalization to higher-J lines was implemented in \cite{Keating2020}.

We use the [CII] emission line model from \citet{2019MNRAS.488.3014P}.  Please refer to this reference for a detailed description.  Although we are aware that other very different [CII] models exist (e.g. see \citet{2015MNRAS.450.3829Y}, \citet{2016MNRAS.461...93P}, and \citet{2022ApJ...929..140Y}), we chose this model because \citet{2019MNRAS.488.3014P} is calibrated to the most recent [CII] LIM measurement \citep{2019MNRAS.489L..53Y} that EXCLAIM is designed to refine. To construct this model, $L_\mathrm{[CII]}(M,z=0)$ is modeled with an analytical formula given below that is then calibrated by abundance matching between a Sheth-Tormen halo mass function and the measured luminosity function for line $X$, $\phi(L_X)$, at $z=0$.  The abundance matching was performed using the [CII] luminosity function from \citet{2017ApJ...834...36H}, while the luminosity is scaled to higher redshifts assuming the relation $L_\mathrm{[CII]}\propto {\rm SFR}^\alpha$ where ${\rm SFR}$ is the star formation rate.  ${\rm SFR}(M,z)$ is scaled in redshift based on measurements of the SFR density measured in \citet{2014ARA&A..52..415M} and $\alpha$ is set based on the [CII] intensity limit at $z=2.6$ in \citet{2019MNRAS.489L..53Y}. The resulting [CII] luminosity-mass relation derived in \citet{2019MNRAS.488.3014P} is given by
\begin{eqnarray}
L_\mathrm{[CII]}(M,z)&=&\left(\frac{M}{M_1}\right)^{0.49}e^{-N_1/M}\nonumber\\
&&\times\left[\frac{(1+z)^{2.7}}{1+[(1+z)/2.9]^{5.6}}\right]^{1.79}L_\odot\, ,
\end{eqnarray}
in units of solar luminosity ($L_\odot$) with $M_1=2.39\times 10^{-5}\,M_\odot$ and $N_1=4.19\times 10^{11}\,M_\odot$ and the redshift dependence comes from the SFRD redshift evolution from \citet{Behroozi2013}.  We will forecast constraints on the parameters for a variant of this model.  Specifically, we substitute the $M_1$ parameter with an amplitude $A$, such that our fiducial [CII] model is
\begin{eqnarray}\label{E:lc2mod}
L_\mathrm{[CII]}(M,z)=A(z)\left(\frac{M}{N_1}\right)^\beta e^{-N_1/M}L_\odot\, ,
\end{eqnarray}
where the parameters to be fitted are $\{A,N_1,\beta\}$.  In this case, we set $f_{\rm duty}=1$, as assumed by \citet{2019MNRAS.488.3014P}.

A more pessimistic model for [CII] emission is one from \citet{2022ApJ...929..140Y}, which is an empirical model for the [CII] halo luminosity-mass relation based on results from the Santa Cruz semi-analytic model (SAM) for galaxy formation \citep{1999MNRAS.310.1087S,2015MNRAS.453.4337S} coupled with the DESPOTIC spectral synthesis model \citep{2014MNRAS.437.1662K,2014MNRAS.442.2398P,2019MNRAS.482.4906P}. The empirical form for this model at $1.0<z<4.0$ is given as
\begin{eqnarray} \label{E:lc2mod2}
L_\mathrm{[CII]}(M,z)&=&2N\left(\frac{M}{M_\odot}\right)\left[\left(\dfrac{M/M_\odot}{M_1}\right)^{-\alpha}\right.\nonumber\\
&&+\left.\left(\frac{M/M_\odot}{M_1}\right)^{\beta}\right]^{-1}L_\odot\, ,
\end{eqnarray}
where $\log[N(z)]=-0.907e^{-z/0.867}-3.04$, $\log[M_1(z)]=12.11z^{-0.04105}$, $\alpha(z)=1.35+0.450z-0.0805z^2$, and $\beta(z)=2.57e^{-z/1.55}+0.0575$.  This model also accounts for a non-unity $f_{\rm duty}$ for $z<4$, given by
\begin{eqnarray}
f_{\rm duty}(M,z)=\left[1+\left(\frac{M/M_\odot}{M_2}\right)^\gamma\right]^{-1}\, ,
\end{eqnarray}
where $\log[M_2(z)]=11.73+0.6634z$ and $\gamma(z)=1.37-0.190z+0.215$.  Although in principle we could allow the parameters in $f_{\rm duty}$ to vary, we will assume those parameters to be fixed in our forecasts, as was assumed for the \citet{2019MNRAS.488.3014P} forecasts.
In our forecasts, we will consider this model a pessimistic alternative to the \citet{2019MNRAS.488.3014P} model.

\subsection{Power Spectrum Models} \label{S:Pkmod}

EXCLAIM baseline analysis employs only the cross-power spectra with galaxies because of its utility in controlling and rejecting the effects of foregrounds. In addition, cross-power spectra do not have noise bias, allowing us to conduct a shallow survey that can access large scales in the linear clustering regime. In this section, we model the auto and cross-power spectra between EXCLAIM intensity maps and galaxies from the surveys we will use in our analysis.  Primary science comes from the cross-power, but the errors of the cross-power depend on the auto-power spectrum and its variance contributions.

\subsubsection{Cross-power spectrum}

Here, we define an X-Tr cross-power spectrum model, where X is the emission line and Tr is the compact object LSS tracer, which is either the quasar or galaxy sample.  The relevant pairs for our forecasts are [CII]-quasar and CO-galaxy. We have

\begin{eqnarray}
    P_\mathrm{X-Tr}(\textbf{k},z) = P_\mathrm{X-Tr}^{\rm clust}(\textbf{k},z) + P_\mathrm{X-Tr}^{\rm shot}(z)
\end{eqnarray}
where $\textbf{k}=(k_\parallel,\textbf{k}_\perp)$ and the clustering cross-power spectrum is given by
\begin{eqnarray}
    P_\mathrm{X-Tr}^{\rm clust}(\textbf{k},z) &=& I_\mathrm{X}(z) b_\mathrm{X}(z) b_{\rm Tr}(z)\nonumber\\
    &&\times F_\mathrm{X}(\textbf{k},z) F_{\rm Tr}(\textbf{k},z) P_m(k,z).
\end{eqnarray}
$P_m(k,z)$ describes the linear matter power spectrum, $k=\sqrt{k_\parallel^2+|\mathbf{k}_\perp|^2}$, and $F_\mathrm{X}$ and $F_\mathrm{Tr}$ describe redshift space distortions in the emission line and LSS tracer overdensities, respectively, given by
\begin{eqnarray}
    F_\mathrm{X} = \left ( 1 + \frac{f_g}{b_\mathrm{X}} \frac{k_{||}^2}{k^2} \right ) {\rm exp} \left [ - \frac{k_{||}^2 \sigma_v^2}{2 H^2(z)} \right ]
\end{eqnarray}
for X and correspondingly for Tr. $f_g = [\Omega_m(z)]^{0.55}$ describes the growth rate and $\sigma_v$ describes velocity dispersion of the source fields. $k_{||}$ describes the wavenumber projected along the line of sight.  $b_\mathrm{X}$ is the clustering bias for the line luminosity, which is given by
\begin{eqnarray}
b_\mathrm{X}(z) = \frac{\int dM\,n(M,z) b_h(M,z) L_\mathrm{X}(M,z)}{\int dM\,n(M,z) L_\mathrm{X}(M,z)}\, ,
\label{eqn:IMbias}
\end{eqnarray}
where $b_h(M,z)$ is the halo clustering bias consistent with the assumed halo model. Note that all power spectra refer to the two-point statistic on the sky, corrected for instrument angular and frequency resolution effects, which appear in the instrument noise through Eq.\,\ref{eqn:resolution}.

The line-emitter $\times$ LSS-tracer shot cross-power is a combination of a one-halo term and a line shot noise term \citep{2017MNRAS.470.3220W, 2021JCAP...05..068S}
\begin{eqnarray}
P_\mathrm{X-Tr}^{\rm shot}(k,z)=P_\mathrm{X-Tr}^{\rm 1-halo}(k,z)+P_\mathrm{X-Tr}^{\rm line\,shot}(z) \, .
\end{eqnarray}
The one-halo term is produced by halos that host both line emitters and the LSS tracers, while the line shot noise term is produced by LSS tracers that are also line emitters.  EXCLAIM will measure power spectra over scales with $k_{\rm max}\sim 0.5 h$/Mpc, which is much less than the scales where the one-halo term decays ($k\sim 10h$/Mpc).  Thus, we are in the regime where the one-halo term is uniform for the scales of interest in our analysis, allowing us to write it as
\begin{eqnarray}
    P_\mathrm{X-Tr}^{\rm 1-halo}(z) = \frac{I_\mathrm{X}}{\bar{n}_{\rm Tr}}\, ,
\end{eqnarray}
where $\bar{n}_{\rm Tr}$ is the number density of the LSS tracer given in Table \ref{tab:gal_surveys}.  The line shot noise term is given by
\begin{eqnarray}
P_\mathrm{X-Tr}^{\rm line\,shot}(z)=\frac{I_\mathrm{X}}{\bar{n}_{\rm Tr-X}}\, ,
\end{eqnarray}
where $\bar{n}_{\rm Tr-X}$ is the number density of LSS tracers that emit line $X$. Since both terms are now scale-independent, we can then write the full shot cross-power as
\begin{eqnarray}
    P_\mathrm{X-Tr}^{\rm shot}(z) = \frac{I_\mathrm{X} f_\mathrm{X-Tr}^{\rm shot}}{\bar{n}_{\rm Tr}}\, ,
\end{eqnarray}
  where $f_\mathrm{X-Tr}^{\rm shot}$ is a number that absorbs both the 1-halo and line shot noise terms which can be written as $1+f_{\rm Tr}$ where $f_{\rm Tr}$ is the fraction of the total surface brightness $I_\mathrm{X}$ produced by the tracer population.  
  %In our fiducial model, we assume the fraction of the total surface brightness produced by the tracers is very small; specifically, we set $f_{\rm Tr}=10^{-6}$. 
  To estimate $f_{\rm Tr}$, we find the ratio of the integrated luminosity down to the effective cutoff mass $M_{\rm cutoff}$ of the redshift survey compared to the integrated luminosity down to $M_{\rm min}$ for the emitting gas. We take $M_{\rm min}$ for the line emission to be $10^{10}\,M_\odot$ for both CO \citep{2016ApJ...817..169L, 2013ApJ...768...15P} and CII \citep{2019MNRAS.488.3014P}, and find mild sensitivity to this assumption. The mass cutoff for the redshift surveys are taken to be $10^{11.6}\,M_\odot$ for SDSS-MAIN \citep{2015MNRAS.453.4368G}, evaluation of Eq. 5 of \citet{2015MNRAS.447..437M} for LOWZ/CMASS (based on $\bar n$ for the survey), and $10^{12}\,M_\odot$ with $f_{\rm duty} {\approx} 0.01$ from \citet{2015MNRAS.453.2779E} for BOSS-QSOs. Numerically these are evaluated as is $\{0.73, 0.3, 0.34, 1.27 \times 10^{-3} \}$ for CO from MAIN, LOWZ, CMASS, and [CII] from QSOs, respectively. 

\subsubsection{Line auto-power spectrum}

The line auto-power spectrum for our analysis is the sum of clustering and shot noise from the line, interlopers at other redshifts, instrument noise, and variance from the Milky Way, as given by
\begin{eqnarray}
    P_\mathrm{X}(\textbf{k},z) &=& P_\mathrm{X}^{\rm clust}(\textbf{k},z) +
    P_\mathrm{X}^{\rm shot}(z) +
    P_\mathrm{X}^\mathrm{IL}(\textbf{k},z)\nonumber\\
    &&+ \frac{P_N(z)}{W_\mathrm{X}(\textbf{k},z)} + P_{\rm MW}(k)\, ,
\end{eqnarray}
with the clustering auto-power spectrum given by
\begin{eqnarray}
    P_\mathrm{X}^{\rm clust}(\textbf{k},z) = I_\mathrm{X}^2(z) b_\mathrm{X}^2(z) F_\mathrm{X}^2(\textbf{k},z) P_m(k,z)\, ,
\end{eqnarray}
and the auto-shot power $P_\mathrm{X}^{\rm shot}$, sourced by both the 1-halo shot noise and the shot noise of the galaxy emitters, given by
\begin{eqnarray}
P_\mathrm{X}^{\rm shot}(z)&=&f_\mathrm{X}^{\rm shot}\left[\frac{1}{4\pi\nu_\mathrm{X}}\frac{c}{H(z_\mathrm{X})}\right]^2\int dM\,f_{\rm duty}(M)n(M)\nonumber\\
&&\times L_\mathrm{X}^2(M)\, ,
\end{eqnarray}
where $f_\mathrm{X}^{\rm shot}$ accounts for both the halo and galaxy shot noise terms. The instrument noise $P_N$ in a given redshift bin is given by 
\begin{eqnarray}\label{eqn:P_N}
    P_N = \frac{{\rm NEI}^2_{\rm eff}V_{\rm surv}}{t_{\rm surv} N_{\rm specs}}.
\end{eqnarray}
$V_{\rm surv}$ and $t_{\rm surv}=10.5$ hours describe the spatial volume per bin and the survey observing time, respectively. $N_{\rm specs}=6$ (for EXCLAIM) describes the number of spectrometers.   The effective NEI is the inverse variance-weighted sum of the squared NEI per channel, which minimizes the variance per pixel in the line intensity map \citep{2021JATIS...7d4004S} and is summarized for each redshift interval in Table\,\ref{tab:EXCLAIM_PARAMS}. 

The resolution window function $W_\mathrm{X}(\textbf{k},z)$ is given by
\begin{eqnarray}
W_\mathrm{X}=\exp\left[-\left(\sigma_\perp^2|\mathbf{k}_\perp|^2+\sigma_\parallel^2k_\parallel^2\right)\right]\, ,
%W_\mathrm{X}=\exp\left\{-k^2\left[\sigma_\perp^2\left(\frac{|\mathbf{k}_\perp|}{k}\right)^2+\sigma_\parallel^2\left(\frac{k_\parallel}{k}\right)^2\right]\right\}\, ,
\label{eqn:resolution}
\end{eqnarray}
where resolutions in the transverse and radial directions are given by
\begin{eqnarray}
\sigma_\perp=\chi(z)\sigma_{\rm b}\, ;\,\,\,\,\,\,\sigma_\parallel=\frac{c(1+z)}{H(z)R}\, ,
\end{eqnarray}
where $\chi$ is the comoving distance, $\sigma_{\rm b} = \theta_{\rm FWHM} / \sqrt{8 \ln 2}$ is the Gaussian width of the telescope PSF, and $R=\nu/\delta\nu$ is the spectral channel resolution \citep{2019PhRvD.100l3522B}. 

We also consider interlopers from other lines.  In particular, mapping [CII] is contaminated by all the CO lines in the auto-power spectrum.  The interloper power spectrum is given by
\begin{equation} \label{E:Pinterlopers}
    P_{\rm CII}^\mathrm{IL}(\textbf{k},z) = \sum_i \frac{1}{\alpha_{||}(z_i) \alpha_{\perp}^2(z_i)} P_i \left (\frac{k_{||}}{\alpha_{||}(z_i)},\frac{k_{\perp}}{\alpha_{\perp}(z_i)} , z_i \right)\, ,
\end{equation}
\citep{2016ApJ...825..143L}. Here $z_i$ is the redshift of the $i$th interloper line emitters for primary line emission at redshift $z$ and $P_i(\mathbf{k},z)$ is the intensity auto-power spectrum for interloper $i$. $\alpha_{||}=\sigma_\parallel(z_i)/\sigma_\parallel(z)$ and $\alpha_{\perp}=\sigma_\perp(z_i)/\sigma_\perp(z)$ are the parallel and perpendicular $k$-mode distortion factors caused by assuming the interloper redshifts as the redshift of the target line emitters. In our analysis, we can treat the interlopers as unknown and marginalize them over the interloper values.  Furthermore, we can compare these to the case where interlopers are not present in the measurement.  We will consider both cases among our power spectrum and intensity forecasts in Sec.~\ref{S:stats}.

\subsubsection{Continuum emission}

Finally, we also know that EXCLAIM maps will contain continuum emission from the Milky Way galaxy and extragalactic emission, specifically the CIB from young, dusty stars. %Recent work has shown how this foreground can be mitigated in a tomographic spherical harmonic analysis \citep{2022MNRAS.tmp.1280A}. 
It was shown in \citet{2019ApJ...872...82S} that while the broadband measurements of [CII] line intensity performed by cross-correlating broadband intensity maps from the Planck satellite with BOSS quasars in \citet{2018MNRAS.478.1911P} and \citet{2019MNRAS.489L..53Y} are limited in the removal of correlated foregrounds like CIB, an instrument with a moderate spectral resolution like EXCLAIM would be able to remove foregrounds by removing the lowest few $k_\parallel$-modes for which continuum emission is dominant.  
%A more optimal way to remove contaminated modes is to perform a principle component analysis (PCA), in which a singular value decomposition (SVD) is performed on a simulated map, including continuum foregrounds, to find the modes with the highest variance \citep{2013MNRAS.434L..46S,2018MNRAS.476.3382A}.  We can then apply this to data by subtracting these modes directly.  However, in our forecast, we will simply subtract the first $k_\parallel$-mode, which is dominated by continuum foregrounds, from our forecast.  

Using the Python version \citep{2017MNRAS.469.2821T} of the Planck Sky Model (PSM) \citep{2013A&A...553A..96D}, we simulate an EXCLAIM map including a finite PSF, instrumental noise, and Galactic continuum emission.  We then perform a Fast Fourier Transform to find the $k_\parallel$ corresponding to the lowest $k_\parallel$ mode.  We find that removing the lowest $k_\parallel$-mode in this simulated EXCLAIM map is sufficient to make the resulting map consistent with white noise, which is expected for a map containing only instrumental noise.  We will perform our forecast with the corresponding range of $k_\parallel$ removed. We will consider forecasts both with and without this cut in our results. We note that instrumental terms such as passband spectral response require high stability, and empirical approaches (e.g. \citet{2015ApJ...815...51S}) can be employed to isolate continuum modes in the real data. \citet{2021JATIS...7d4004S} describes calibration plans for the EXCLAIM instrument, which employs a reference emitter for monitoring stability.
\iffalse
\begin{figure*}
    \centering
    \includegraphics[width=0.45\textwidth]{figs/PySM_map_original.pdf}
    \includegraphics[width=0.45\textwidth]{figs/PySM_map_rm_lowest_kpar.pdf}\\
    %\includegraphics[width=0.45\textwidth]{figs/PySM_map_sum_per_kpar.pdf}
    \caption{A map over XXX$\times$XXX degrees containing the PSM-simulated Milky Way emission and instrumental noise and PSF smoothing from EXCLAIM before (left) and after (right) subtracting the lowest $k_\parallel$-mode.  These panels show that removing the lowest $k_\parallel$-mode should be sufficient to remove continuum foreground emission. \arp{Need to confirm map areas, add units, and change xy tick labels.}}
    \label{fig:psm_map}
\end{figure*}
%  Bottom panel: The total intensity ascribed to each $k_\parallel$-mode.
\fi

\subsubsection{Tracer auto-power spectrum}

The LSS tracer auto-power spectrum is given by
\begin{eqnarray}
P_{\rm Tr}(\mathbf{k},z)=P_{\rm Tr}^{\rm clust}(\mathbf{k},z)+\frac{1}{\bar{n}_{\rm Tr}}\, ,
\end{eqnarray}
where the first term is the quasar clustering auto-power spectrum given by
\begin{eqnarray}
P_{\rm Tr}^{\rm clust}(\mathbf{k},z)=b_{\rm Tr}^2(z)F_{\rm Tr}^2(\mathbf{k},z)P_m(k,z)\, ,
\end{eqnarray}
and the second term is the shot noise.  Note that the expressions are similar for both the quasar and galaxy surveys.

\subsubsection{Power spectrum analysis model}

\begin{figure*}
    \centering
    \includegraphics[width=0.45\textwidth]{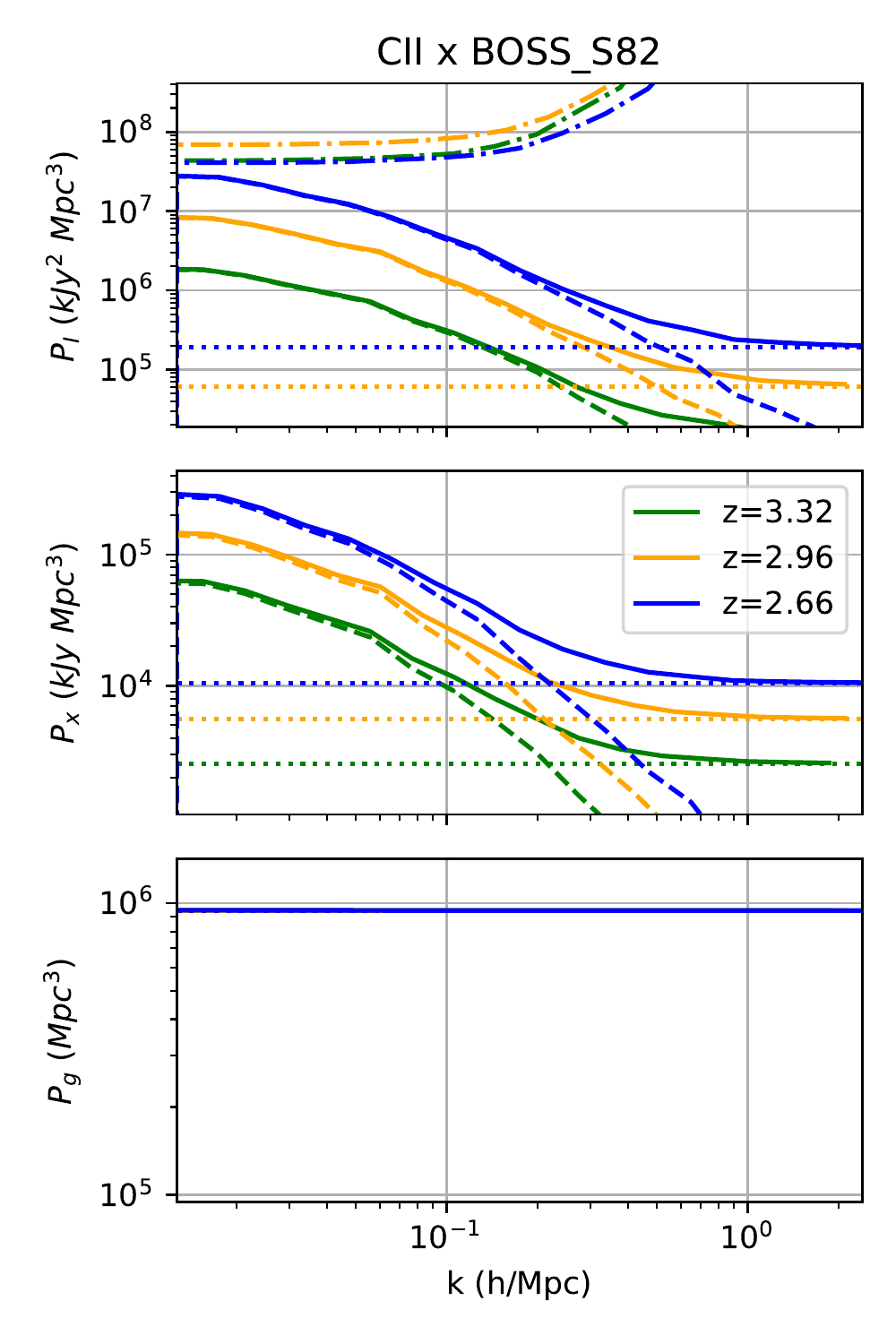}
    \includegraphics[width=0.45\textwidth]{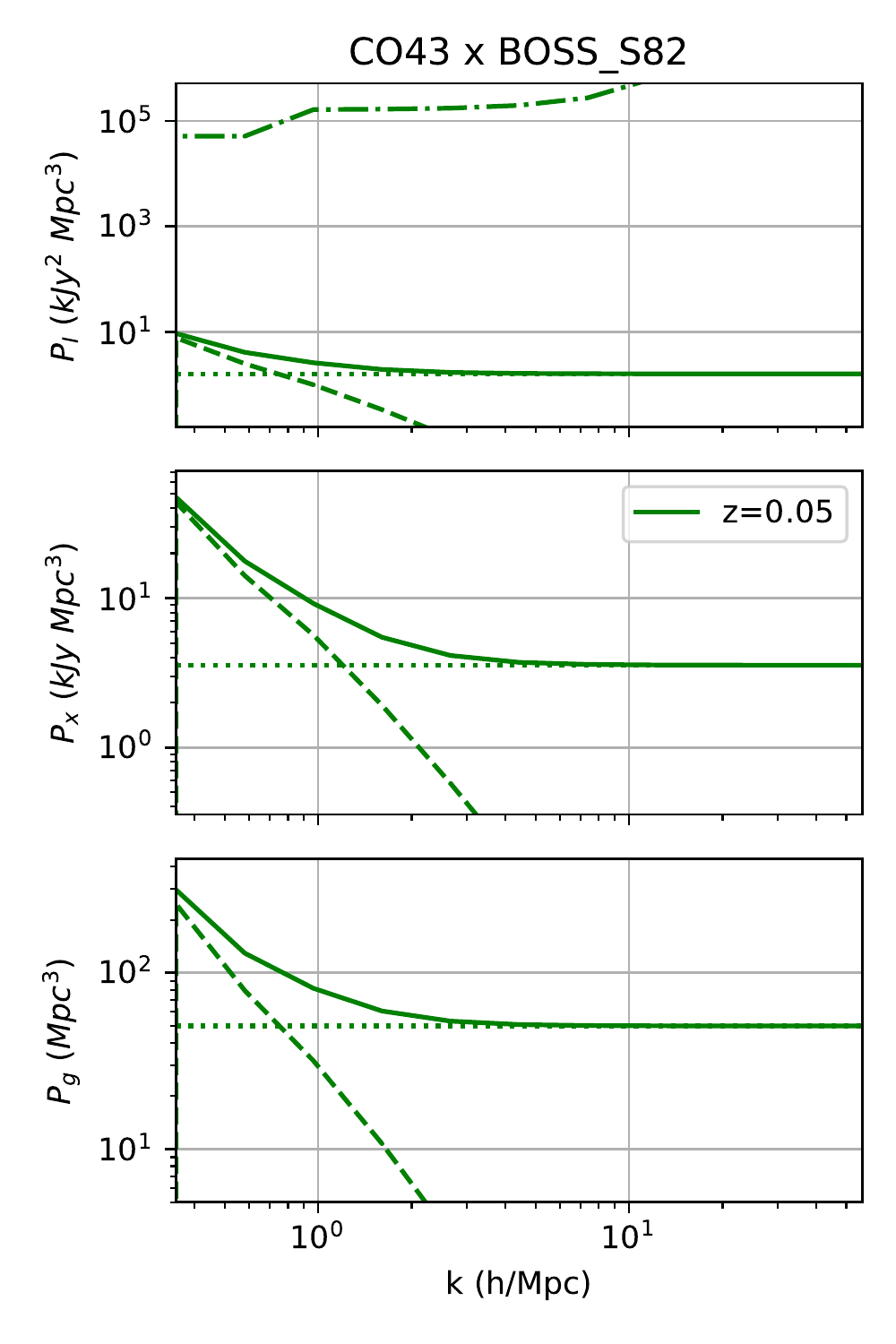}
    \caption{EXCLAIM-tracer auto and cross power spectra with [CII]-quasar on the left and CO(4-3)-galaxy on the right using Stripe 82.  The top, middle, and bottom panels show the intensity auto, intensity-tracer cross, and tracer auto-power, respectively.  All power spectra are spherically averaged in $k$-space.  The dashed (dotted) curves show the clustering (shot) power spectra, the solid curves show the total power spectra, and the dash-dotted curve shows instrument noise. The quasar power spectrum (bottom left) is dominated by shot noise, while the [CII] (top left) and CO(4-3) (top right) auto-power spectra are dominated by instrument noise, though less for [CII] than CO(4-3).  Note that the discontinuity at $k\sim 1h$/Mpc in the CO(4-3) instrumental noise curve is caused by the rectangular EXCLAIM field, in that its short side corresponds to this wavenumber.}
    \label{fig:pk}
\end{figure*}

Fig.~\ref{fig:pk} shows fiducial models of the auto- and cross-power spectrum for both the [CII] $\times$ S82 SDSS/BOSS quasars joint analysis and the CO(4-3) $\times$ S82 SDSS/BOSS galaxies joint analysis.  These plots show that for the [CII] analysis, the intensity auto and cross-power spectra are dominated by clustering for wavenumbers $k\lesssim 0.2\,h$/Mpc, while the quasar auto-power is primarily shot noise. Alternately, for the CO(4-3) analysis, the intensity auto-power is dominated by shot noise, while the galaxy auto and cross-power spectra show the clustering contributions for wavenumbers $k\lesssim 1\,h$/Mpc.

EXCLAIM's primary science constraint employs the cross-power between the intensity data cube and a tracer population. This choice is made to simplify the treatment of foregrounds and interlopers, which additively bias the auto-power. In the cross-power, these terms add noise, so they need a best-effort minimization or marginalization rather than airtight rejection. Additionally, the large survey area is optimized for cross rather than auto-power \citep{2021PhRvD.104h3501O}. The fiducial analysis assumes a data vector consisting of only the cross-power $\bar{P}_\mathrm{X-Tr}(k)$ for each redshift bin, where $\bar{P}(k)$ is the spherically averaged power spectrum
\begin{eqnarray}
\bar{P}(k)=\frac{1}{2}\int_{-1}^1d\mu\,P(k,\mu)\, ,
\end{eqnarray}
where $\mu=\cos\theta$ with $\theta$ being the angle of the wavevector $\mathbf{k}$ from the line of sight.  We perform a standard Fisher analysis in each redshift bin to predict errors in $b_\mathrm{X}I_\mathrm{X}$ for all the [CII] and CO lines.  In the Fisher analysis, we define a covariance matrix using standard methods  \citep{2019PhRvD.100l3522B}.  The parameters used when constructing the Fisher matrix for measuring intensities are $\{bI,\,bf_{\rm shot},\,\bar{P}_\mathrm{IL}(k_i)\}$.  For the two cases, we consider (see Sec.~\ref{S:stats}), ``no interlopers'' sets $\bar{P}_\mathrm{IL}(k_i)=0$ and ``marginalized interlopers'' treats each $\bar{P}_\mathrm{IL}$ for binned $k_i$ as an independent parameter that is marginalized.  Note that this case is more conservative than treating the scale-dependence of $\bar{P}_\mathrm{IL}$ as a known quantity to then marginalize over the intensity.  We avoid attempting this since the redshift dependence of the emission line intensities is not well known. 

\subsection{CVID Models}
Here we review the formalism for the voxel intensity distribution (VID) along with the CVID, the natural extension of the VID formalism that is implemented in our forecasts. This is slightly modified from \citet{Breysse2017} and \citet{Breysse2019}.

Because an intensity map is not generically a purely Gaussian random field, a substantial amount of information is not captured by the power spectrum.  In particular, information about the mass-luminosity relation of the target galaxies can be inferred by examining the one-point probability distribution of brightness temperature measurements.  This quantity has been referred to as the Voxel Intensity Distribution, or VID.  We can look at either the one-dimensional PDF of a single random field or the conditional PDF of two fields.

\subsubsection{Auto-VID}

Let us begin with the assumption that we know exactly how many line emitters there are in each voxel.  For voxels containing no galaxy redshift measurements, the PDF $\p_0(I)$ of intensity $I$ is simply
\be
\p_0(I)=\delta_D(I),
\ee
where $\delta_D$ is the Dirac delta function, and we refer to probability distributions by $\p$ to distinguish them from power spectra.  For voxels that contain exactly one emitter, we have
\be
\p_1(I)\propto\frac{1}{\overline{n}}\frac{dn}{dL},
\ee
where $dn/dL$ is the luminosity function, which can be computed from $L(M,z)$, the mass function, and $f_{\rm{duty}}$.  Voxels with two or more sources have PDFs computed from the recursion relation
\be
\p_N(I)=\int \p_{N-1}(I')\p_1(I-I')dI'=\p_{N-1}\circ\p_1(I),
\label{PN_conv}
\ee
where the circle operator denotes a convolution.  The total VID is then
\be
\p(I)=\sum_{N=0}^\infty \p_N(I)\p(N),
\ee
where $\p(N)$ is the probability of observing a voxel that contains $N$ sources.  In the classical $P(D)$ analysis of unclustered sources, $\p(N)$ is simply the Poisson distribution.  However, we expect a distribution with a long tail for cosmological sources. \citet{Breysse2017} used a mixed lognormal and Poisson distribution for $\p(N)$ which we expect to be a reasonably accurate approximation, though better forms likely exist (e.g., \citealt{Leicht2019}).

In practice, there will be several sources of emission in a given voxel in addition to the target emission line.  Assuming these different components are uncorrelated, we can compute the full VID by understanding that Eq. (\ref{PN_conv}) holds for the sum of any two fields.  For example, if we have two lines with PDFs $\p_{\rm{CII}}(I)$ and $\p_{\rm{ CO}}(I)$, then the full observed VID will be
\be
\p(I)=\p_{\rm{CII}}\circ\p_{\rm{CO}}(I).
\label{VID_FG}
\ee
Note that Eq. (\ref{VID_FG}) only holds for \emph{uncorrelated} fields.  ``Foreground" emission from sources like the CIB will correlate with the signal and need to be modeled more carefully.

\subsubsection{Conditional VID}

As seen in Eq. (\ref{VID_FG}), VID measurements using a single field suffer from the same foreground issues as auto-power spectrum measurements.  \cite{Breysse2017} showed that this issue can be mitigated to an extent using masking methods commonly used for foreground cleaning. However, it would be preferable to have an analogy to the cross-correlation, which can be used to isolate specific redshifts. \citet{Breysse2019} describe such a one-point analogy, termed the ``conditional voxel intensity distribution" or CVID.

As in Section \ref{S:Pkmod}, assume that each voxel has a measurement of the total intensity $I$ and the number of directly detected galaxies $N_{\rm{det}}$ at our target redshift.  Let us further assume that our map consists of a signal CII line and some foreground CO lines.  The intensity of the CII line will be correlated with $N_{\rm{det}}$, as both are coming from the same redshift, while the intensity of the CO line will be uncorrelated.  We can then examine the \emph{conditional} PDF
\be
\p(I|\Ndet)=\p_{\rm{CII}}(I|\Ndet)\circ\p_{\rm{CO}}(I).
\label{CVID_conv}
\ee
The conditional PDF $\p_{\rm{CII}}(I|\Ndet)$ for a given line can be computed similarly to the standard VID.  One possible prescription for doing so can be found in \citet{Breysse2019}.

Eq. (\ref{CVID_conv}) is difficult to work with on its own, but we can put it in a more useful form by applying the Fourier convolution theorem:
\be
\pf(\If|\Ndet)=\pf_{\rm{CII}}(\If|\Ndet)\pf_{\rm{CO}}(\If),
\label{FT_CVID}
\ee
where $\If$ is the Fourier conjugate of intensity (units (Jy/sr)$^{-1}$) and
\be
\pf(\If)\equiv\int_{-\infty}^\infty\p(I)e^{\im I\If}dI.
\ee
If we examine the intensity PDFs of voxels with two different, known, $\Ndet$ values, we can define the quantity
\be
\R_{\Ndet^1,\Ndet^2}\equiv\frac{\pf(\If|\Ndet^1)}{\pf(\If|\Ndet^2)}.
\label{Rdef}
\ee
We can show by combining Eqs. (\ref{FT_CVID}) and (\ref{Rdef}) that $\R$ depends only on the signal PDFs, not on the foreground contamination:
\be
\R_{\Ndet^1,\Ndet^2}=\frac{\pf_{\rm{CII}}(\If|\Ndet^1)\pf_{\rm{CO}}(\If)}{\pf_{\rm{CII}}(\If|\Ndet^2)\pf_{\rm{CO}}(\If)}=\frac{\pf_{\rm{CII}}(\If|\Ndet^1)}{\pf_{\rm{CII}}(\If|\Ndet^2)}.
\ee
Thus, just like the cross-power spectrum, $\R$ is a quantity which isolates only emission from the same redshift as a given galaxy population.

\subsection{VID Measurements}
In practice, we cannot measure the true, continuous VID from a real map, as we only have access to a finite number of voxels.  We instead can estimate the VID by constructing histograms of voxel intensities.  Measurements therefore will work with the quantity
\be
B_i=N_{\rm{vox}}\int_{I_i-\Delta I/2}^{I_i+\Delta I/2}\p(I)dI\approx \p(I_i)N_{\rm{vox}}\Delta I,
\ee
where $B_i$ is the number of voxels in an intensity bin centered at $I_i$ with width $\Delta I$, assuming a map with $N_{\rm{vox}}$ total voxels.  The observed $\p(I)$ will be the convolution of the PDF of the total sky emission (e.g. Eq. \ref{VID_FG}) with that of the instrumental noise.  For uncorrelated white noise of width $\sigma_N$, we have a Gaussian noise PDF
\be
\p_{\rm{Noise}}=(2\pi\sigma_N^2)^{-1/2}e^{-I^2/2\sigma_N^2}.
\label{noise_VID}
\ee

If every map voxel is independent from all of the others, then each $B_i$ can be modeled as a draw from a binomial distribution \citep{Ihle2019}, with variance
\be
\rm{Var}(B_i)=B_i(1+B_i/N_{\rm{vox}})\approx B_i,
\label{VID_error}
\ee
where the second equality holds for sufficiently narrow intensity bins.

In the unconditional VID, we thus have two regimes recognizable to those familiar with power spectrum analysis.  Instrument noise smooths the PDF on scales smaller than $\sigma_N$, limiting our ability to measure small-scale variations.  Larger scale, smooth variations in the PDF are limited instead by Eq. (\ref{VID_error}) and the finite number of voxels in the map.  Thus, depending on the noise level and the ``scale" in intensity space of the feature we are trying to measure, we can find ourselves in either a noise- or sample-variance dominated regime.

In the case of the CVID, error computation becomes more difficult.  Simulations suggest that assigning independent errors to each histogram bin $B_i$ is reasonably accurate. However, Fourier transforming to compute $\R$ will mix these errors over every Fourier space bin, leading to highly correlated errors on $R$.  However, \citet{Breysse2019} argued that if we use the measured $B_i$ to estimate the histogram errors from Eq. (\ref{VID_error}), we can get a good estimate of the full covariance matrix of $\R$, even if we do not know \emph{a priori} the PDFs of the signal and foregrounds.  

\section{Intensity Statistics Forecasts} \label{S:stats}

Considering the models for [CII] and CO line emission that we developed in the previous section, we present forecasts for intensity measurements and constraints of galaxy emission models using both power spectra and the CVID.  We also consider the effects of interlopers and foreground continuum emission.  For our forecasts, we will divide the spectral range into three brackets, unless otherwise indicated, shown in Table \ref{tab:EXCLAIM_PARAMS}, to forecast the potential of EXCLAIM to detect redshift evolution in models of emission and galaxy properties.  For most forecast results, we define the survey area to be Stripe 82 unless stated otherwise.

\begin{table*}
    \centering
    \begin{tabular}{|c||c|c|c|c|c|c|c|c|c|c|}
    \hline \rowcolor{lightgray}
    $f_{\rm mid}$ & $N_{\rm ch}$ & $\Delta f$ & ${\rm NEI}_{\rm eff}$ & $P_N$%$P_N/V_{\rm vox}$
    & $z_{\rm CII}$ & $z_{\rm CO(6-5)}$ & $z_{\rm CO(5-4)}$ & $z_{\rm CO(4-3)}$ & $\sigma_{\rm b}$ \\
    \rowcolor{lightgray} (GHz) & & (GHz) & (kJy/sr-s$^{1/2}$) & (kJy/sr/Mpc$^3)^2$ & & & & & (arcmin)\\
    \hline
    \hline
    440 & 110 & 39.8 & 1951.8 & 5772.3%12025
    & 3.32 & 0.573 & 0.311 & 0.05 & 1.97 \\
    480 & 110 & 39.7 & 2649.5 & 10636%22158
    & 2.96 & 0.441 & 0.201 & - & 1.81 \\
    520 & 110 & 39.8 & 2194.3 & 7295.5%15177
    & 2.66 & 0.330 & 0.109 & - & 1.67 \\
    \hline
    \end{tabular}
    \caption{Instrument parameters for the EXCLAIM band separated into three redshift bins.  For each redshift bin, the columns correspond to the middle frequency ($f_{\rm mid}$), the number of channels ($N_{\rm ch}$), the bandwidth ($\Delta f$), the effective NEI, the noise power spectrum per Fourier mode ($P_N$), the redshifts of the various lines corresponding to $f_{\rm mid}$, and the PSF Gaussian width ($\sigma_{\rm b}$). }
    \label{tab:EXCLAIM_PARAMS}
\end{table*}

\begin{figure*}
    \centering
    \includegraphics[width=.9\textwidth]{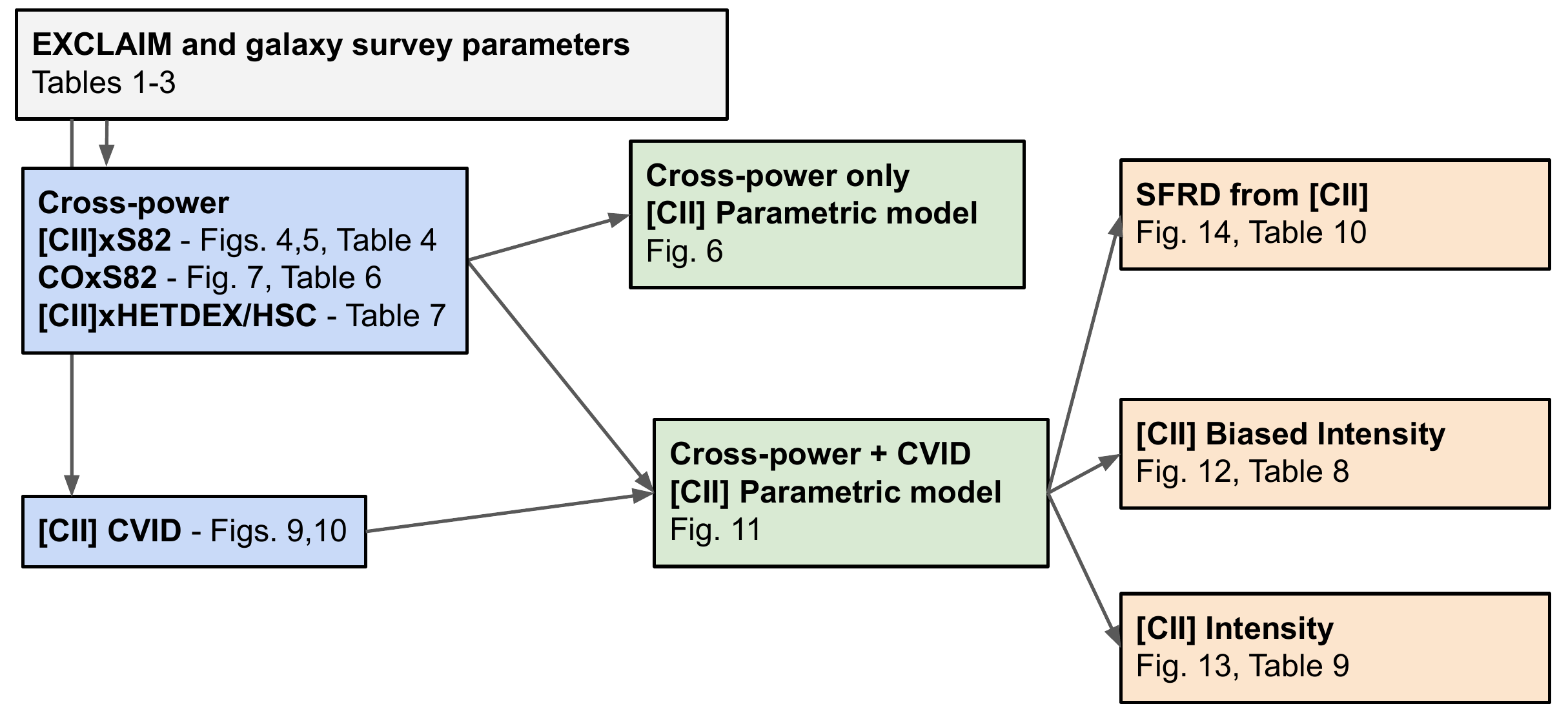}
    \caption{Flow of survey parameters to forecasts.  We include the cross-power and CVID observable forecasts (blue) which depend on the EXCLAIM and galaxy survey parameters.  These lead to parametric model forecasts (green), which then lead to forecasts on intensity and SFRD measurements (orange).  Each box lists the figures and tables that present these forecasts, which mostly appear in Secs.~\ref{S:stats} and \ref{S:science}.}
    \label{fig:diagram}
\end{figure*}

\subsection{Power spectrum forecasts} \label{S:Pkfore}

%Here we present power spectrum forecasts for [CII] and CO emission lines.

\subsubsection{[CII] cross-power sensitivity}

\begin{figure}
    \centering
    \includegraphics[width=.45\textwidth]{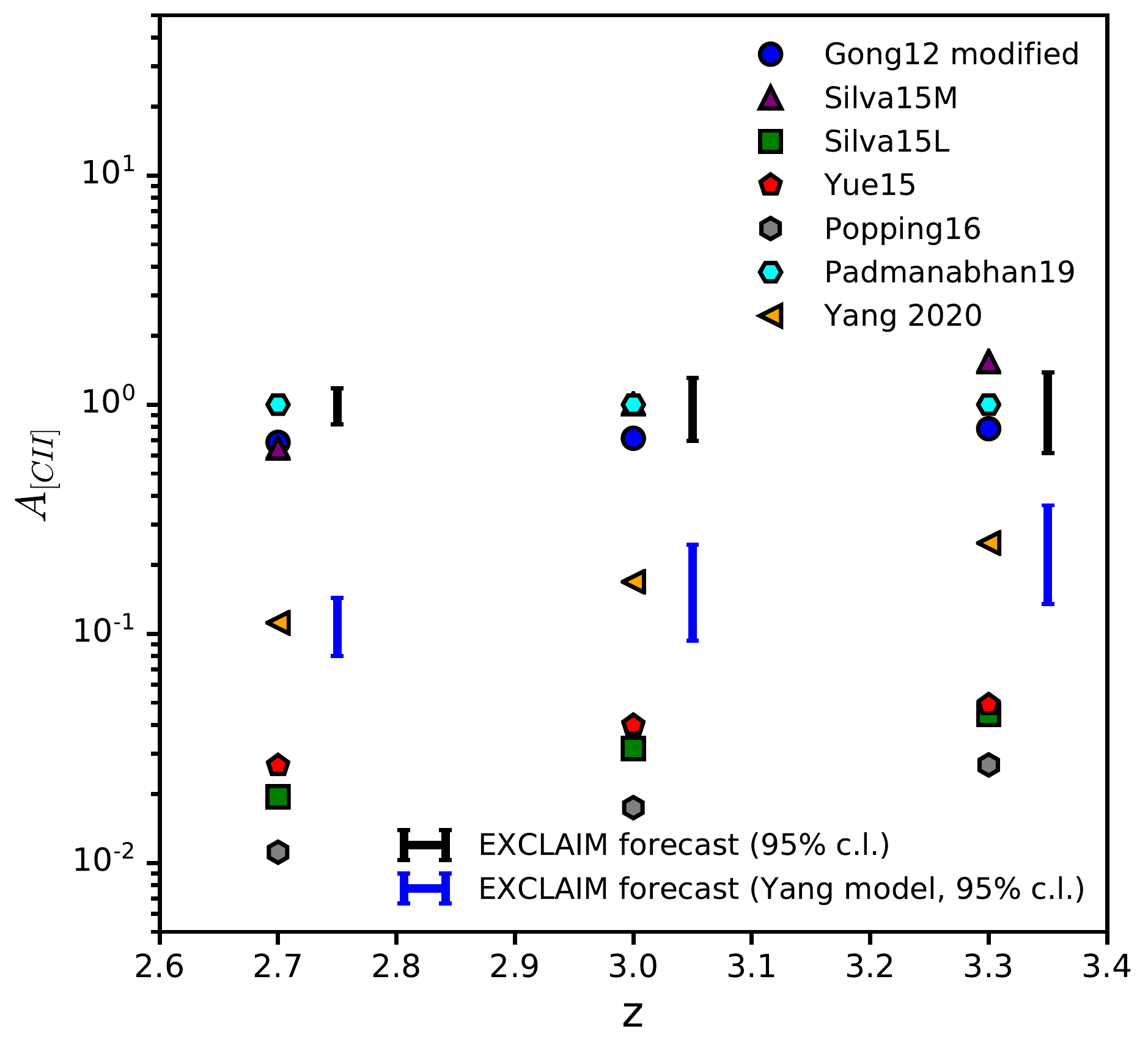}
    \caption{Various [CII] models for the power spectrum amplitude $A_\mathrm{[CII]}$ at redshifts corresponding to the quasar sample.  We also plot 2$\sigma$ error bars corresponding to the EXCLAIM power spectrum measurement uncertainty assuming both the \citet{2019MNRAS.488.3014P} and \citet{2022ApJ...929..140Y} intensity models.}
    \label{fig:ACII_models}
\end{figure}

\begin{figure*}
    \centering
    \includegraphics[width=0.4\textwidth]{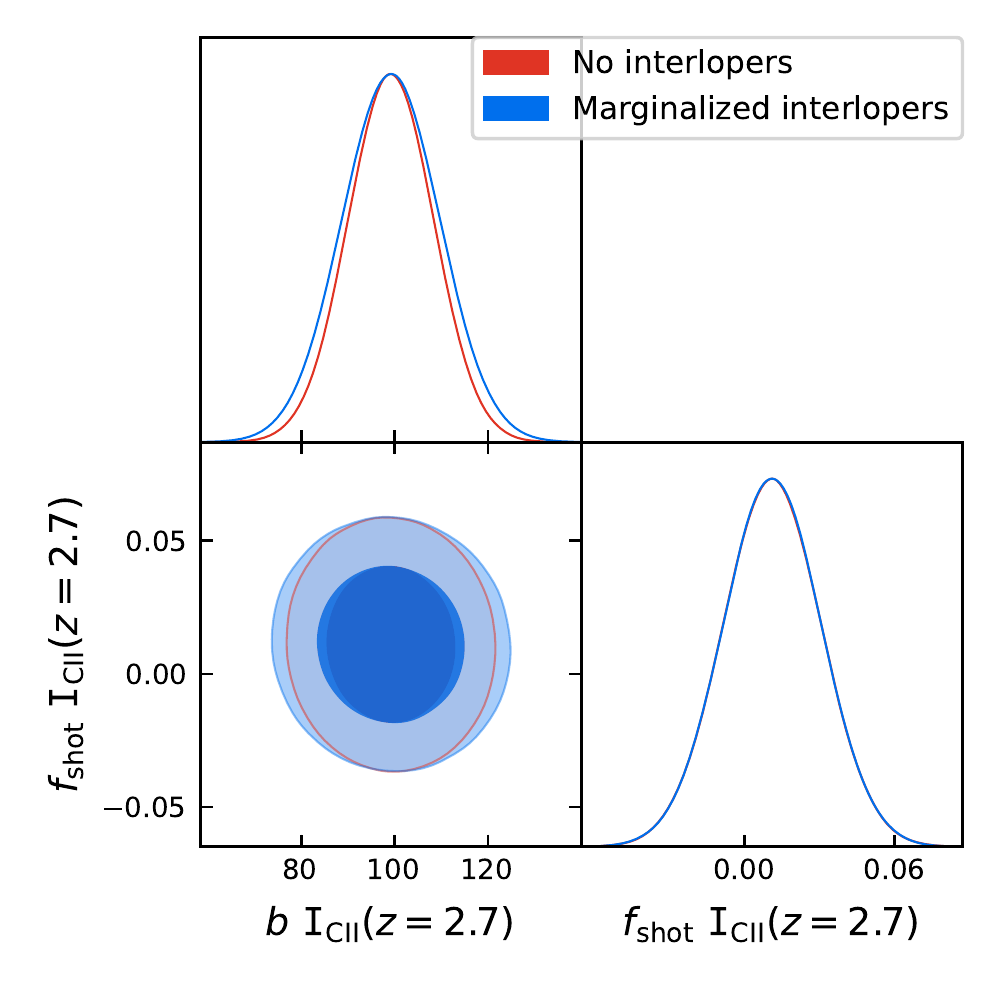}
    \includegraphics[width=0.4\textwidth]{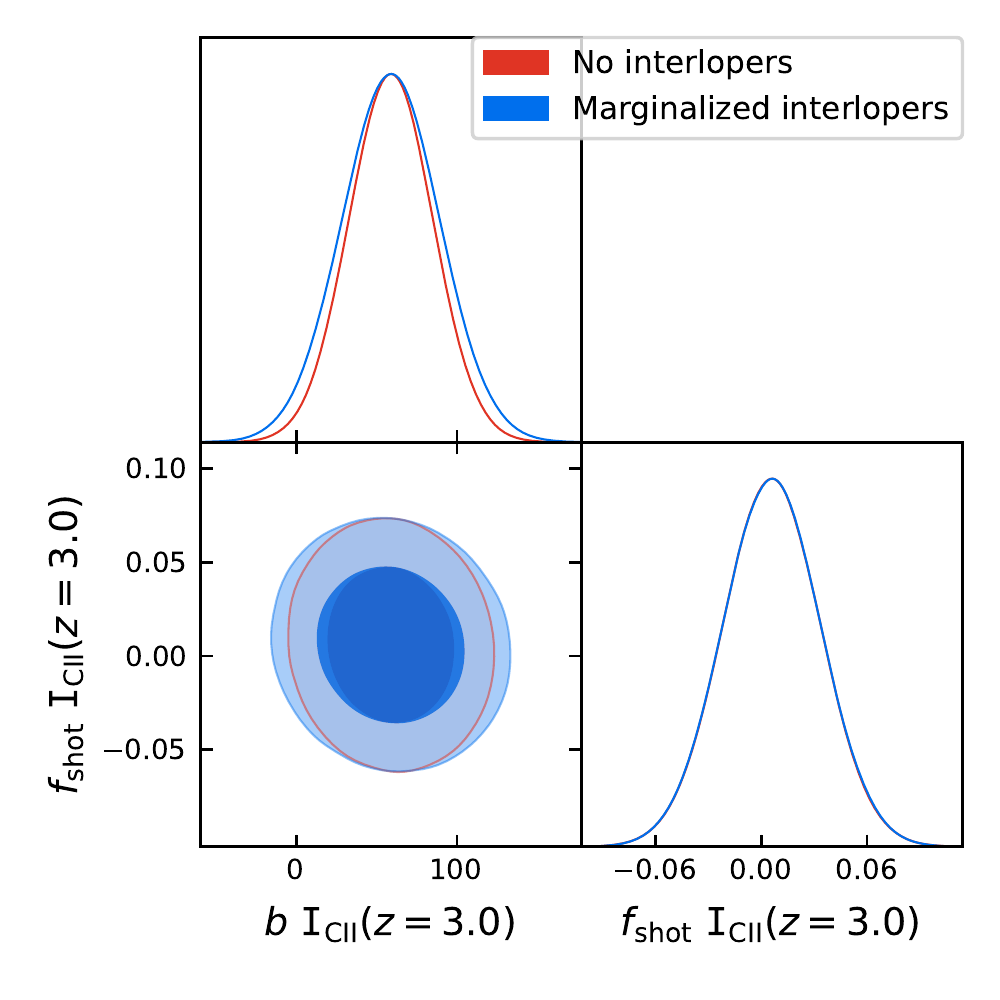}
    \includegraphics[width=0.4\textwidth]{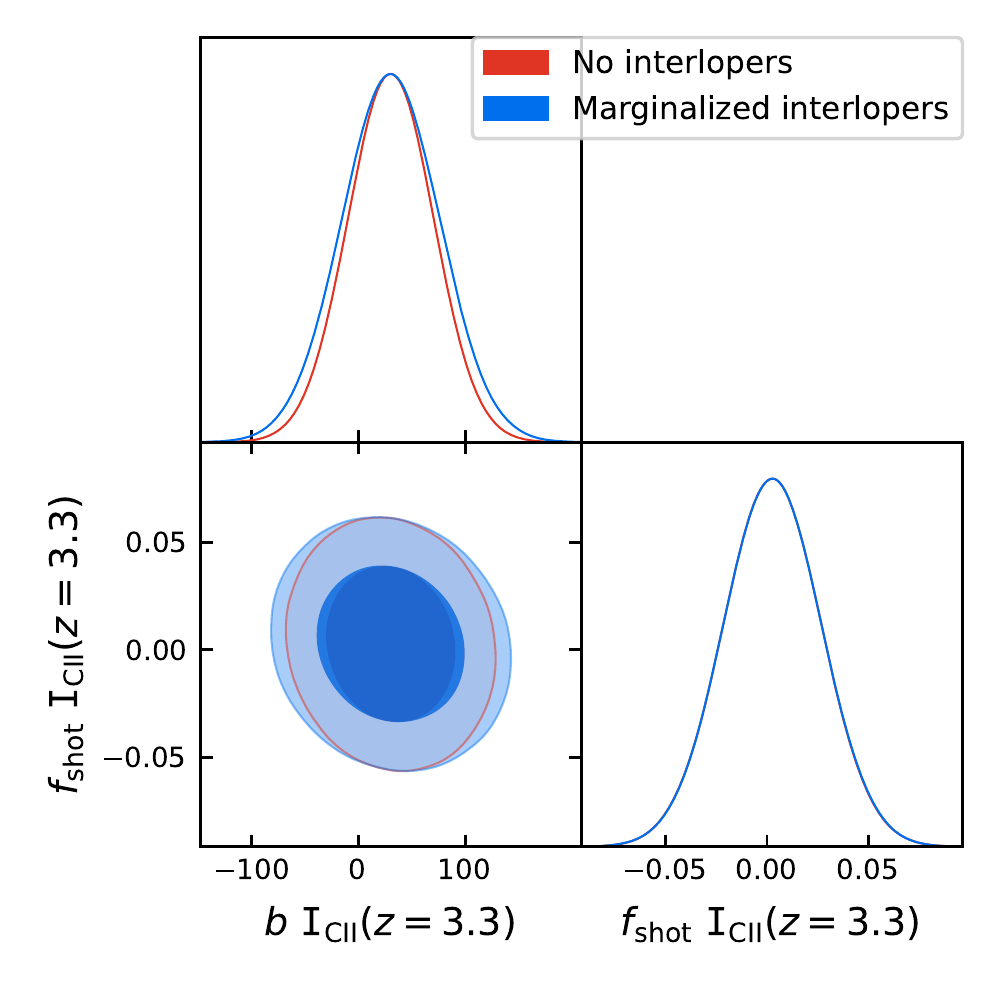}
    \caption{[CII] intensity forecasts from EXCLAIM cross power spectra with S82 BOSS quasars.  The panels correspond to $z=2.66$ (top left), $z=2.96$ (top right), and $z=3.32$ (bottom).  The darker and lighter shades denote 68\% and 95\% confidence intervals, respectively.}
    \label{fig:c2pk}
\end{figure*}

We present forecasts for [CII] intensity in Figs.~\ref{fig:ACII_models} and \ref{fig:c2pk} and Tables \ref{tab:CIIPklim} and \ref{tab:CIIlim} for several different scenarios.  We find for the fiducial [CII] intensity model from \citet{2019MNRAS.488.3014P}, the cross-power spectrum $P_\mathrm{CII-Q}(k)$ could be detected in each of the three redshift bins with signal-to-noise ratios (SNR) ranging from 5-11.  Note that keeping or removing the lowest $k_\parallel$ mode to mitigate foreground variance does not significantly affect the SNR.  In Fig.~\ref{fig:ACII_models} we present the power spectrum amplitude $A_\mathrm{[CII]}$ for several [CII] models \citep{2012ApJ...745...49G,2015ApJ...806..209S,2015MNRAS.450.3829Y,2016MNRAS.461...93P,2019MNRAS.488.3014P,2022ApJ...929..140Y}, with the \citet{2019MNRAS.488.3014P} model normalized to $A_\mathrm{[CII]}=1$ at all redshifts.  We also show the expected errors from EXCLAIM measurements of [CII] power spectra.  These results show that EXCLAIM can differentiate between the cluster of high intensity models, including \citet{2019MNRAS.488.3014P}, and the more pessimistic \citet{2022ApJ...929..140Y} model. If the cluster of low intensity models is more accurate, then the EXCLAIM measurement would give us strong constraints on [CII] that could potentially rule out the high intensity models.  

Due to relatively high instrumental noise in the EXCLAIM maps and shot noise from both the [CII] galaxies and the quasar sample, the amplitude of clustering alone, $b_\mathrm{[CII]}I_\mathrm{[CII]}$ has lower SNR than the full power spectrum.  For the $z=2.7$ redshift bin, $b_\mathrm{[CII]}I_\mathrm{[CII]}$ can be detected with an SNR of 8 after removing the lowest $k_\parallel$ mode and marginalizing over interlopers, while for the other two redshift bins has a $b_\mathrm{[CII]}I_\mathrm{[CII]}$ SNR less than 2.  

These values change when the [CII] line intensity model from \citet{2022ApJ...929..140Y} is assumed.  The SNRs for the total power spectrum $P_\mathrm{CII-Q}(k)$ reduce to the range 4-7, but are still able to differentiate between models, as shown in Fig.~\ref{fig:ACII_models}.  However, in this case the clustering amplitude $b_\mathrm{[CII]}I_\mathrm{[CII]}$ is then undetectable with SNRs $\lesssim$ 1.  Thus, we expect detection of $P_\mathrm{CII-Q}(k)$ in all three redshift bins under a range of physical scenarios. However, we are not guaranteed a measurement of $b_\mathrm{[CII]}I_\mathrm{[CII]}$ from power spectrum measurements alone, though it is possible for more optimistic models of [CII] emission at $z=2.7$. Note that we should generally expect a measurement of $I_\mathrm{[CII]}$ independently from the power spectrum shot noise.  However, the angular resolution of EXCLAIM is not high enough to sample the relevant scales.  Thus, all the information on [CII] intensity comes from the clustering signal in the power spectrum, which gives us $b_\mathrm{[CII]}I_\mathrm{[CII]}$.

\begin{table*}
    \centering
    \begin{tabular}{|c||c|c|}
    \hline \rowcolor{lightgray}
    $z$ & $P_\mathrm{[CII]-Q}(k)/\sigma[P_\mathrm{[CII]-Q}(k)]$ (all $k_\parallel$ modes) & $P_\mathrm{CII-Q}(k)/\sigma[P_\mathrm{CII-Q}(k)]$ (lowest $k_\parallel$ mode removed)\\
    \hline
    \hline
    2.7 & 11.4 & 11.2 / 7.04 \\
    3.0 & 6.71& 6.57 / 4.47 \\
    3.3 & 5.31 & 5.20 / 4.35 \\
%    2.7 & 99.2 / 27.7 & 166 / 30.6 & 116 / 22.2 \\
%    3.0 & 58.9 / 24.4 & 58.8 / 13.5 & 40.8 / 9.45 \\
%    3.3 & 30.2 / 19.8 & 23.9 / 11.2 & 16.6  / 7.82 \\
    \hline
    \end{tabular}
    \caption{[CII] power spectrum forecasts from the EXCLAIM-tracer cross-power spectra with S82 BOSS quasars for multiple line models.  Specifically, we show the signal-to-noise ratio (SNR) for $P(k)$ averaged over a bandwidth of scales.  The two major columns give the SNR over all scales and with the lowest $k_\parallel$ mode removed, respectively.  For the left column, the SNR assumes the [CII] model from \citet{2019MNRAS.488.3014P}.  For the right column, the left and right entries correspond to the [CII] models from \citet{2019MNRAS.488.3014P}  and \citet{2022ApJ...929..140Y}, respectively.}
    \label{tab:CIIPklim}
\end{table*}

\begin{table*}
    \centering
    \begin{tabular}{|c||c|c|c|c|}
    \hline \rowcolor{lightgray}
    $z$ & $b_\mathrm{[CII]}I_\mathrm{[CII]}$ [kJy/sr]& $bI/\sigma(bI)$ [CII] (all $k_\parallel$ modes) & $bI/\sigma(bI)$ [CII] (no int.) & $bI/\sigma(bI)$ [CII] (marg. int.)\\
    \hline
    \hline
    2.7 & 98.4 & 10.9 & 9.18 & 8.00 \\
    3.0 & 55.8 & 2.24 & 1.87 & 1.53 \\
    3.3 & 30.3 & 0.76 & 0.60 & 0.53 \\
%    2.7 & 98.4 / 26.8 & 10.9 & 9.18 / 1.02 & 8.00 / 0.90 \\
%    3.0 & 55.8 / 24.0 & 2.24 & 1.87 / 0.42 & 1.53 / 0.36 \\
%    3.3 & 30.3 / 20.5 & 0.76 & 0.60 / 0.34 & 0.53  / 0.30 \\
%    2.7 & 99.2 / 27.7 & 166 / 30.6 & 116 / 22.2 \\
%    3.0 & 58.9 / 24.4 & 58.8 / 13.5 & 40.8 / 9.45 \\
%    3.3 & 30.2 / 19.8 & 23.9 / 11.2 & 16.6  / 7.82 \\
    \hline
    \end{tabular}
    \caption{[CII] biased intensity $b_\mathrm{[CII]}I_\mathrm{[CII]}$ forecasts from EXCLAIM cross-power spectra with S82 BOSS quasars for multiple interloper foreground treatments and line models.  The second column includes all $k_\parallel$ modes with no interlopers.  The two latter columns remove the first $k_\parallel$ mode and assume no interlopers and marginalized interlopers, respectively.  All entries correspond to the [CII] model from \citet{2019MNRAS.488.3014P}; the \citet{2022ApJ...929..140Y} model finds SNR$\lesssim 1$ for $b_\mathrm{[CII]}I_\mathrm{[CII]}$.}
    \label{tab:CIIlim}
\end{table*}

\subsubsection{Constraints on parametric models of [CII] emission}

\begin{figure}
    \centering
    \includegraphics[width=0.45\textwidth]{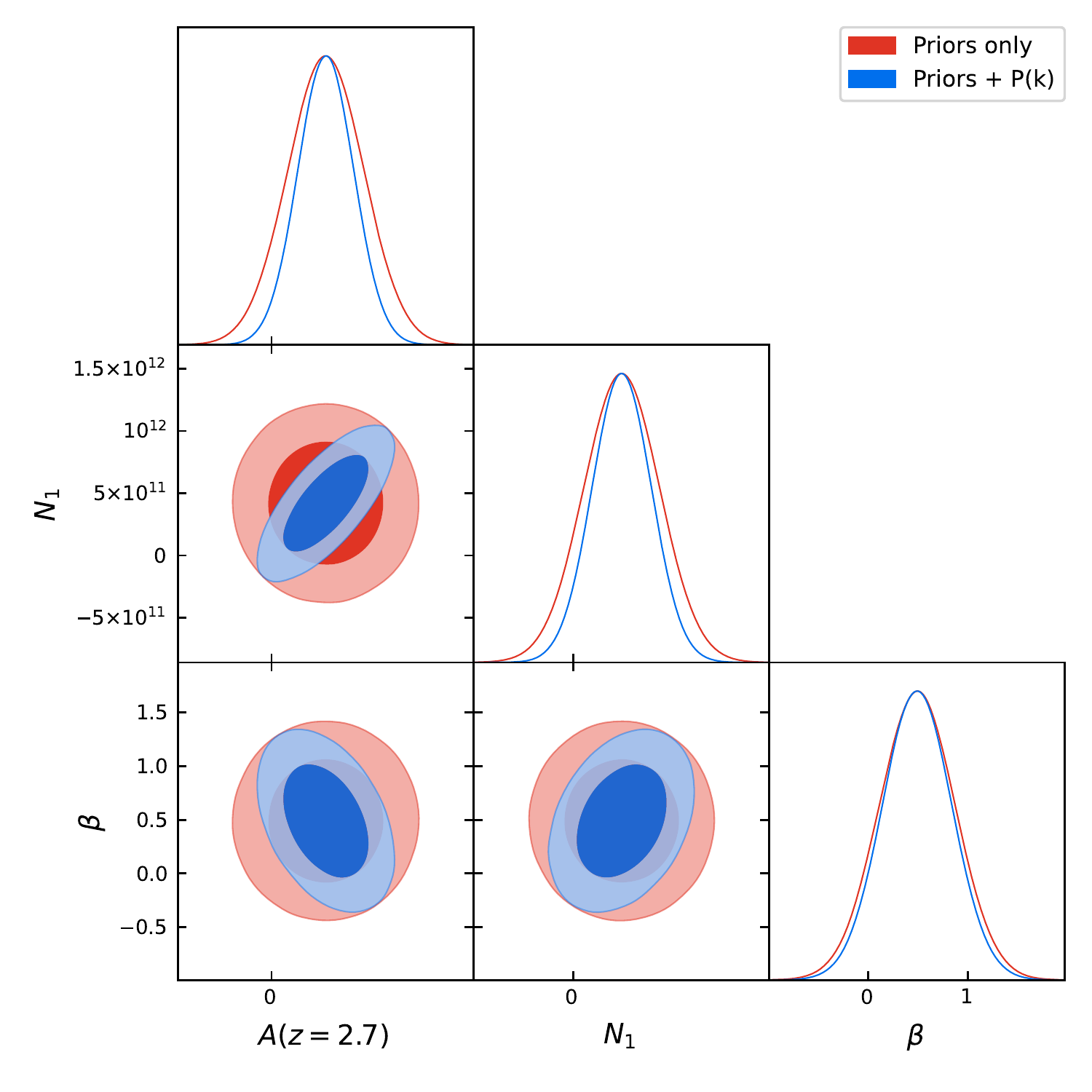}\\
    \caption{Forecasts for constraints on the Padmanabhan [CII] model parameters (see Eq.~\ref{E:lc2mod}) from EXCLAIM $P(k)$ measurements at redshift $z=2.7$.  The red ellipse assumes only the priors from previous measurements in \citet{2019MNRAS.488.3014P}, while the blue curve includes both priors and EXCLAIM power spectrum data.  The EXCLAIM data assumes marginalized interlopers.  The darker and lighter shades denote 68\% and 95\% confidence intervals, respectively.}
    \label{fig:lc2fore}
\end{figure}

With these forecasts, we can also consider how well this measurement could constrain the [CII] $L$-$M$ models given in Eqs.~\ref{E:lc2mod} and \ref{E:lc2mod2}.  Specifically, we use the forecast for $b_\mathrm{[CII]}I_\mathrm{[CII]}$ at each redshift to construct a Fisher matrix $F_{ij}$ that forecasts errors in the parameters constituting the $L_\mathrm{[CII]}(M)$ model,
\begin{eqnarray}
F_{ij}=\frac{\partial_{p_i}(bI)|_{\mathbf{p}_0}\partial_{p_j}(bI)|_{\mathbf{p}_0}}{\sigma^2[b_\mathrm{[CII]}I_\mathrm{[CII]}]}\, ,
\end{eqnarray}
where $\mathbf{p}$ is a vector containing the parameters (described below) and $\mathbf{p}_0$ denotes the expected values of the parameters, with the halo bias $b_h(M)$ and the halo mass function $n(M)$ fixed.  This Fisher matrix is noninvertible, and measurements of these parameters using only intensity data from EXCLAIM will be highly degenerate. Thus, we include priors for the parameters.  For the \citet{2019MNRAS.488.3014P} model, the parameter vector is $\mathbf{p}=\{A,N_1,\beta\}$ with the prior based on the limits given by $M_1=(2.39\pm1.86)\times 10^{-5}$ M$_\odot$, $N_1=(4.19\pm3.27)\times10^{11}$ M$_\odot$, and $\beta=0.49\pm0.38$.  We conservatively assume the prior on $A$ will only depend on the errors in $M_1$ and $\alpha$ while being orthogonal to the errors in $N_1$ and $\beta$, which yields $A=(2.24\pm 0.66)\times 10^9$ L$_\odot$.  For the \citet{2022ApJ...929..140Y} model, the parameter vector is $\mathbf{p}=\{N,M_1,\alpha,\beta\}$ with a prior assuming 100\% errors on the parameter values.  The resulting forecasts are shown in Fig.~\ref{fig:lc2fore}.  These forecasts assume that interlopers are present with marginalized intensities.  These forecasts show that measurements of both $A$ and $N_1$ from EXCLAIM will improve significantly on current measurements.  These parameters correspond to the overall amplitude and mass-scaling of the luminosity-mass law.

\subsubsection{Impact on prior knowledge}

Using this parameter covariance matrix, we can then construct new errors on $b_\mathrm{[CII]}I_\mathrm{[CII]}$ from power spectrum measurements that include the effect of the prior, using the formula
\begin{eqnarray} \label{E:sigbI}
\sigma^2[b_\mathrm{[CII]}I_\mathrm{[CII]}] = [\nabla_{\mathbf{p}}(bI)|_{\mathbf{p}_0}]^T\cdot\mathbf{C}\cdot\nabla_{\mathbf{p}}(bI)|_{\mathbf{p}_0}\, ,
\end{eqnarray}
where $\mathbf{C}=\mathbf{F}^{-1}$ is the parameter covariance matrix.  Adding the parameter priors slightly increases the SNR for $b_\mathrm{[CII]}I_\mathrm{[CII]}$ with the most dramatic effect taking place for $z=3.3$, where the SNR increases from 0.5 to 1.  Except for this highest redshift, EXCLAIM data provide significant information relative to the prior. We present the full results in Table \ref{tab:bICIIlim}.  

\subsubsection{Separation of clustering bias and intensity}

We can also use this formalism to forecast measurements of (unbiased) intensity with the error in $I_\mathrm{[CII]}$ given by
\begin{eqnarray} \label{E:sigI}
\sigma^2[I_\mathrm{[CII]}] = (\nabla_{\mathbf{p}}I|_{\mathbf{p}_0})^T\cdot\mathbf{C}\cdot\nabla_{\mathbf{p}}I|_{\mathbf{p}_0}\, .
\end{eqnarray}
We present the forecasts in Table \ref{tab:ICIIlim}. We find that the SNR for intensity is significant for $z=2.7$ with a value of 4.6, while the other redshift bins are undetectable.  Note that this SNR value is much higher than the value assuming only the prior, which is 0.8. This means that the EXCLAIM $P(k)$ measurement should provide significant information on the [CII] intensity, depending on the accuracy of the assumed halo model, though the SNR can be slightly affected by the prior.  For example, we find a 10\% increase in the prior widths can reduce the SNR by approximately a few percent, while doubling the prior widths can reduce the SNR by 20\%. Information about the bias in the constraint system comes from Eq.\,\ref{eqn:IMbias} and the parametric form of $L(M,z)$. Observational data from redshift space distortions of the line intensity may ultimately provide a constraint from within intensity data, which is not included here.

%\nb{We can consider a diagram showing the different phases of incorporation of information. This would clarify some of the differences in assumptions and inputs as more pieces are combined, and help the reader keep track of which tables show results at a given level. A left block could have ``EXCLAIM x QSO" or ``EXCLAIM two-point" pointing at it, and (Fig 4, Table 4, 5, Eq. TBD) in a box.  In this case, Eq. TBD would be the equation for the sensitivity from observations, similar to Eq. 38 etc. Then the next box could show CO (though maybe that's in the first box because the data vector and sensitivity setup is the same?). Other boxes would include CVID and prior information, and then SFRD extraction. There could be horizontal arrows showing how two-point constraints flow into a higher level of integration. We'll probably want to iterate on this a bit to tell the story of how different sorts of information are combined.}

\subsubsection{CO Forecasts}

\begin{figure}
    \centering
    \includegraphics[width=.45\textwidth]{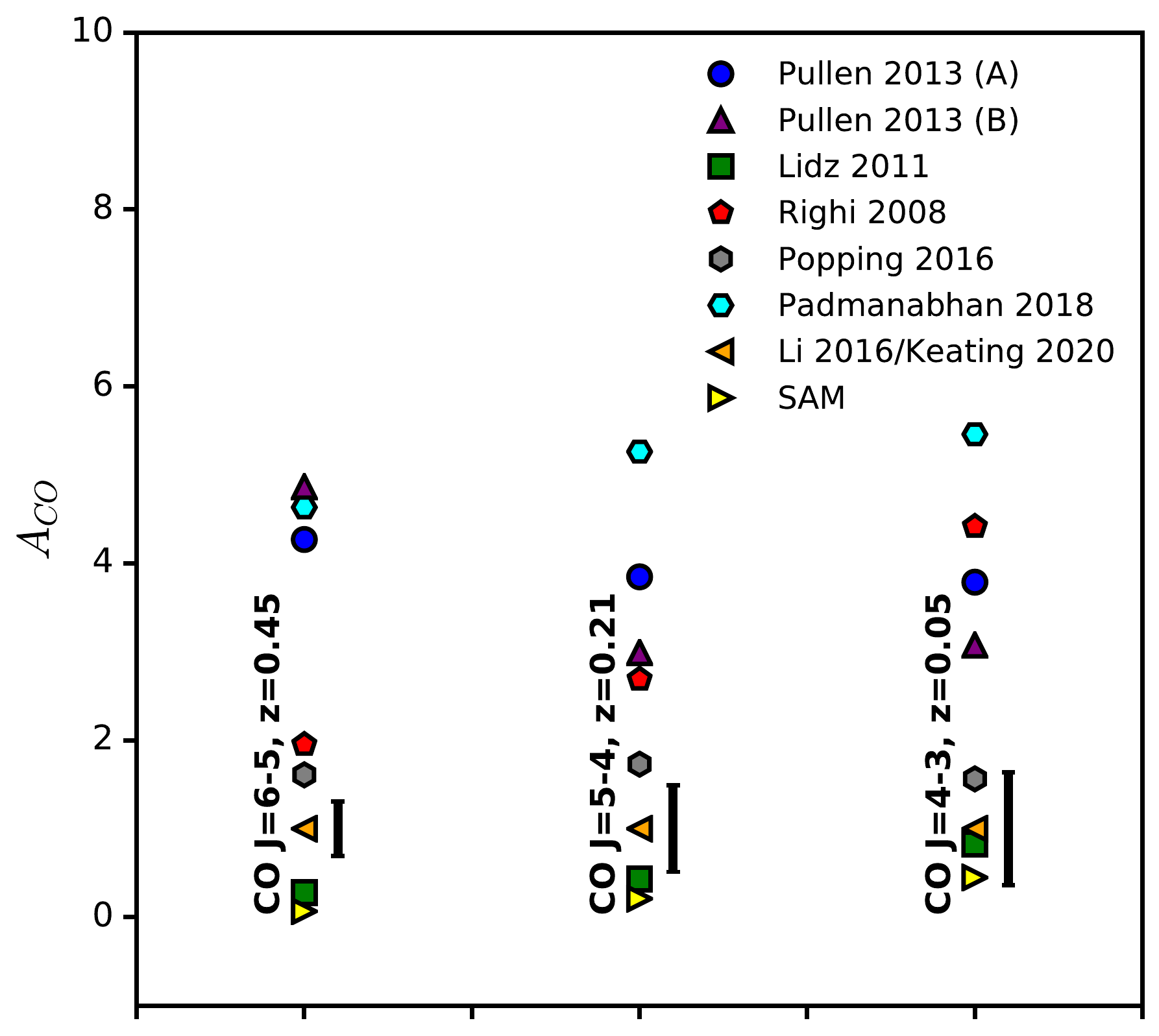}
    \caption{Various CO(4-3), CO(5-4), and CO(6-5) models for the power spectrum amplitude $A_\mathrm{CO}$ at redshifts for specific galaxy samples.  We also plot 2$\sigma$ error bars corresponding to the EXCLAIM power spectrum measurement uncertainty assuming the Li/Keating intensity model.}
    \label{fig:CO_models}
\end{figure}

Our forecasts for lines CO(4-3) at $z=0.05$, CO(5-4) at $z=0.21$, and CO(6-5) at $0.45$ are presented in Fig.~\ref{fig:CO_models} and Table \ref{tab:COlim}.  We predict lower but detectable signal-to-noise ratios (SNRs) of the cross-power spectrum $P_\mathrm{CO-gal}$, with values ranging from 3-6 depending on the redshift bin.  In Fig.~\ref{fig:CO_models}, we present the power spectrum amplitude $A_\mathrm{CO}$ of CO(4-3), CO(5-4), and CO(6-5) for both the model from \citet{Li2016} and \citet{Keating2020} as well as other models \citep{2013ApJ...768...15P, 2011ApJ...741...70L, 2008A&A...489..489R, 2016MNRAS.461...93P, 2018MNRAS.475.1477P, 2022ApJ...929..140Y} at the redshifts of the galaxies for which we will cross-correlate with the Li/Keating model normalized to $A_\mathrm{CO}=1$ at all $J$ and redshifts.  We also show the expected errors from EXCLAIM measurements of CO with the indicated galaxy samples.  These results show that EXCLAIM can differentiate the various CO models.  However, we show in Table \ref{tab:COlim} that the $b_{\rm CO}I_{\rm CO}$ SNRs from clustering are all less than one for each redshift bin assuming the Li/Keating model.
%, meaning EXCLAIM will not be able to measure $b_{\rm CO}I_{\rm CO}$ to constrain galaxy properties, such as the star formation rate density for models at or below $A_\mathrm{CO}=1$.  However, if the more optimistic models with $A_\mathrm{CO}>1$ are more accurate, then it is possible that intensity or biased intensity could be measured.
% \ref{fig:copk} and

\iffalse
\begin{figure*}
    \centering
    \begin{subfigure}[t]{0.4\textwidth}
        \centering
        \includegraphics[width=\linewidth]{figs/CO43z_0.05_compare_marg_2.pdf}
        \subcaption{CO(4-3) (z=0.05)}
%    \end{subfigure}
    \begin{subfigure}[t]{0.4\textwidth}
        \centering
        \includegraphics[width=\linewidth]{figs/CO54z_0.21_compare_marg_2.pdf}
        \subcaption{CO(5-4) (z=0.21)}
    \end{subfigure}
    \begin{subfigure}[t]{0.4\textwidth}
        \centering
        \includegraphics[width=\linewidth]{figs/CO65z_0.45_compare_marg_2.pdf}
        \subcaption{CO(6-5) (z=0.45)}
    \end{subfigure}
    \caption{CO intensity forecasts from EXCLAIM cross power spectra with S82 SDSS/BOSS galaxies.  The panels correspond to CO(4-3) at $z=0.05$ (top left), CO(5-4) at $z=0.21$ (top right), and CO(6-5) at $z=0.45$ (bottom). The darker and lighter shades denote 68\% and 95\% confidence intervals, respectively.}
    \label{fig:copk}
\end{figure*}
\fi

\begin{table*}
    \centering
    \begin{tabular}{|c|c||c|c|c|c|}
    \hline \rowcolor{lightgray}
    line & $z$ & $b_{\rm CO}I_{\rm CO}$ [kJy/sr] & $P_\mathrm{CO-gal}(k)/\sigma[P_\mathrm{CO-gal}(k)]$ & $bI/\sigma(bI)$ CO (marg. int.)\\
    \hline
    \hline
    CO(4-3) & 0.05 & 0.178 & 3.12 & 0.083 \\
    CO(5-4) & 0.21 & 0.283 & 4.10 & 0.43 \\
    CO(6-5) & 0.45 & 0.391 & 6.45 & 0.25\\
%    CO(4-3) & 0.05 & 0.178 & 0.755 & 0.741 \\
%    CO(5-4) & 0.21 & 0.283 & 14.0 & 6.20 \\
%    CO(6-5) & 0.45 & 0.391 & 12.9 & 5.99\\
    \hline
    \end{tabular}
    \caption{CO forecasts from EXCLAIM cross-power spectra with S82 SDSS galaxies.  The fourth column gives the SNR for $P_\mathrm{CO-gal}(k)$ averaged over a bandwidth of scales.  The fifth column provides the SNR for $b_\mathrm{CO}I_\mathrm{CO}$ with marginalized interlopers and the lowest $k_\parallel$ mode removed.  These forecasts assume the CO model from \citet{Li2016} and \citet{Keating2020}.}
    \label{tab:COlim}
\end{table*}

\subsubsection{Forecasts including alternate galaxy surveys}

\emph{Hyper-Suprime CAM Forecasts:} We also forecast an alternative measurement where we replace S82 BOSS quasars with tracers from the HSC North field.  The forecasts for $bI$ limits are in Table \ref{tab:altgal}.  Assuming the Padmanabhan [CII] model, we find that %SNR values for the [CII] cross-power spectrum are lower than the values using BOSS quasars, in the range of 2-3.5 depending on the redshift.  This behavior is caused by a reduction in the cross-power shot noise reducing the total cross-power signal, which lowers the SNR.  However, 
the SNR for $b_\mathrm{[CII]}I_\mathrm{[CII]}$ is slightly higher than the BOSS quasar survey with values between 2-10 since the lower shot noise allows the clustering signal to be more distinguishable. The CO intensities from HSC all have SNRs < 1.  This shows that including measurements from EXCLAIM-HSC cross-power spectra could increase the sensitivity to [CII], while it would not improve the CO sensitivity.

\emph{HETDEX Forecasts:} Table \ref{tab:altgal} shows forecasts for $bI$ limits for HETDEX Fall and SHELA fields.  Although the SHELA survey has a much smaller area than the Fall survey, this does not affect the SNRs significantly, especially for cases where the SNR > 1.  Assuming the Padmanabhan [CII] model, we find an SNR range for %the [CII] cross-power spectrum of 5-9, depending on the redshift.  The SNR for
$b_\mathrm{[CII]}I_\mathrm{[CII]}$ is 2-10, similar to that for the HSC survey.  For CO, the biased intensity SNRs < 1 mostly, with none greater than 1.5.  This shows that the EXCLAIM-HETDEX cross-power spectra would be significantly helpful for the [CII] survey, while only modestly improving the sensitivity to the CO power spectrum.

\begin{table*}
    \centering
    \begin{tabular}{|c||c|c|}
    \hline \rowcolor{lightgray}
    Survey & $bI/\sigma(bI)$ [CII] & $bI/\sigma(bI)$ CO \\
%    Survey & $P_\mathrm{[CII]-gal}(k)/\sigma[P_\mathrm{[CII]-gal}(k)]$ & $bI/\sigma(bI)$ [CII] & $P_\mathrm{CO-gal}(k)/\sigma[P_\mathrm{CO-gal}(k)]$ & $bI/\sigma(bI)$ CO \\
    \hline
    \hline
    HSC & 9.59 / 4.15 / 2.29 & $4\times 10^{-4}$ / 0.32 / 0.40  \\
    HETDEX-F & 11.8 / 4.64 / 2.30 & 0.04 / 0.31 / 0.35 \\
    HETDEX-SHELA & 7.75 / 3.21 / 1.64 & 0.56 / 1.14 / 0.21\\
%    HSC & 3.53 / 2.28 / 1.79 & 9.59 / 4.15 / 2.29 & $9\times 10^{-3}$ / 0.67 / 0.79 & $4\times 10^{-4}$ / 0.32 / 0.40  \\
%    HETDEX-F & 8.99 / 5.52 / 4.49 & 11.8 / 4.64 / 2.30 & 3.50 / 4.24 / 3.79 & 0.04 / 0.31 / 0.35 \\
%    HETDEX-SHELA & 9.00 / 5.52 / 4.48 & 7.75 / 3.21 / 1.64 & 2.80 / 4.72 / 3.53 & 0.56 / 1.14 / 0.21\\
    \hline
    \end{tabular}
    \caption{Forecasts from EXCLAIM cross-power spectra with HSC and HETDEX tracers.  The  columns provide the SNRs for $b_\mathrm{[CII]}I_\mathrm{[CII]}$ and $b_\mathrm{CO}I_\mathrm{CO}$, respectively, with marginalized interlopers and the lowest $k_\parallel$ mode removed.  In each cell, the multiple entries correspond to increasing redshift bins (centered on $\{ 2.7, 3.0, 3.3 \}$) for the [CII] columns and CO(4-3) / CO(5-4) / CO(5-6) for the CO columns centered on $\{0.05, 0.21, 0.45\}$.   These forecasts assume the CII model from \citet{2019MNRAS.488.3014P} and the CO model from \citet{Li2016} and \citet{Keating2020}.}
    \label{tab:altgal}
\end{table*}

\subsection{Conditional VID forecasts}

\begin{figure}
\centering
\includegraphics[width=\columnwidth]{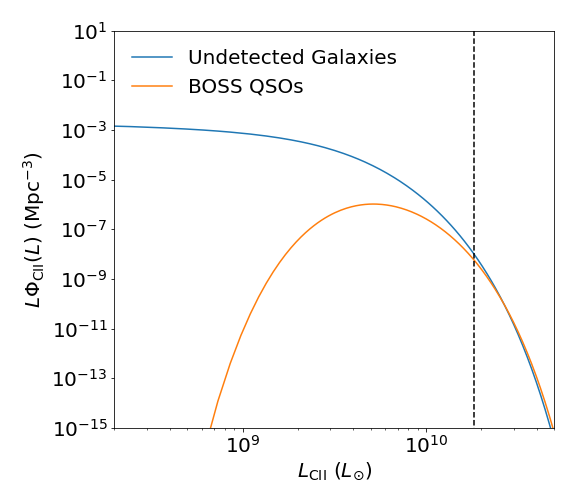}
\caption{CII luminosity function from the \citet{2019MNRAS.488.3014P} CII emission model separated into emission from BOSS QSO's (orange) and the larger undetected population, which appear only in the intensity map (blue).  Curves are plotted for the $z=3.3$ EXCLAIM redshift bin.  The black dashed line marks the location of $L_{\rm{cut}}$ as defined in Eq. (\ref{eq:Lcut}).}
\label{fig:dndL_CVID}
\end{figure}

For the CVID analysis, we need to separate CII emitted by the individually-detected BOSS quasars from that emitted by fainter sources that do not appear in the galaxy survey.  Due to the substantial uncertainty in the amplitude of the overall CII signal, we chose an intentionally simple form for this separation where
\be
\left.\frac{dn}{dL}\right|_{\rm{QSO}}=\frac{dn}{dL}e^{-L_{\rm{cut}}/L},
\label{eq:Lcut}
\ee
and
\be
\left.\frac{dn}{dL}\right|_{un} = \frac{dn}{dL}-\left.\frac{dn}{dL}\right|_{\rm{QSO}}.
\ee
Since we know the number density of BOSS sources, we set the value of $L_{\rm{cut}}$ in each redshift bin to enforce $\int (dn/dL)_{\rm{QSO}}dL=n_{\rm{QSO}}$.  The resulting separated luminosity functions are shown in Figure \ref{fig:dndL_CVID}.

\begin{figure*}
\centering
\includegraphics[width=\textwidth]{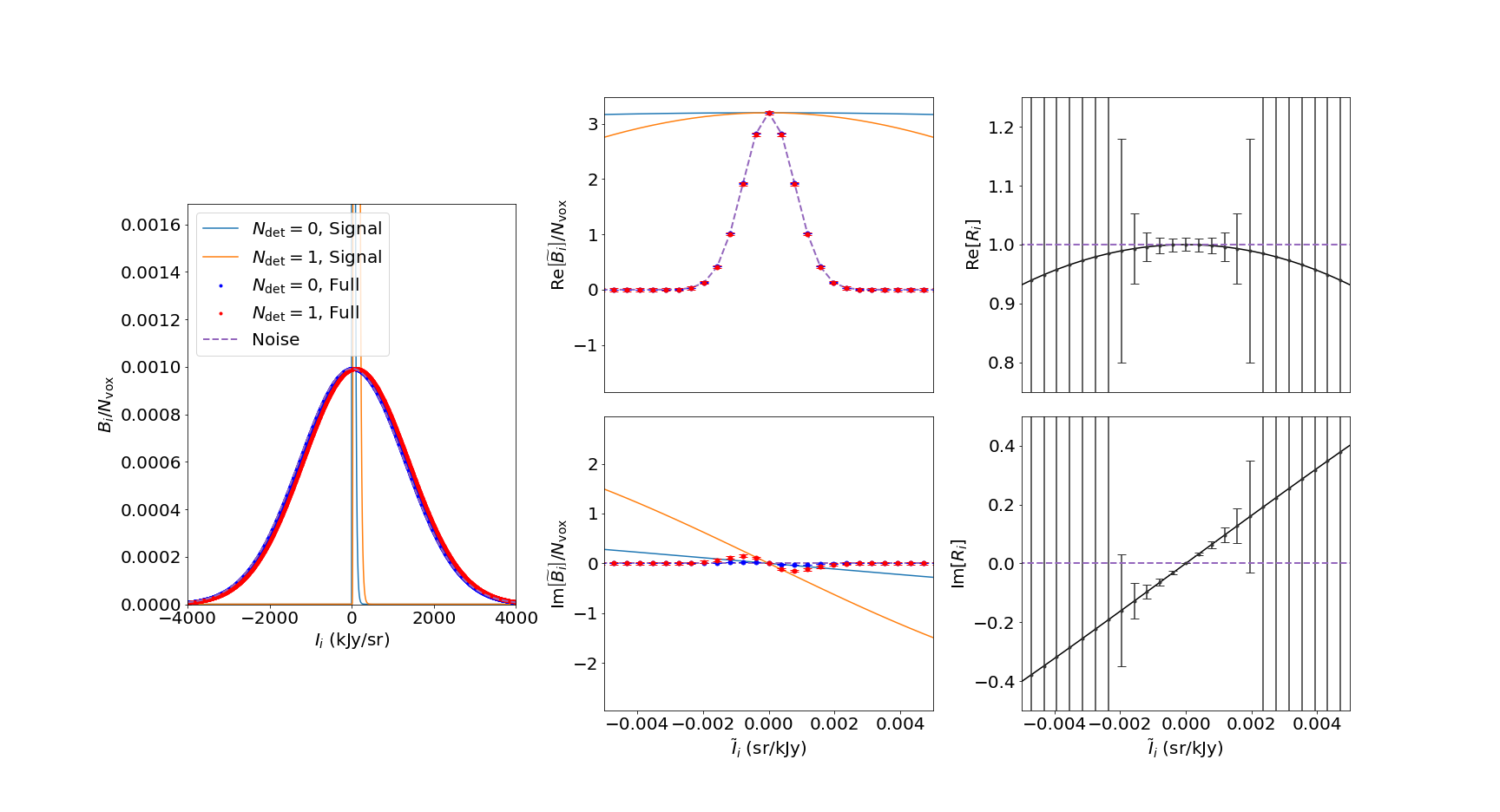}
\caption{(Left column) Predicted histograms from the luminosity functions plotted in Figure \ref{fig:dndL_CVID} for the $z=3.3$ redshift bin.  Thin curves show the signal-only predictions, thick curves show the effect of convolving in the EXCLAIM instrumental noise.  Blue curves are for voxels that do not contain a BOSS QSO, while red/orange curves are for those which do.  Histograms are normalized by the total number of voxels in each group, intensity bins have width $\Delta I=1.6$ kJy/sr.  The noise-only histogram is shown as a purple dashed curve, and is mostly indistinguishable from the $N_{\rm{det}}=0$ histogram.  (Middle column) Real (top) and imaginary (bottom) parts of the Fourier transformed histograms, colors are the same as in the left column.  Error bars are propagated assuming binomial errors on the original histograms. Note that these display the Fourier-conjugate of the intensity distribution, so they have units of sr/kJy, analogous to $x \leftrightarrow k$. (Right column) Predicted $\mathcal{R}_{10}$ values for the signal-only case (black curves) and the case including noise (points and error bars).  The purple dashed curves again indicate the noise-only prediction.}
\label{fig:CVID_summary}
\end{figure*}

Figure \ref{fig:CVID_summary} summarizes forecasting the CVID from the above luminosity functions.  The full procedure is described in detail in \citet{Breysse2019}.  First, we convolve the signal with a PDF of the Gaussian instrumental noise and bin it into a predicted histogram.  We can immediately see that voxels containing a detected QSO are noticeably brighter on average than those that do not.  The overall offset between the $N_{\rm{QSO}}=1$ and $N_{\rm{QSO}}=0$ distributions is the average difference we would obtain from a simple stacking measurement.  We then Fourier-transform the two histograms to produce the center column of Fig. \ref{fig:CVID_summary}.  Note that we are now working with complex variables, so we have twice as many data points as in the original histograms.  However, because our original data was real, these curves and their measured errors are symmetric about $\tilde{I}=0$.

\begin{figure}
\centering
\includegraphics[width=\columnwidth]{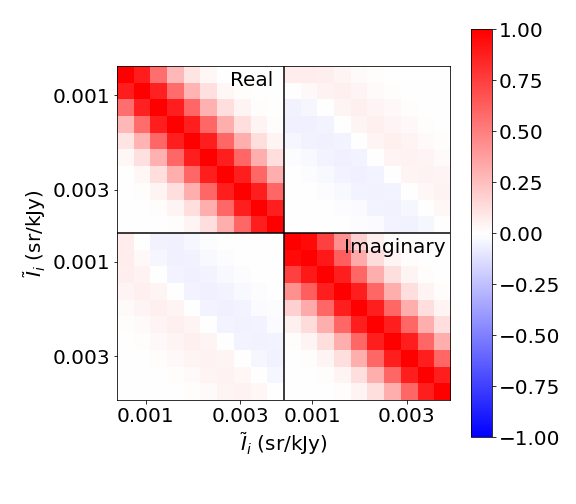}
\caption{Correlation matrix for the real and imaginary parts of the CVID estimator from Figure \ref{fig:CVID_summary}.  Errors are shown for the first eight positive $\tilde{I}$ bins.  Negative $\tilde{I}$ bins are not included as they are exactly correlated with the positive bins, higher $\tilde{I}$ bins are not included because, as shown in Figure \ref{fig:CVID_summary}, they contribute negligible signal to noise.  These bins and this covariance matrix are used in the CVID Fisher forecast.}
\label{fig:CVID_cov}
\end{figure}

Finally, we take the ratio of the two Fourier transforms to produce a forecast for $\mathcal{R}_{10}$.  We can immediately see that the result with noise included falls exactly on the signal-only prediction.  Note that, again, points at positive and negative $\tilde{I}$ have exactly correlated errors.  We can see a "window" in $\tilde{I}$ space set by the overall noise level wherein the error bars on $\mathcal{R}_{10}$ are small, and we get useful information.  Because information from each of the original histogram bins is mixed into all Fourier-space bins, these uncertainties are highly correlated.  Figure \ref{fig:CVID_cov} shows the correlation matrix for the positive-$\tilde{I}$ bins of $\mathcal{R}_{10}$ computed using the procedure from \citet{Breysse2019}.

Given the forecasted CVID, we can estimate an SNR assuming either the Padmanabhan or Yang [CII] models.  We account for the highly-correlated error bars in the CVID by writing the SNR as
\begin{eqnarray}
{\rm SNR}^2 = \sum_{ij}(\mathcal{R}_{10,i}-1)[\mathbf{C}^{-1}]_{ij}(\mathcal{R}_{10,j}-1)\, ,
\end{eqnarray}
where $\mathbf{C}$ is the covariance matrix of the CVID.  The $\mathcal{R}-1$ form comes from the fact that, by Eq. (\ref{Rdef}) $\mathcal{R}=1$ in the absence of any signal.  For $z=\{2.7,3.0,3.3\}$, we find SNRs of $\{24.5, 10.1, 5.8\}$ for the Padmanabhan [CII] model and $\{13.8, 9.3, 8.8\}$ for the Yang [CII] model.  These SNRs are significantly higher than the SNRs for the cross-power spectrum measurements.  To some extent, this is caused by our choice to assign all of the BOSS quasars the very highest [CII] luminosities, which is a very optimistic one that makes the CVID extremely sensitive to the bright tail of the model.  This is why, for example, we get a better SNR for the Yang [CII] model at $z=3.3$ than the Padmanabhan model; the Yang model is fainter on average (smaller $bI$), but because it includes a scatter factor, it has a stronger tail at the bright end, which is what appears during the stacking.  While this assumption is not necessarily wrong, this may not be as well-founded as the details of the power spectrum forecasts.

\subsection{Joint forecasts for [CII] emission}

Finally, we combine the EXCLAIM forecasts from both [CII] power spectra and the [CII] CVID to predict the potential measurements of the Padmanabhan [CII] model.  We assume for this paper that uncertainties on our one- and two-point estimators are entirely uncorrelated.  Figure 2 of \citet{Ihle2019} shows that, for a similar intensity mapping model, the power spectrum and VID are mostly uncorrelated at low signal-to-noise, with nontrivial correlations appearing as the sample variance limit is approached.  We assume the former case will hold for early EXCLAIM measurements and leave a full exploration of the optimal combination of these estimators for future work.

\begin{figure*}
    \centering
        \includegraphics[width=0.45\textwidth]{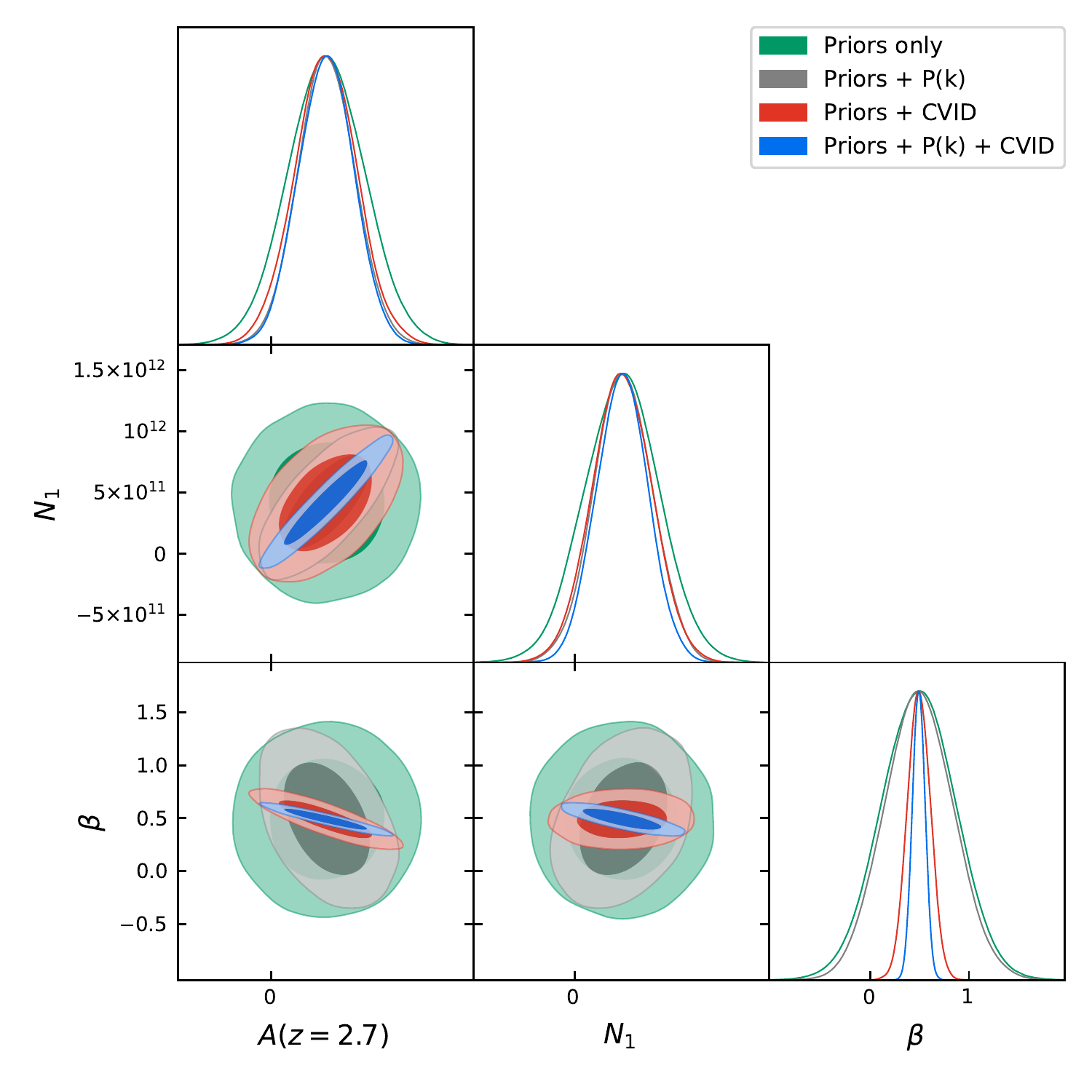}
    \includegraphics[width=0.45\textwidth]{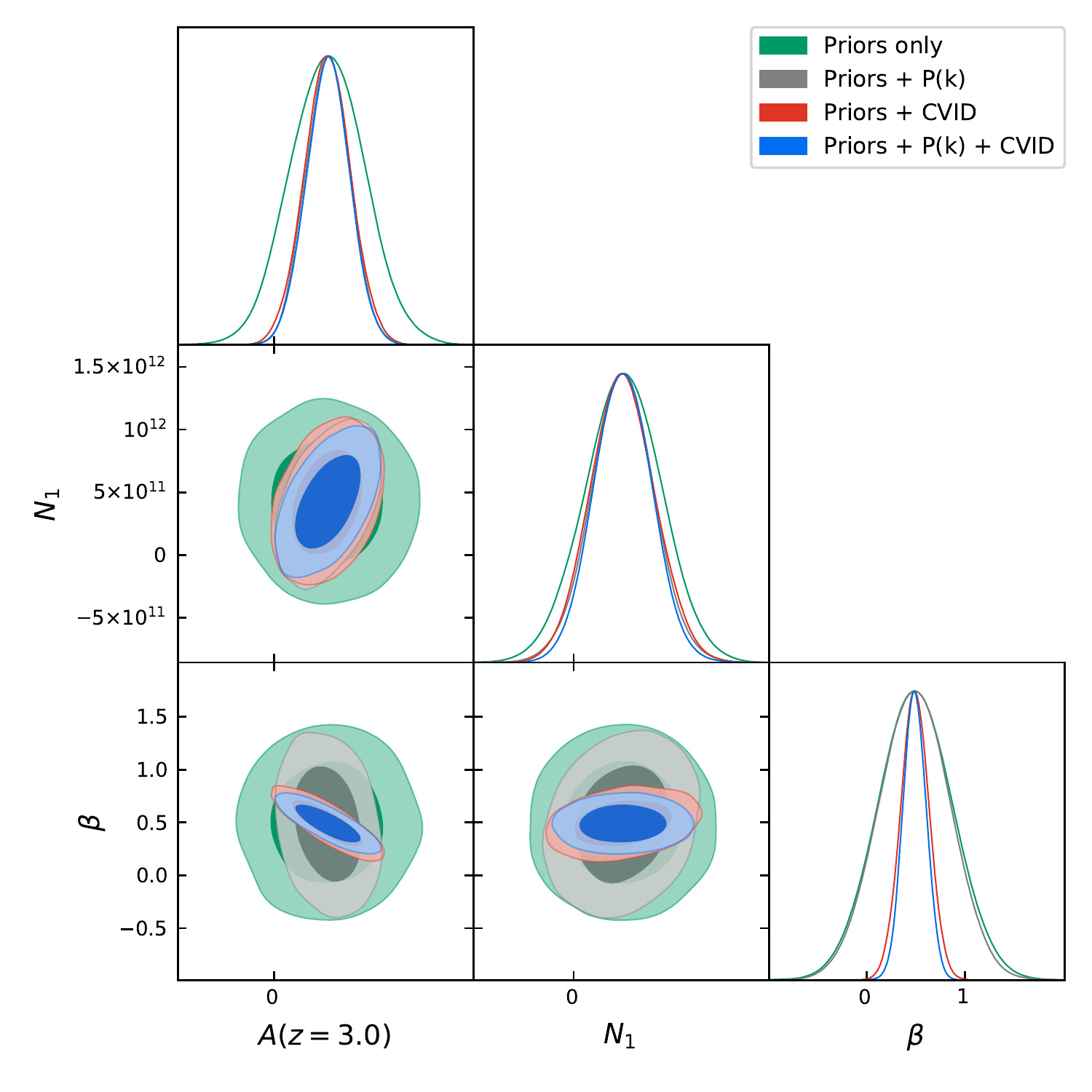}\\
    \includegraphics[width=0.45\textwidth]{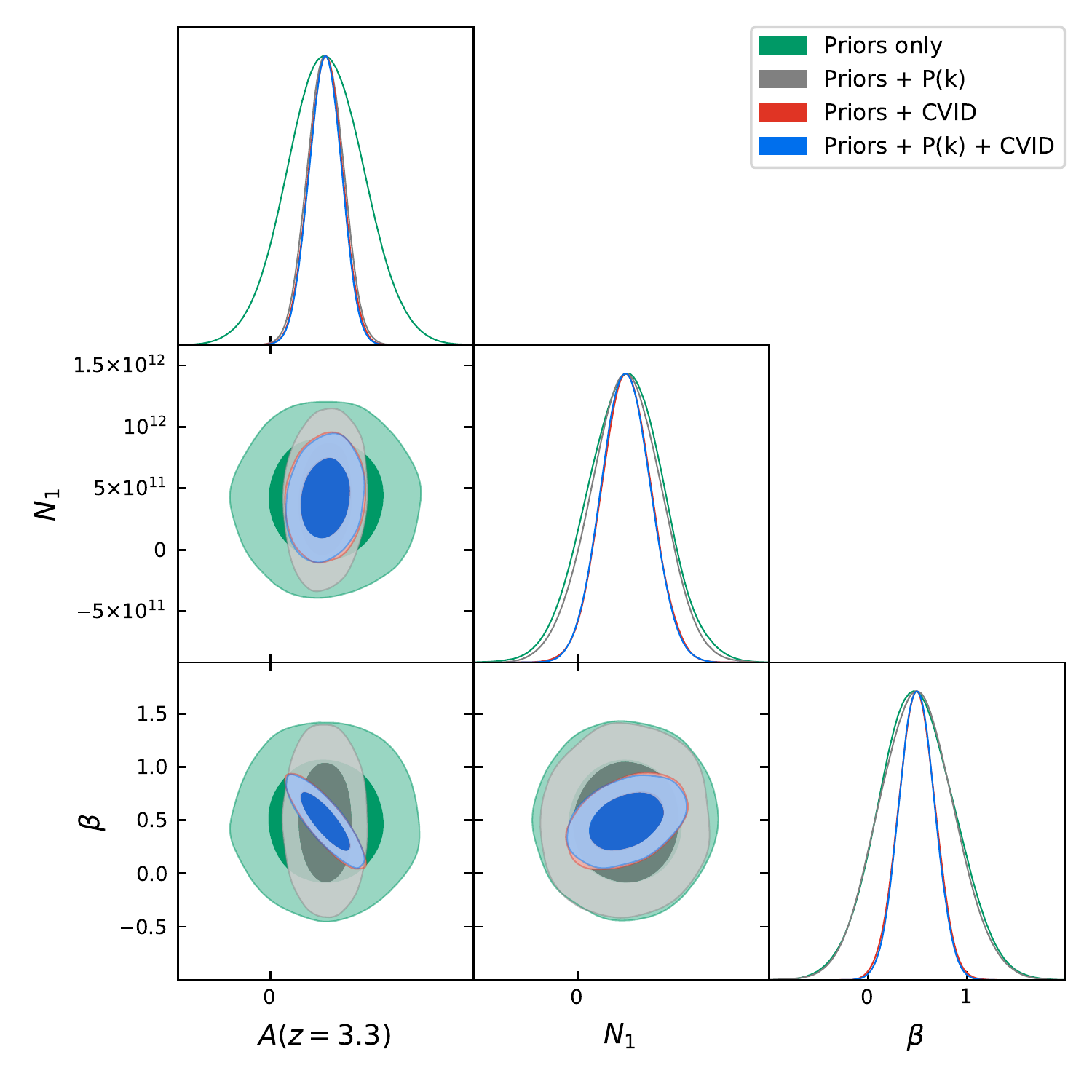}
    \caption{Forecasts on the Padmanabhan [CII] model parameters from both the [CII] power spectrum and the [CII] conditional voxel intensity distribution.  The gray shaded ellipses include only the priors from previous measurements in \citet{2019MNRAS.488.3014P}, while the red (blue) shaded ellipses add the power spectrum (CVID) with the prior.  The darker and lighter shades denote 68\% and 95\% confidence intervals, respectively.}
    \label{fig:joint}
\end{figure*}

Fig.~\ref{fig:joint} shows the constraints on the model parameters $A$, $\beta$, and $N_1$ from observables, individually and jointly.  Our forecasts predict that the CVID will be more instrumental than the power spectrum in constraining $\beta$. In contrast, the CVID and the power spectrum contribute roughly equally to constraining both $A$ and $N_1$. Under this model, the bright end of the CII luminosity function is strongly dependent on the $\beta$ parameter.  VID measurements, in general, are most sensitive to the tail of the luminosity function \citep{Ihle2019}, so it makes sense that the CVID constrains $\beta$ most strongly. We also see that for larger redshifts, the uncertainties from the CVID reduce much more than those from the power spectrum.  

Tables \ref{tab:bICIIlim} and \ref{tab:ICIIlim}, present SNRs for $b_\mathrm{[CII]}I_\mathrm{[CII]}$ and $I_\mathrm{[CII]}$ including information from the CVID only and a joint CVID \& power spectra analysis, both including parameter priors, using the formalism from Eqs.~\ref{E:sigbI} and \ref{E:sigI}.  We find that the CVID alone gives $b_\mathrm{[CII]}I_\mathrm{[CII]}$ SNRs$\sim$1.5 for all redshift bins assuming the Padmanabhan [CII] model, which is enough to increase the joint $P(k)$+CVID SNR for $b_\mathrm{[CII]}I_\mathrm{[CII]}$ to 2 at $z=3.0$. For intensity $I_\mathrm{[CII]}$, the CVID SNRs are approximately 1.3 for all redshift bins, which is more significant than for $b_\mathrm{[CII]}I_\mathrm{[CII]}$ due to the lower SNR from $P(k)$.  This raises the joint SNR to 6.6 at $z=2.7$ while the other bins' SNRs remain less than 2. For the Yang [CII] model, all measurements for $b_\mathrm{[CII]}I_\mathrm{[CII]}$ and $I_\mathrm{[CII]}$ have SNR < 2 for all $P(k)$ and CVID combinations, limiting this science yield if a more pessimistic model of [CII] is accurate.

\iffalse
\begin{figure*}
    \centering
     \includegraphics[width=0.45\textwidth]{figs/LM_limit_z2.7.pdf}
    \includegraphics[width=0.45\textwidth]{figs/LM_limit_z3.0.pdf}\\
    \includegraphics[width=0.45\textwidth]{figs/LM_limit_z3.3.pdf}
    \caption{Forecasts on the Padmanabhan (2019) $L_{\rm [CII]}(M)$ model (blue) and the Yang (2021) model (green) from both the [CII] power spectrum and the [CII] conditional voxel intensity distribution.  The darker and lighter shades denote 68\% and 95\% confidence intervals, respectively. \arp{Aaron suggests we improve this plot, possibly by replacing the curves with lines, making a light color for 95\% and lines for 68\%, or changing green to orange.  Same for SFR(M) figure.}}
    \label{fig:sigLM}
\end{figure*}
\fi

\begin{table*}
    \centering
    \begin{tabular}{|c||c|c|c|c|}
    \hline \rowcolor{lightgray}
    $z$ & $b_\mathrm{[CII]}I_\mathrm{[CII]}$ [kJy/sr] & $bI/\sigma(bI)$ [CII] ($P(k)$) & $bI/\sigma(bI)$ [CII] (CVID) & $bI/\sigma(bI)$ [CII] ($P(k)$ + CVID)\\
    \hline
    \hline
    2.7 & 98.4 / 26.8 & 8.06 / 1.25 & 1.75 / 1.8 & 8.20 / 2.00 \\
    3.0 & 55.8 / 24.0 & 1.75 / 0.87 & 1.39 / 1.90 & 2.07 / 1.92 \\
    3.3 & 30.3 / 20.5 & 0.99 / 0.73 & 1.41 / 1.38 & 1.51  / 1.40 \\
    \hline
    \end{tabular}
    \caption{[CII] biased intensity $b_\mathrm{[CII]}I_\mathrm{[CII]}$ forecasts from joint analyses including cross-power spectra and the CVID from EXCLAIM maps and S82 BOSS quasars for multiple line models.  For the power spectra, interlopers are marginalized, and the lowest $k_\parallel$ mode is removed.  The third column includes only power spectrum measurements, the fourth column includes only the CVID measurement, and the fifth column includes both power spectra and CVID measurements.  In all forecasts, the parameter priors are included.  In each cell, the entry on the left [right] corresponds to the [CII] model from \citet{2019MNRAS.488.3014P} [\citet{2022ApJ...929..140Y}].}
    \label{tab:bICIIlim}
\end{table*}

\begin{table*}
    \centering
    \begin{tabular}{|c||c|c|c|c|}
    \hline \rowcolor{lightgray}
    $z$ & $I_\mathrm{[CII]}$ [kJy/sr] & $I/\sigma(I)$ [CII] ($P(k)$) & $I/\sigma(I)$ [CII] (CVID) & $I/\sigma(I)$ [CII] ($P(k)$ + CVID)\\
    \hline
    \hline
    2.7 & 32.1 / 8.96 & 4.64 / 1.20 & 1.48 / 1.48 & 6.64 / 1.64 \\
    3.0 & 17.1 / 7.13 & 1.69 / 0.75 & 1.26 / 1.41 & 1.87 / 1.43 \\
    3.3 & 7.8 / 5.17 & 0.91 / 0.56 & 1.21 / 0.91 & 1.30  / 0.93 \\
    \hline
    \end{tabular}
    \caption{[CII] intensity $I_\mathrm{[CII]}$ forecasts from joint analyses including cross-power spectra and the CVID from EXCLAIM maps and S82 BOSS quasars for multiple line models.  For the power spectra, interlopers are marginalized and the lowest $k_\parallel$ mode is removed.  The third column includes only power spectrum measurements, the fourth column includes only the CVID measurement, and the fifth column includes both power spectra and CVID measurements.  In all forecasts, the parameter priors are included.  In each cell, the entry on the left [right] corresponds to the [CII] model from \citet{2019MNRAS.488.3014P} [\citet{2022ApJ...929..140Y}].}
    \label{tab:ICIIlim}
\end{table*}

\begin{figure}
    \centering
    \includegraphics[width=.45\textwidth]{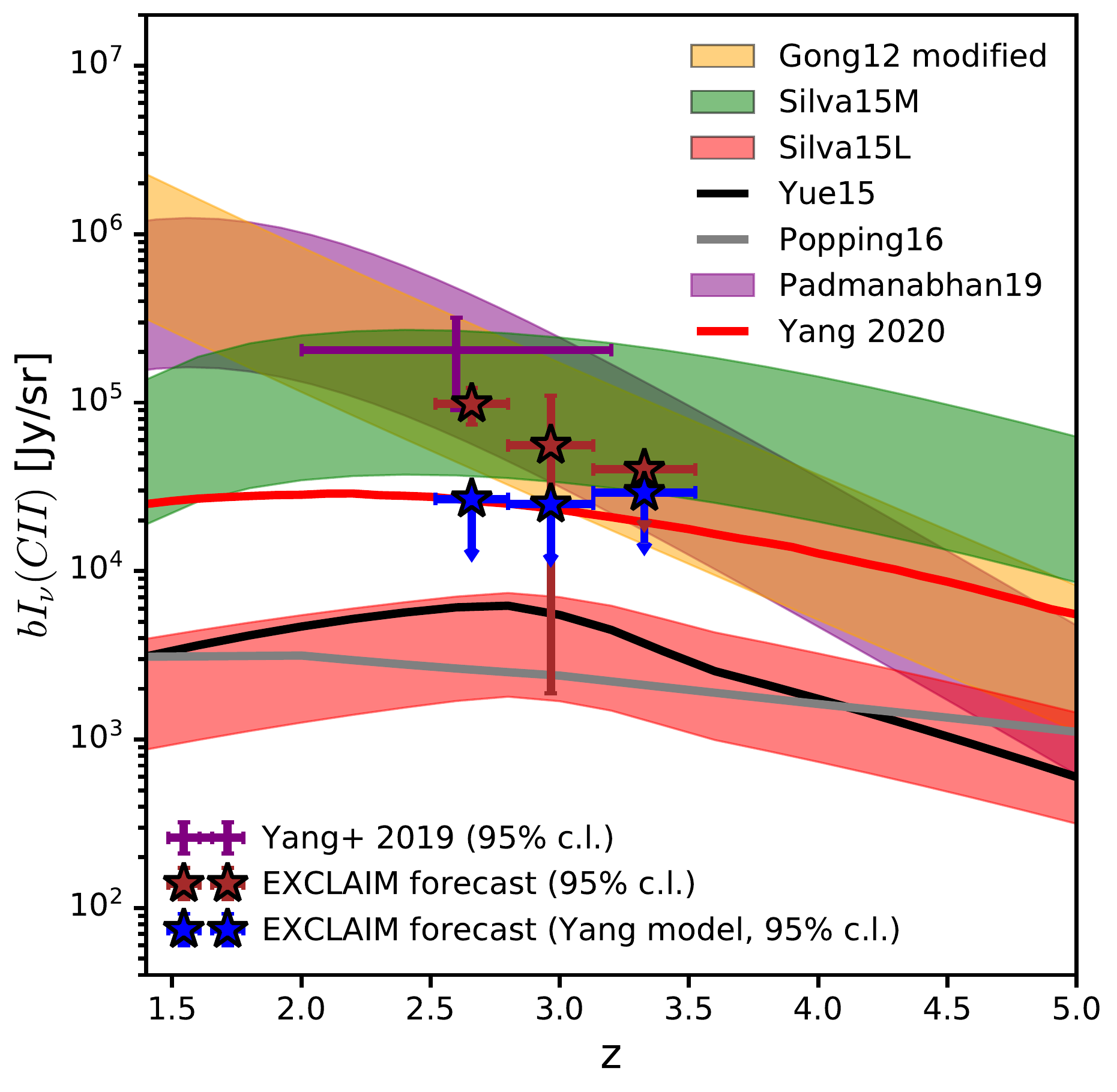}
    \caption{Various [CII] models for the quantity $bI_\nu$ as functions of redshift, along with a previous measurement from \citet{2019MNRAS.489L..53Y} and EXCLAIM forecasts for both fiducial [CII] models, including joint constraints from $P(k)$ and CVID measurements with a prior.  The shaded regions for most of the models correspond to the range of predictions associated with those models.  Crosses denote detection, while arrows denote upper limits.}
    \label{fig:CII_models}
\end{figure}

\begin{figure}
    \centering
    \includegraphics[width=.45\textwidth]{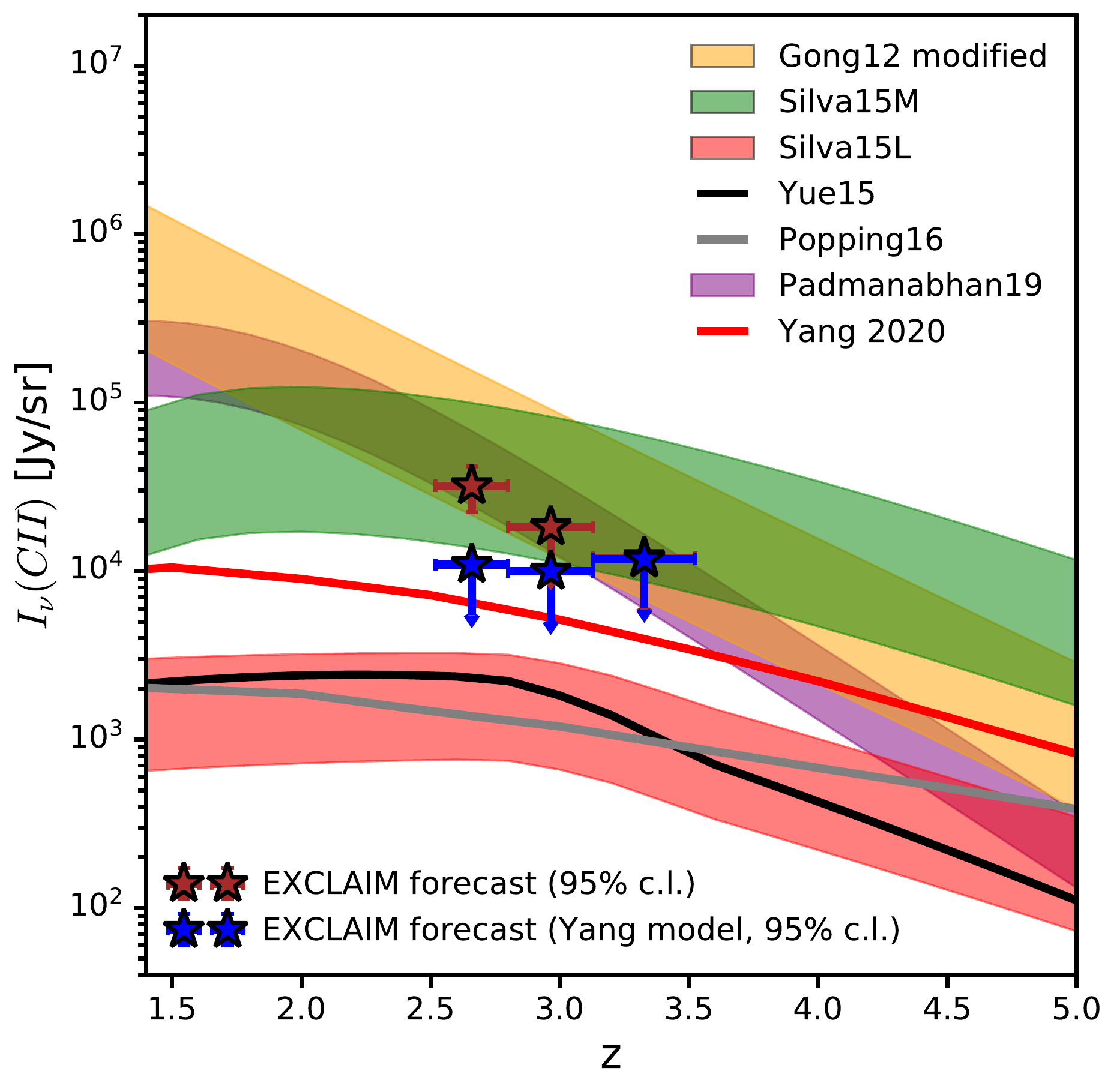}
    \caption{Various [CII] models for the intensity $I_\nu$ as functions of redshift, along with EXCLAIM forecasts for both fiducial [CII] models including joint constraints from $P(k)$ and CVID measurements with a prior.  The shaded regions for most of the models correspond to the range of predictions associated with those models.  Crosses denote detection, while arrows denote upper limits.}
    \label{fig:CII_models_nob}
\end{figure}

We plot the intensities of [CII] for the fiducial models we present, along with alternative models for both lines in the literature. In Fig.~\ref{fig:CII_models}, we plot the [CII] intensity multiplied by the clustering bias (see Sec.~\ref{S:Pkmod}) as functions of redshift for both the fiducial \citet{2019MNRAS.488.3014P} model and several other models \citep{2012ApJ...745...49G,2015ApJ...806..209S,2015MNRAS.450.3829Y,2016MNRAS.461...93P,2022ApJ...929..140Y}.  In Fig.~\ref{fig:CII_models_nob}, we plot just the [CII] intensity for the same models.  In both figures, we also show the expected errors from cross-power spectra of EXCLAIM measurements of [CII] with Stripe 82 quasars.  These plots show how the errors from EXCLAIM could compare to the range of models in the literature and confirm that [CII] intensity at $z=2.7$ should be measurable.

\iffalse
\subsection{Common parameters for $P(k)$, $C_\ell$, and $VID$ model comparison}

\begin{itemize}
    \item First step: CII only, Padmanabhan 2019 model.  Three fitted parameters: $\beta$, $M_1$, amplitude (combination of $M_1$ and $\alpha$).
    \item Three redshift bins in the instrument model shown in Table \ref{tab:EXCLAIM_PARAMS}
    \item BOSS-S82 survey parameters for CII redshifts: $\bar{n} = 1.06\times10^{-6} ~ Mpc^{-3}$, $b = 3.5$, $(\Delta RA, \Delta DEC) = (105\deg,2.6\deg)$
    \item Two cases: with and without foregrounds from Tony Zhou's PySM model
    \item Assuming fixed values for cosmological parameters
    \item Anderson and Oxholm: velocity dispersion $500~km/s$ (not included in VID)
    \item Assume low-mass halo cutoffs of $M_{\rm CII}^{min} = 10^{10}~M_{\sun}$ and $M_{gal}^{min} = 10^{12}~M_{\sun}$ for CII and galaxy clustering, respectively.
    \item Assume Tinker2010 halo mass function
    \item Assume eBOSS QSO HOD for cross-shot power from Table 2 in arXiv: 2007.09003 -- ambiguity in which model to choose.
    \item Not including interlopers (Oxholm will show they are negligible for CII) or CIB
\end{itemize}

\fi

\section{SFRD Forecasts for EXCLAIM} \label{S:science}

EXCLAIM's sensitivity to [CII] intensity at $z=2.7$ enables a measurement of the star formation rate density (SFRD), for which the [CII] line is known to be a strong tracer.  Specifically, [CII] luminosity has been shown to be related to the star formation rate (SFR) by a power law, $L_\mathrm{[CII]}\propto {\rm SFR}^\alpha$ \citep{2016ApJ...829L..11P,2016ApJ...833...71A}.  To keep this model consistent with our fiducial model \citep{2019MNRAS.488.3014P}, we set $\alpha=1.79$ and the redshift-dependent power law proportionality constant such that the resulting SFR density (SFRD), assuming a Sheth-Tormen halo mass function, matches the SFRD fitting function from \citet{Behroozi2013}.  We also construct SFRD forecasts for the \citet{2022ApJ...929..140Y} model.  This SAM-based model provides the [CII] luminosity and SFR for each halo, which we use to fit a power law $L=C(SFR)^\alpha$ for the $L-SFR$ relation.  We find $\log C=(6.85,6.81,6.80)$ and $\alpha=(1.29,1.26,1.25)$ for $z=(2.7,3.0,3.3)$, respectively.  We then use this relation to create a forecast.  Using both models, we forecast limits on the $SFRD$ assuming the Sheth-Tormen halo model mass function $n(M)$.  We forecast the uncertainty in the SFRD for both models using the earlier computed covariance matrix for each model's parameters, according to the formula
\begin{eqnarray} \label{E:sigSFRD}
\sigma^2[{\rm SFRD}] = [\nabla_{\mathbf{p}}{\rm SFRD}|_{\mathbf{p}_0}]^T\cdot\mathbf{C}\cdot\nabla_{\mathbf{p}}{\rm SFRD}|_{\mathbf{p}_0}\, .
\end{eqnarray}
Although the $L-{\rm SFR}$ proportionality constant must vary with the parameters for the Padmanabhan [CII] model, we fix it when we compute $\nabla_{\mathbf{p}}{\rm SFRD}|_{\mathbf{p}_0}$.  We treat it as we would in a typical measurement, where the constant is set to give a reasonable result for the SFRD based on previous data, but we do not know \emph{a priori} the exact value of the SFRD when constructing the $L-{\rm SFR}$ model.  We give the constraints for both models assuming various data combinations in Table \ref{tab:SFRDCIIlim}.  We see that only the $z=2.7$ measurement using both $P(k)$ and CVID data would be sensitive to the SFRD, and that is only if the true [CII] model is closer to the Padmanabhan model.  Note that the SFRD values shown in the second column of Table \ref{tab:SFRDCIIlim} for both models are not equal.  In particular, the SFRD prediction for the \citet{2022ApJ...929..140Y} model is not consistent with the model from \citet{Behroozi2013}.  This difference is because the SAM captures very low-mass halos that are not detected in the measurements used to construct the Behroozi model, potentially causing this model to under-predict the global SFRD.  We plot the forecasted constraint for the SFRD at $z=2.7$ for the Padmanabhan [CII] model along with known SFRD constraints from galaxy surveys, given in Table C3 of \citet{2019MNRAS.488.3143B}, in Fig.~\ref{fig:sfrd}.  The SNR for this SFRD measurement is 4, which is competitive with measurements from direct imaging; note this is significantly higher than the measurement assuming only the priors, which has an SNR of 1.  In addition, it gives a measurement over all star-forming galaxies without bias due to selection effects or sample variance. However, the uncertainty of the SFR to [CII] luminosity conversion contributes systematic errors in these measurements, though they may be reduced using results from hydrodynamic simulations and semi-analytic star formation models \citep{1999MNRAS.310.1087S,2015MNRAS.453.4337S,2014MNRAS.437.1662K,2014MNRAS.442.2398P,2019MNRAS.482.4906P}. 

\iffalse
\begin{figure*}
    \centering
     \includegraphics[width=0.45\textwidth]{figs/SFRM_limit_unnorm_z2.7.pdf}
    \includegraphics[width=0.45\textwidth]{figs/SFRM_limit_unnorm_z3.0.pdf}\\
    \includegraphics[width=0.45\textwidth]{figs/SFRM_limit_unnorm_z3.3.pdf}
    \caption{Forecasts on the ${\rm SFR}(M)$ model based on $L_{[CII]}\propto {\rm SFR}^\alpha$ and the $L_{\rm [CII]}(M)$ forecasts presented in Fig.~\ref{fig:sigLM}.  The darker and lighter shades denote 68\% and 95\% confidence intervals, respectively.  Note that both of these models result in equal values of the star-formation-rate density.}
    \label{fig:sfrm}
\end{figure*}
\fi

\begin{table*}
    \centering 
    \begin{tabular}{|c||c|c|c|c|}
    \hline \rowcolor{lightgray}
    $z$ & ${\rm SFRD}$ [$M_\odot$/yr] & ${\rm SFRD}/\sigma({\rm SFRD})$ ($P(k)$) & ${\rm SFRD}/\sigma({\rm SFRD})$ (CVID) & ${\rm SFRD}/\sigma({\rm SFRD})$ ($P(k)$ + CVID)\\
    \hline
    \hline
    2.7 & 0.104 / 0.285 & 2.26 / 1.15 & 1.47 / 1.37 & 4.22 / 1.48 \\
    3.0 & 0.090 / 0.302 & 1.58 / 0.96 & 1.24 / 1.49 & 1.78 / 1.52 \\
    3.3 & 0.076 / 0.281 & 1.04 / 0.92 & 1.34 / 1.21 & 1.42  / 1.23 \\
    \hline
    \end{tabular}
    \caption{SFRD forecasts from joint analyses including [CII] cross-power spectra and the CVID from EXCLAIM maps and S82 BOSS quasars for multiple line models.  For the power spectra, interlopers are marginalized and the lowest $k_\parallel$ mode is removed.  The third column includes only power spectrum measurements, the fourth column includes only the CVID measurement, and the fifth column includes both power spectra and CVID measurements.  In all forecasts, the parameter priors are included.  In each cell, the entry on the left [right] corresponds to the [CII] model from \citet{2019MNRAS.488.3014P} [\citet{2022ApJ...929..140Y}].}
    \label{tab:SFRDCIIlim}
\end{table*}

\begin{figure}
    \centering
     \includegraphics[width=0.45\textwidth]{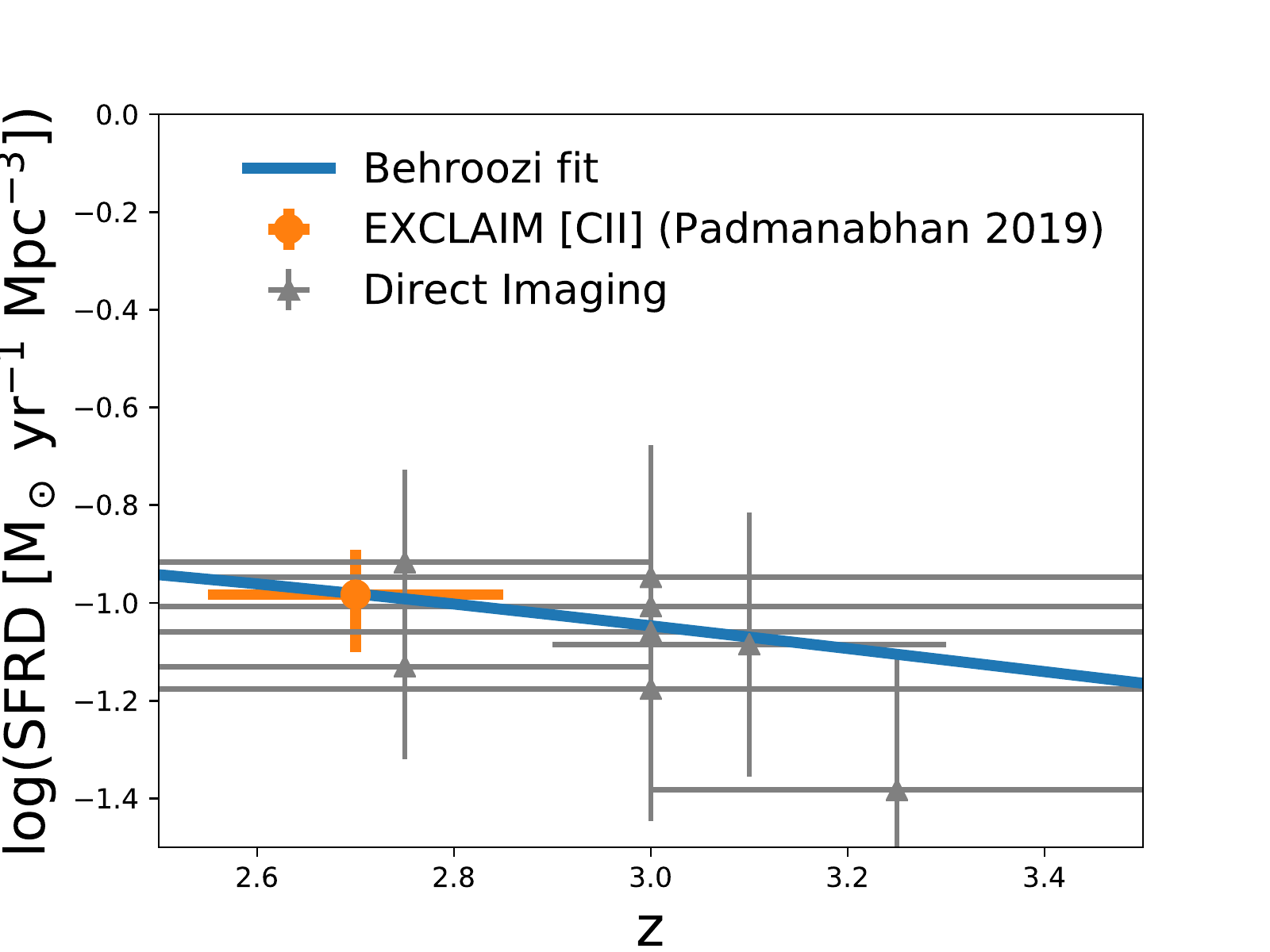}
    \caption{Forecast on the ${\rm SFRD}(z)$ model at $z=2.7$ from [CII] based on the Padmanabhan [CII] model.  The orange bars denote 68\% confidence intervals EXCLAIM forecast, while the gray bars denote current limits from direct imaging results from galaxy surveys, tabulated in Table C3 of \citet{2019MNRAS.488.3143B}.  The blue line is the SFRD fit from \citet{Behroozi2013}.}
    \label{fig:sfrd}
\end{figure}
\iffalse
We will also use the CO(5-4) and CO(6-5) measurements to constrain the star formation rate.  CO emission lines from high-$J$ transitions are known to be strong tracers of star formation.  In particular, CO luminosity for high-$J$ lines is proportional to far-infrared luminosity \citep{2015A&A...577A..46D}, which is proportional to the SFR \citep{2012ARA&A..50..531K, 2011ApJ...741..124H, 2011ApJ...737...67M}.  A linear $L_{\rm CO}$-SFR relation makes measuring the SFRD much simpler since we can directly relate the CO intensity $I_{\rm CO}$ to the SFRD without introducing model uncertainties in measuring $L(M)$, which are required when using [CII].  The signal-to-noise ratio (SNR) for the SFRD will be equal to the SNR for the CO intensities, which are approximately 6 for both CO lines.  We plot forecasted SFRD measurements using CO line emission in Fig.~\ref{fig:sfrd}.
\fi

\section{Conclusions} \label{S:conclude}

EXCLAIM is poised to provide a unique probe for galaxy evolution across cosmic time using the line intensity mapping technique.  In this paper, we outline the instrumentation of the balloon telescope and the survey plan.  We then construct forecasts for EXCLAIM's upcoming measurements of the cross-power spectra for both [CII] at $z=3$ and CO lines for $J=4-7$ at $z<1$ with quasar and galaxy surveys, respectively. We show that EXCLAIM will potentially measure the cross-power spectrum for both lines with high sensitivity for a range of emission models and for different galaxy surveys.  These emission effects also include the effects of expected line interloper and continuum foregrounds.  These measurements will allow EXCLAIM to distinguish between widely varying CO and [CII] models in the literature.

We also show that the conditional VID will be measured by EXCLAIM for [CII] with high precision.  Using both the cross-power spectra and the CVID, EXCLAIM can constrain the [CII] luminosity-mass relation $L_\mathrm{[CII]}(M)$ parameters.  Then, using known $L({\rm SFR})$ relations for [CII], we can forecast measurements for the star formation rate density, predicting that EXCLAIM can potentially measure this quantity with an SNR$\sim$4 at $z=2.7$.  This LIM measurement of the SFRD would be unique from measurements using bright galaxy spectra because LIM provides a global measurement including emission from less massive galaxies that may not be detected in traditional surveys of individual galaxies.

EXCLAIM has the potential to be the first LIM survey to provide precise measurements of the SFRD.  This mission will serve as a pathfinder for future balloon-borne and space-based LIM surveys with greater capabilities, and this paper lays out the items to consider when assessing the power of these surveys.  In addition, as the relation of [CII] to galaxy properties such as the SFRD and metallicity become better understood, potentially from galaxy evolution simulations and semi-analytic models, EXCLAIM may be able to measure the distribution of multiple galaxy properties across cosmic time.

Finally, the [CII] power spectrum from EXCLAIM at $z=2.7$ may be able to measure the Hubble expansion rate $H(z)$ through its BAO signature.  The SNR for the [CII] power spectrum at $z=2.7$ is predicted to be within the range of 7-11, depending on the true [CII] model.  \citet{2019PhRvL.123y1301B} forecast that COMAP, which is predicted to have an SNR for its power spectrum of a similar value, could measure $H(z)$ with the SNR=5.  Thus, we expect that EXCLAIM could measure $H(z)$ with a similar sensitivity at $z=2.7$.  We also expect this measurement to be insensitive to uncertainties in the [CII] intensity since the BAO frequency is unaffected by the height of the power spectrum.  Of course, this measurement would not be competitive with $z>2$ current high-precision measurements of $H(z)$ from the Ly$\alpha$ forest \citep{2014JCAP...05..027F} or upcoming measurements from the Roman Space Telescope High Latitude Survey \citep{2022ApJ...928....1W}.  However, the EXCLAIM measurement could serve as a proof-of-concept for a future LIM mission with better sensitivity that can measure $H(z)$ over volumes and redshifts inaccessible to upcoming galaxy surveys and place significant constraints on models of dark energy and early dark energy \citep{2016PhRvD..94j3523K, 2021MNRAS.505.2285B}.

\section*{Acknowledgements}

EXCLAIM began in April 2019 as a 5-year NASA Astrophysics Research and Analysis (APRA 1263 17-APRA17-0077) grant. ARP was supported by NASA under award numbers 80NSSC18K1014, NNH17ZDA001N, and 80NSSC22K0666, and by the NSF under award number 2108411. ARP was also supported by the Simons Foundation.  PCB was supported by the James Arthur Postdoctoral Fellowship. AY is supported by an appointment to the NASA Postdoctoral Program (NPP) at NASA Goddard Space Flight Center, administered by Oak Ridge Associated Universities under contract with NASA.

%\arp{Please add other sources of funding}

%%%%%%%%%%%%%%%%%%%%%%%%%%%%%%%%%%%%%%%%%%%%%%%%%%



\bibliographystyle{mnras}

\begin{thebibliography}{}
\makeatletter
\relax
\def\mn@urlcharsother{\let\do\@makeother \do\$\do\&\do\#\do\^\do\_\do\%\do\~}
\def\mn@doi{\begingroup\mn@urlcharsother \@ifnextchar [ {\mn@doi@}
  {\mn@doi@[]}}
\def\mn@doi@[#1]#2{\def\@tempa{#1}\ifx\@tempa\@empty \href
  {http://dx.doi.org/#2} {doi:#2}\else \href {http://dx.doi.org/#2} {#1}\fi
  \endgroup}
\def\mn@eprint#1#2{\mn@eprint@#1:#2::\@nil}
\def\mn@eprint@arXiv#1{\href {http://arxiv.org/abs/#1} {{\tt arXiv:#1}}}
\def\mn@eprint@dblp#1{\href {http://dblp.uni-trier.de/rec/bibtex/#1.xml}
  {dblp:#1}}
\def\mn@eprint@#1:#2:#3:#4\@nil{\def\@tempa {#1}\def\@tempb {#2}\def\@tempc
  {#3}\ifx \@tempc \@empty \let \@tempc \@tempb \let \@tempb \@tempa \fi \ifx
  \@tempb \@empty \def\@tempb {arXiv}\fi \@ifundefined
  {mn@eprint@\@tempb}{\@tempb:\@tempc}{\expandafter \expandafter \csname
  mn@eprint@\@tempb\endcsname \expandafter{\@tempc}}}

\bibitem[\protect\citeauthoryear{{Abazajian} et~al.,}{{Abazajian}
  et~al.}{2009}]{2009ApJS..182..543A}
{Abazajian} K.~N.,  et~al., 2009, \mn@doi [\apjs]
  {10.1088/0067-0049/182/2/543}, \href
  {https://ui.adsabs.harvard.edu/abs/2009ApJS..182..543A} {182, 543}

\bibitem[\protect\citeauthoryear{{Ade} et~al.,}{{Ade}
  et~al.}{2020}]{2020JLTP..199.1027A}
{Ade} P.~A.~R.,  et~al., 2020, \mn@doi [Journal of Low Temperature Physics]
  {10.1007/s10909-019-02320-5}, \href
  {https://ui.adsabs.harvard.edu/abs/2020JLTP..199.1027A} {199, 1027}

\bibitem[\protect\citeauthoryear{{Ahumada} et~al.,}{{Ahumada}
  et~al.}{2020}]{2020ApJS..249....3A}
{Ahumada} R.,  et~al., 2020, \mn@doi [\apjs] {10.3847/1538-4365/ab929e}, \href
  {https://ui.adsabs.harvard.edu/abs/2020ApJS..249....3A} {249, 3}

\bibitem[\protect\citeauthoryear{{Aihara} et~al.,}{{Aihara}
  et~al.}{2019}]{2019PASJ...71..114A}
{Aihara} H.,  et~al., 2019, \mn@doi [\pasj] {10.1093/pasj/psz103}, \href
  {https://ui.adsabs.harvard.edu/abs/2019PASJ...71..114A} {71, 114}

\bibitem[\protect\citeauthoryear{{Alam} et~al.,}{{Alam}
  et~al.}{2015}]{2015ApJS..219...12A}
{Alam} S.,  et~al., 2015, \mn@doi [\apjs] {10.1088/0067-0049/219/1/12}, \href
  {https://ui.adsabs.harvard.edu/abs/2015ApJS..219...12A} {219, 12}

\bibitem[\protect\citeauthoryear{{Anderson} et~al.,}{{Anderson}
  et~al.}{2018}]{2018MNRAS.476.3382A}
{Anderson} C.~J.,  et~al., 2018, \mn@doi [\mnras] {10.1093/mnras/sty346}, \href
  {https://ui.adsabs.harvard.edu/abs/2018MNRAS.476.3382A} {476, 3382}

\bibitem[\protect\citeauthoryear{{Aravena} et~al.,}{{Aravena}
  et~al.}{2016}]{2016ApJ...833...71A}
{Aravena} M.,  et~al., 2016, \mn@doi [\apj] {10.3847/1538-4357/833/1/71}, \href
  {https://ui.adsabs.harvard.edu/abs/2016ApJ...833...71A} {833, 71}

\bibitem[\protect\citeauthoryear{{Barrentine} et~al.,}{{Barrentine}
  et~al.}{2016}]{uSpec_Barrentine_2016}
{Barrentine} E.~M.,  et~al., 2016, in {Holland} W.~S.,  {Zmuidzinas} J.,  eds,
  Society of Photo-Optical Instrumentation Engineers (SPIE) Conference Series
  Vol. 9914, Millimeter, Submillimeter, and Far-Infrared Detectors and
  Instrumentation for Astronomy VIII. p. 99143O, \mn@doi{10.1117/12.2234462}

\bibitem[\protect\citeauthoryear{Behroozi, Wechsler  \& Conroy}{Behroozi
  et~al.}{2013}]{Behroozi2013}
Behroozi P.~S.,  Wechsler R.~H.,   Conroy C.,  2013, \mn@doi [Astrophysical
  Journal] {10.1088/0004-637X/770/1/57}, 770

\bibitem[\protect\citeauthoryear{{Behroozi}, {Wechsler}, {Hearin}  \&
  {Conroy}}{{Behroozi} et~al.}{2019}]{2019MNRAS.488.3143B}
{Behroozi} P.,  {Wechsler} R.~H.,  {Hearin} A.~P.,   {Conroy} C.,  2019,
  \mn@doi [\mnras] {10.1093/mnras/stz1182}, \href
  {https://ui.adsabs.harvard.edu/abs/2019MNRAS.488.3143B} {488, 3143}

\bibitem[\protect\citeauthoryear{{Bernal} \& {Kovetz}}{{Bernal} \&
  {Kovetz}}{2022}]{2022arXiv220615377B}
{Bernal} J.~L.,  {Kovetz} E.~D.,  2022, arXiv e-prints, \href
  {https://ui.adsabs.harvard.edu/abs/2022arXiv220615377B} {p. arXiv:2206.15377}

\bibitem[\protect\citeauthoryear{{Bernal}, {Breysse}, {Gil-Mar{\'\i}n}  \&
  {Kovetz}}{{Bernal} et~al.}{2019a}]{2019PhRvD.100l3522B}
{Bernal} J.~L.,  {Breysse} P.~C.,  {Gil-Mar{\'\i}n} H.,   {Kovetz} E.~D.,
  2019a, \mn@doi [\prd] {10.1103/PhysRevD.100.123522}, \href
  {https://ui.adsabs.harvard.edu/abs/2019PhRvD.100l3522B} {100, 123522}

\bibitem[\protect\citeauthoryear{{Bernal}, {Breysse}  \& {Kovetz}}{{Bernal}
  et~al.}{2019b}]{2019PhRvL.123y1301B}
{Bernal} J.~L.,  {Breysse} P.~C.,   {Kovetz} E.~D.,  2019b, \mn@doi [\prl]
  {10.1103/PhysRevLett.123.251301}, \href
  {https://ui.adsabs.harvard.edu/abs/2019PhRvL.123y1301B} {123, 251301}

\bibitem[\protect\citeauthoryear{{Bolatto}, {Wolfire}  \& {Leroy}}{{Bolatto}
  et~al.}{2013}]{2013ARA&A..51..207B}
{Bolatto} A.~D.,  {Wolfire} M.,   {Leroy} A.~K.,  2013, \mn@doi [\araa]
  {10.1146/annurev-astro-082812-140944}, \href
  {https://ui.adsabs.harvard.edu/abs/2013ARA&A..51..207B} {51, 207}

\bibitem[\protect\citeauthoryear{{Breysse}, {Kovetz}, {Behroozi}, {Dai}  \&
  {Kamionkowski}}{{Breysse} et~al.}{2017}]{Breysse2017}
{Breysse} P.~C.,  {Kovetz} E.~D.,  {Behroozi} P.~S.,  {Dai} L.,
  {Kamionkowski} M.,  2017, \mn@doi [\mnras] {10.1093/mnras/stx203}, \href
  {https://ui.adsabs.harvard.edu/abs/2017MNRAS.467.2996B} {467, 2996}

\bibitem[\protect\citeauthoryear{{Breysse}, {Anderson}  \& {Berger}}{{Breysse}
  et~al.}{2019}]{Breysse2019}
{Breysse} P.~C.,  {Anderson} C.~J.,   {Berger} P.,  2019, \mn@doi [\prl]
  {10.1103/PhysRevLett.123.231105}, \href
  {https://ui.adsabs.harvard.edu/abs/2019PhRvL.123w1105B} {123, 231105}

\bibitem[\protect\citeauthoryear{{Bull}, {White}  \& {Slosar}}{{Bull}
  et~al.}{2021}]{2021MNRAS.505.2285B}
{Bull} P.,  {White} M.,   {Slosar} A.,  2021, \mn@doi [\mnras]
  {10.1093/mnras/stab1338}, \href
  {https://ui.adsabs.harvard.edu/abs/2021MNRAS.505.2285B} {505, 2285}

\bibitem[\protect\citeauthoryear{{CHIME Collaboration} et~al.,}{{CHIME
  Collaboration} et~al.}{2022}]{2022arXiv220201242C}
{CHIME Collaboration} et~al., 2022, arXiv e-prints, \href
  {https://ui.adsabs.harvard.edu/abs/2022arXiv220201242C} {p. arXiv:2202.01242}

\bibitem[\protect\citeauthoryear{Carilli \& Walter}{Carilli \&
  Walter}{2013}]{carilli2013}
Carilli C.~L.,  Walter F.,  2013, \mn@doi [Annual Review of Astronomy and
  Astrophysics] {10.1146/annurev-astro-082812-140953}, 51, 105

\bibitem[\protect\citeauthoryear{{Cataldo}, {Hsieh}, {Huang}, {Moseley},
  {Stevenson}  \& {Wollack}}{{Cataldo} et~al.}{2014}]{uSpec_Cataldo_2014}
{Cataldo} G.,  {Hsieh} W.-T.,  {Huang} W.-C.,  {Moseley} S.~H.,  {Stevenson}
  T.~R.,   {Wollack} E.~J.,  2014, \mn@doi [\ao] {10.1364/AO.53.001094}, \href
  {https://ui.adsabs.harvard.edu/abs/2014ApOpt..53.1094C} {53, 1094}

\bibitem[\protect\citeauthoryear{{Cataldo}, {Moseley}  \& {Wollack}}{{Cataldo}
  et~al.}{2015}]{2015AcAau.114...54C}
{Cataldo} G.,  {Moseley} S.~H.,   {Wollack} E.~J.,  2015, \mn@doi [Acta
  Astronautica] {10.1016/j.actaastro.2015.04.002}, \href
  {https://ui.adsabs.harvard.edu/abs/2015AcAau.114...54C} {114, 54}

\bibitem[\protect\citeauthoryear{{Cataldo} et~al.,}{{Cataldo}
  et~al.}{2018}]{uSpec_Second_Gen_Cataldo_2018}
{Cataldo} G.,  et~al., 2018, \mn@doi [Journal of Low Temperature Physics]
  {10.1007/s10909-018-1902-7}, \href
  {https://ui.adsabs.harvard.edu/abs/2018JLTP..193..923C} {193, 923}

\bibitem[\protect\citeauthoryear{{Cataldo} et~al.,}{{Cataldo}
  et~al.}{2020}]{2020SPIE11445E..24C}
{Cataldo} G.,  et~al., 2020, in Society of Photo-Optical Instrumentation
  Engineers (SPIE) Conference Series. p. 1144524, \mn@doi{10.1117/12.2561069}

\bibitem[\protect\citeauthoryear{{Chang}, {Pen}, {Bandura}  \&
  {Peterson}}{{Chang} et~al.}{2010}]{2010Natur.466..463C}
{Chang} T.-C.,  {Pen} U.-L.,  {Bandura} K.,   {Peterson} J.~B.,  2010, \mn@doi
  [\nat] {10.1038/nature09187}, \href
  {https://ui.adsabs.harvard.edu/abs/2010Natur.466..463C} {466, 463}

\bibitem[\protect\citeauthoryear{{Cucciati} et~al.,}{{Cucciati}
  et~al.}{2012}]{2012A&A...539A..31C}
{Cucciati} O.,  et~al., 2012, \mn@doi [\aap] {10.1051/0004-6361/201118010},
  \href {https://ui.adsabs.harvard.edu/abs/2012A&A...539A..31C} {539, A31}

\bibitem[\protect\citeauthoryear{{Cunnington} et~al.,}{{Cunnington}
  et~al.}{2022}]{2022arXiv220601579C}
{Cunnington} S.,  et~al., 2022, arXiv e-prints, \href
  {https://ui.adsabs.harvard.edu/abs/2022arXiv220601579C} {p. arXiv:2206.01579}

\bibitem[\protect\citeauthoryear{{De Looze} et~al.,}{{De Looze}
  et~al.}{2014}]{DeLooze2014}
{De Looze} I.,  et~al., 2014, \mn@doi [A\&A] {10.1051/0004-6361/201322489},
  \href {https://ui.adsabs.harvard.edu/abs/2014A&A...568A..62D} {568, A62}

\bibitem[\protect\citeauthoryear{{Delabrouille} et~al.,}{{Delabrouille}
  et~al.}{2013}]{2013A&A...553A..96D}
{Delabrouille} J.,  et~al., 2013, \mn@doi [\aap] {10.1051/0004-6361/201220019},
  \href {https://ui.adsabs.harvard.edu/abs/2013A&A...553A..96D} {553, A96}

\bibitem[\protect\citeauthoryear{{Dunne} et~al.,}{{Dunne}
  et~al.}{2009}]{2009MNRAS.394....3D}
{Dunne} L.,  et~al., 2009, \mn@doi [\mnras] {10.1111/j.1365-2966.2008.13900.x},
  \href {https://ui.adsabs.harvard.edu/abs/2009MNRAS.394....3D} {394, 3}

\bibitem[\protect\citeauthoryear{Eftekharzadeh et~al.,}{Eftekharzadeh
  et~al.}{2015a}]{Eftekharzadeh2015}
Eftekharzadeh S.,  et~al., 2015a, Monthly Notices of the Royal Astronomical
  Society, 453, 2779

\bibitem[\protect\citeauthoryear{{Eftekharzadeh} et~al.,}{{Eftekharzadeh}
  et~al.}{2015b}]{2015MNRAS.453.2779E}
{Eftekharzadeh} S.,  et~al., 2015b, \mn@doi [\mnras] {10.1093/mnras/stv1763},
  \href {https://ui.adsabs.harvard.edu/abs/2015MNRAS.453.2779E} {453, 2779}

\bibitem[\protect\citeauthoryear{{Ellis} et~al.,}{{Ellis}
  et~al.}{2013}]{2013ApJ...763L...7E}
{Ellis} R.~S.,  et~al., 2013, \mn@doi [\apjl] {10.1088/2041-8205/763/1/L7},
  \href {https://ui.adsabs.harvard.edu/abs/2013ApJ...763L...7E} {763, L7}

\bibitem[\protect\citeauthoryear{{Essinger-Hileman} et~al.,}{{Essinger-Hileman}
  et~al.}{2020}]{2020SPIE11453E..0HE}
{Essinger-Hileman} T.,  et~al., 2020, in Society of Photo-Optical
  Instrumentation Engineers (SPIE) Conference Series. p. 114530H (\mn@eprint
  {arXiv} {2012.10481}), \mn@doi{10.1117/12.2576254}

\bibitem[\protect\citeauthoryear{{Fixsen} et~al.,}{{Fixsen}
  et~al.}{2011}]{ARCADE2_Fixsen_Results_2011}
{Fixsen} D.~J.,  et~al., 2011, \mn@doi [\apj] {10.1088/0004-637X/734/1/5},
  \href {https://ui.adsabs.harvard.edu/abs/2011ApJ...734....5F} {734, 5}

\bibitem[\protect\citeauthoryear{{Font-Ribera} et~al.,}{{Font-Ribera}
  et~al.}{2014}]{2014JCAP...05..027F}
{Font-Ribera} A.,  et~al., 2014, \mn@doi [\jcap]
  {10.1088/1475-7516/2014/05/027}, \href
  {https://ui.adsabs.harvard.edu/abs/2014JCAP...05..027F} {2014, 027}

\bibitem[\protect\citeauthoryear{{Gong}, {Cooray}, {Silva}, {Santos}, {Bock},
  {Bradford}  \& {Zemcov}}{{Gong} et~al.}{2012}]{2012ApJ...745...49G}
{Gong} Y.,  {Cooray} A.,  {Silva} M.,  {Santos} M.~G.,  {Bock} J.,  {Bradford}
  C.~M.,   {Zemcov} M.,  2012, \mn@doi [\apj] {10.1088/0004-637X/745/1/49},
  \href {http://adsabs.harvard.edu/abs/2012ApJ...745...49G} {745, 49}

\bibitem[\protect\citeauthoryear{{Grogin} et~al.,}{{Grogin}
  et~al.}{2011}]{2011ApJS..197...35G}
{Grogin} N.~A.,  et~al., 2011, \mn@doi [\apjs] {10.1088/0067-0049/197/2/35},
  \href {https://ui.adsabs.harvard.edu/abs/2011ApJS..197...35G} {197, 35}

\bibitem[\protect\citeauthoryear{{Guo} et~al.,}{{Guo}
  et~al.}{2015b}]{2015MNRAS.453.4368G}
{Guo} H.,  et~al., 2015b, \mn@doi [\mnras] {10.1093/mnras/stv1966}, \href
  {https://ui.adsabs.harvard.edu/abs/2015MNRAS.453.4368G} {453, 4368}

\bibitem[\protect\citeauthoryear{Guo et~al.,}{Guo et~al.}{2015a}]{guo2015}
Guo H.,  et~al., 2015a, Monthly Notices of the Royal Astronomical Society, 453,
  4368

\bibitem[\protect\citeauthoryear{{Hemmati}, {Yan}, {Diaz-Santos}, {Armus},
  {Capak}, {Faisst}  \& {Masters}}{{Hemmati}
  et~al.}{2017}]{2017ApJ...834...36H}
{Hemmati} S.,  {Yan} L.,  {Diaz-Santos} T.,  {Armus} L.,  {Capak} P.,  {Faisst}
  A.,   {Masters} D.,  2017, \mn@doi [\apj] {10.3847/1538-4357/834/1/36}, \href
  {https://ui.adsabs.harvard.edu/abs/2017ApJ...834...36H} {834, 36}

\bibitem[\protect\citeauthoryear{{Herrera-Camus} et~al.,}{{Herrera-Camus}
  et~al.}{2015}]{Herrera-Camus2015}
{Herrera-Camus} R.,  et~al., 2015, \mn@doi [ApJ] {10.1088/0004-637X/800/1/1},
  \href {https://ui.adsabs.harvard.edu/abs/2015ApJ...800....1H} {800, 1}

\bibitem[\protect\citeauthoryear{{Herrera-Camus} et~al.,}{{Herrera-Camus}
  et~al.}{2018}]{Herrera-Camus2018}
{Herrera-Camus} R.,  et~al., 2018, \mn@doi [ApJ] {10.3847/1538-4357/aac0f9},
  \href {https://ui.adsabs.harvard.edu/abs/2018ApJ...861...95H} {861, 95}

\bibitem[\protect\citeauthoryear{{Hill} et~al.,}{{Hill}
  et~al.}{2008}]{2008ASPC..399..115H}
{Hill} G.~J.,  et~al., 2008, in {Kodama} T.,  {Yamada} T.,   {Aoki} K.,  eds,
  Astronomical Society of the Pacific Conference Series Vol. 399, Panoramic
  Views of Galaxy Formation and Evolution. p.~115 (\mn@eprint {arXiv}
  {0806.0183})

\bibitem[\protect\citeauthoryear{Hill et~al.,}{Hill et~al.}{2021}]{Hill2021}
Hill G.~J.,  et~al., 2021, The Astronomical Journal, 162, 298

\bibitem[\protect\citeauthoryear{{Ihle} et~al.,}{{Ihle}
  et~al.}{2019}]{Ihle2019}
{Ihle} H.~T.,  et~al., 2019, \mn@doi [\apj] {10.3847/1538-4357/aaf4bc}, \href
  {https://ui.adsabs.harvard.edu/abs/2019ApJ...871...75I} {871, 75}

\bibitem[\protect\citeauthoryear{Kamenetzky, Rangwala, Glenn, Maloney  \&
  Conley}{Kamenetzky et~al.}{2016}]{Kamenetzky2016}
Kamenetzky J.,  Rangwala N.,  Glenn J.,  Maloney P.~R.,   Conley A.,  2016,
  \mn@doi [The Astrophysical Journal] {10.3847/0004-637x/829/2/93}, 829, 93

\bibitem[\protect\citeauthoryear{{Karwal} \& {Kamionkowski}}{{Karwal} \&
  {Kamionkowski}}{2016}]{2016PhRvD..94j3523K}
{Karwal} T.,  {Kamionkowski} M.,  2016, \mn@doi [\prd]
  {10.1103/PhysRevD.94.103523}, \href
  {https://ui.adsabs.harvard.edu/abs/2016PhRvD..94j3523K} {94, 103523}

\bibitem[\protect\citeauthoryear{{Keating}, {Marrone}, {Bower}, {Leitch},
  {Carlstrom}  \& {DeBoer}}{{Keating} et~al.}{2016}]{2016ApJ...830...34K}
{Keating} G.~K.,  {Marrone} D.~P.,  {Bower} G.~C.,  {Leitch} E.,  {Carlstrom}
  J.~E.,   {DeBoer} D.~R.,  2016, \mn@doi [\apj] {10.3847/0004-637X/830/1/34},
  \href {https://ui.adsabs.harvard.edu/abs/2016ApJ...830...34K} {830, 34}

\bibitem[\protect\citeauthoryear{Keating, Marrone, Bower  \& Keenan}{Keating
  et~al.}{2020}]{Keating2020}
Keating G.~K.,  Marrone D.~P.,  Bower G.~C.,   Keenan R.~P.,  2020, \mn@doi
  [The Astrophysical Journal] {10.3847/1538-4357/abb08e}, 901, 141

\bibitem[\protect\citeauthoryear{{Keenan}, {Keating}  \& {Marrone}}{{Keenan}
  et~al.}{2022}]{2022ApJ...927..161K}
{Keenan} R.~P.,  {Keating} G.~K.,   {Marrone} D.~P.,  2022, \mn@doi [\apj]
  {10.3847/1538-4357/ac4888}, \href
  {https://ui.adsabs.harvard.edu/abs/2022ApJ...927..161K} {927, 161}

\bibitem[\protect\citeauthoryear{{Kennicutt}}{{Kennicutt}}{1998}]{1998ApJ...498..541K}
{Kennicutt} Robert~C. J.,  1998, \mn@doi [\apj] {10.1086/305588}, \href
  {https://ui.adsabs.harvard.edu/abs/1998ApJ...498..541K} {498, 541}

\bibitem[\protect\citeauthoryear{{Koekemoer} et~al.,}{{Koekemoer}
  et~al.}{2013}]{2013ApJS..209....3K}
{Koekemoer} A.~M.,  et~al., 2013, \mn@doi [\apjs] {10.1088/0067-0049/209/1/3},
  \href {https://ui.adsabs.harvard.edu/abs/2013ApJS..209....3K} {209, 3}

\bibitem[\protect\citeauthoryear{Kogut, Essinger-Hileman, Switzer, Wollack,
  Fixsen, Lowe  \& Mirel}{Kogut et~al.}{2021}]{doi:10.1063/5.0048800}
Kogut A.,  Essinger-Hileman T.,  Switzer E.,  Wollack E.,  Fixsen D.,  Lowe L.,
    Mirel P.,  2021, \mn@doi [Review of Scientific Instruments]
  {10.1063/5.0048800}, 92, 064501

\bibitem[\protect\citeauthoryear{{Kovetz} et~al.,}{{Kovetz}
  et~al.}{2017}]{2017arXiv170909066K}
{Kovetz} E.~D.,  et~al., 2017, arXiv e-prints, \href
  {https://ui.adsabs.harvard.edu/abs/2017arXiv170909066K} {p. arXiv:1709.09066}

\bibitem[\protect\citeauthoryear{{Kovetz} et~al.,}{{Kovetz}
  et~al.}{2019}]{2019BAAS...51c.101K}
{Kovetz} E.,  et~al., 2019, \baas, \href
  {https://ui.adsabs.harvard.edu/abs/2019BAAS...51c.101K} {51, 101}

\bibitem[\protect\citeauthoryear{{Krumholz}}{{Krumholz}}{2014}]{2014MNRAS.437.1662K}
{Krumholz} M.~R.,  2014, \mn@doi [\mnras] {10.1093/mnras/stt2000}, \href
  {https://ui.adsabs.harvard.edu/abs/2014MNRAS.437.1662K} {437, 1662}

\bibitem[\protect\citeauthoryear{{Lagache}, {Cousin}  \& {Chatzikos}}{{Lagache}
  et~al.}{2018}]{2018A&A...609A.130L}
{Lagache} G.,  {Cousin} M.,   {Chatzikos} M.,  2018, \mn@doi [\aap]
  {10.1051/0004-6361/201732019}, \href
  {https://ui.adsabs.harvard.edu/abs/2018A&A...609A.130L} {609, A130}

\bibitem[\protect\citeauthoryear{{Lazear} et~al.,}{{Lazear}
  et~al.}{2014}]{PIPER_Lazear_2014}
{Lazear} J.,  et~al., 2014, in {Holland} W.~S.,  {Zmuidzinas} J.,  eds,
  Society of Photo-Optical Instrumentation Engineers (SPIE) Conference Series
  Vol. 9153, Millimeter, Submillimeter, and Far-Infrared Detectors and
  Instrumentation for Astronomy VII. p. 91531L (\mn@eprint {arXiv}
  {1407.2584}), \mn@doi{10.1117/12.2056806}

\bibitem[\protect\citeauthoryear{{Leicht}, {Uhlemann}, {Villaescusa-Navarro},
  {Codis}, {Hernquist}  \& {Genel}}{{Leicht} et~al.}{2019}]{Leicht2019}
{Leicht} O.,  {Uhlemann} C.,  {Villaescusa-Navarro} F.,  {Codis} S.,
  {Hernquist} L.,   {Genel} S.,  2019, \mn@doi [\mnras]
  {10.1093/mnras/sty3469}, \href
  {https://ui.adsabs.harvard.edu/abs/2019MNRAS.484..269L} {484, 269}

\bibitem[\protect\citeauthoryear{Li, Wechsler, Devaraj  \& Church}{Li
  et~al.}{2016a}]{Li2016}
Li T.~Y.,  Wechsler R.~H.,  Devaraj K.,   Church S.~E.,  2016a, \mn@doi [The
  Astrophysical Journal] {10.3847/0004-637x/817/2/169}, 817, 169

\bibitem[\protect\citeauthoryear{{Li}, {Wechsler}, {Devaraj}  \& {Church}}{{Li}
  et~al.}{2016b}]{2016ApJ...817..169L}
{Li} T.~Y.,  {Wechsler} R.~H.,  {Devaraj} K.,   {Church} S.~E.,  2016b, \mn@doi
  [\apj] {10.3847/0004-637X/817/2/169}, \href
  {https://ui.adsabs.harvard.edu/abs/2016ApJ...817..169L} {817, 169}

\bibitem[\protect\citeauthoryear{{Lidz} \& {Taylor}}{{Lidz} \&
  {Taylor}}{2016}]{2016ApJ...825..143L}
{Lidz} A.,  {Taylor} J.,  2016, \mn@doi [\apj] {10.3847/0004-637X/825/2/143},
  \href {https://ui.adsabs.harvard.edu/abs/2016ApJ...825..143L} {825, 143}

\bibitem[\protect\citeauthoryear{{Lidz}, {Furlanetto}, {Oh}, {Aguirre},
  {Chang}, {Dor{\'e}}  \& {Pritchard}}{{Lidz}
  et~al.}{2011}]{2011ApJ...741...70L}
{Lidz} A.,  {Furlanetto} S.~R.,  {Oh} S.~P.,  {Aguirre} J.,  {Chang} T.-C.,
  {Dor{\'e}} O.,   {Pritchard} J.~R.,  2011, \mn@doi [\apj]
  {10.1088/0004-637X/741/2/70}, \href
  {http://adsabs.harvard.edu/abs/2011ApJ...741...70L} {741, 70}

\bibitem[\protect\citeauthoryear{{Madau} \& {Dickinson}}{{Madau} \&
  {Dickinson}}{2014}]{2014ARA&A..52..415M}
{Madau} P.,  {Dickinson} M.,  2014, \mn@doi [\araa]
  {10.1146/annurev-astro-081811-125615}, \href
  {https://ui.adsabs.harvard.edu/abs/2014ARA&A..52..415M} {52, 415}

\bibitem[\protect\citeauthoryear{Manera et~al.,}{Manera
  et~al.}{2015a}]{manera2015}
Manera M.,  et~al., 2015a, Monthly Notices of the Royal Astronomical Society,
  447, 437

\bibitem[\protect\citeauthoryear{{Manera} et~al.,}{{Manera}
  et~al.}{2015b}]{2015MNRAS.447..437M}
{Manera} M.,  et~al., 2015b, \mn@doi [\mnras] {10.1093/mnras/stu2465}, \href
  {https://ui.adsabs.harvard.edu/abs/2015MNRAS.447..437M} {447, 437}

\bibitem[\protect\citeauthoryear{{Masui} et~al.,}{{Masui}
  et~al.}{2013}]{2013ApJ...763L..20M}
{Masui} K.~W.,  et~al., 2013, \mn@doi [\apjl] {10.1088/2041-8205/763/1/L20},
  \href {https://ui.adsabs.harvard.edu/abs/2013ApJ...763L..20M} {763, L20}

\bibitem[\protect\citeauthoryear{{Mirzaei} et~al.,}{{Mirzaei}
  et~al.}{2020}]{2020SPIE11453E..0MM}
{Mirzaei} M.,  et~al., 2020, in Society of Photo-Optical Instrumentation
  Engineers (SPIE) Conference Series. p. 114530M, \mn@doi{10.1117/12.2562446}

\bibitem[\protect\citeauthoryear{Nishizawa, Hsieh, Tanaka  \& Takata}{Nishizawa
  et~al.}{2020}]{Nishizawa2020}
Nishizawa A.~J.,  Hsieh B.-C.,  Tanaka M.,   Takata T.,  2020, arXiv preprint
  arXiv:2003.01511

\bibitem[\protect\citeauthoryear{Noroozian et~al.,}{Noroozian
  et~al.}{2015}]{uSpec_Noroozian_2015}
Noroozian O.,  et~al., 2015, in 26th International Symposium on Space Terahertz
  Technology.

\bibitem[\protect\citeauthoryear{{Oxholm} \& {Switzer}}{{Oxholm} \&
  {Switzer}}{2021}]{2021PhRvD.104h3501O}
{Oxholm} T.~M.,  {Switzer} E.~R.,  2021, \mn@doi [\prd]
  {10.1103/PhysRevD.104.083501}, \href
  {https://ui.adsabs.harvard.edu/abs/2021PhRvD.104h3501O} {104, 083501}

\bibitem[\protect\citeauthoryear{{Padmanabhan}}{{Padmanabhan}}{2018}]{2018MNRAS.475.1477P}
{Padmanabhan} H.,  2018, \mn@doi [\mnras] {10.1093/mnras/stx3250}, \href
  {https://ui.adsabs.harvard.edu/abs/2018MNRAS.475.1477P} {475, 1477}

\bibitem[\protect\citeauthoryear{{Padmanabhan}}{{Padmanabhan}}{2019}]{2019MNRAS.488.3014P}
{Padmanabhan} H.,  2019, \mn@doi [\mnras] {10.1093/mnras/stz1878}, \href
  {https://ui.adsabs.harvard.edu/abs/2019MNRAS.488.3014P} {488, 3014}

\bibitem[\protect\citeauthoryear{{Papovich} et~al.,}{{Papovich}
  et~al.}{2016}]{2016ApJS..224...28P}
{Papovich} C.,  et~al., 2016, \mn@doi [\apjs] {10.3847/0067-0049/224/2/28},
  \href {https://ui.adsabs.harvard.edu/abs/2016ApJS..224...28P} {224, 28}

\bibitem[\protect\citeauthoryear{{Pawlyk} et~al.,}{{Pawlyk}
  et~al.}{2018}]{PIPER_Flight_Pawlyk_2018}
{Pawlyk} S.,  et~al., 2018, in {Zmuidzinas} J.,  {Gao} J.-R.,  eds,  Society of
  Photo-Optical Instrumentation Engineers (SPIE) Conference Series Vol. 10708,
  Millimeter, Submillimeter, and Far-Infrared Detectors and Instrumentation for
  Astronomy IX. p. 1070806, \mn@doi{10.1117/12.2313874}

\bibitem[\protect\citeauthoryear{{Pen}, {Staveley-Smith}, {Peterson}  \&
  {Chang}}{{Pen} et~al.}{2009}]{2009MNRAS.394L...6P}
{Pen} U.-L.,  {Staveley-Smith} L.,  {Peterson} J.~B.,   {Chang} T.-C.,  2009,
  \mn@doi [\mnras] {10.1111/j.1745-3933.2008.00581.x}, \href
  {https://ui.adsabs.harvard.edu/abs/2009MNRAS.394L...6P} {394, L6}

\bibitem[\protect\citeauthoryear{{Pentericci} et~al.,}{{Pentericci}
  et~al.}{2016}]{2016ApJ...829L..11P}
{Pentericci} L.,  et~al., 2016, \mn@doi [\apjl] {10.3847/2041-8205/829/1/L11},
  \href {https://ui.adsabs.harvard.edu/abs/2016ApJ...829L..11P} {829, L11}

\bibitem[\protect\citeauthoryear{{Popping}, {Somerville}  \&
  {Trager}}{{Popping} et~al.}{2014}]{2014MNRAS.442.2398P}
{Popping} G.,  {Somerville} R.~S.,   {Trager} S.~C.,  2014, \mn@doi [\mnras]
  {10.1093/mnras/stu991}, \href
  {https://ui.adsabs.harvard.edu/abs/2014MNRAS.442.2398P} {442, 2398}

\bibitem[\protect\citeauthoryear{{Popping}, {van Kampen}, {Decarli}, {Spaans},
  {Somerville}  \& {Trager}}{{Popping} et~al.}{2016}]{2016MNRAS.461...93P}
{Popping} G.,  {van Kampen} E.,  {Decarli} R.,  {Spaans} M.,  {Somerville}
  R.~S.,   {Trager} S.~C.,  2016, \mn@doi [\mnras] {10.1093/mnras/stw1323},
  \href {https://ui.adsabs.harvard.edu/abs/2016MNRAS.461...93P} {461, 93}

\bibitem[\protect\citeauthoryear{{Popping}, {Narayanan}, {Somerville}, {Faisst}
   \& {Krumholz}}{{Popping} et~al.}{2019a}]{2019MNRAS.482.4906P}
{Popping} G.,  {Narayanan} D.,  {Somerville} R.~S.,  {Faisst} A.~L.,
  {Krumholz} M.~R.,  2019a, \mn@doi [\mnras] {10.1093/mnras/sty2969}, \href
  {https://ui.adsabs.harvard.edu/abs/2019MNRAS.482.4906P} {482, 4906}

\bibitem[\protect\citeauthoryear{{Popping} et~al.,}{{Popping}
  et~al.}{2019b}]{2019ApJ...882..137P}
{Popping} G.,  et~al., 2019b, \mn@doi [\apj] {10.3847/1538-4357/ab30f2}, \href
  {https://ui.adsabs.harvard.edu/abs/2019ApJ...882..137P} {882, 137}

\bibitem[\protect\citeauthoryear{{Pullen}, {Chang}, {Dor{\'e}}  \&
  {Lidz}}{{Pullen} et~al.}{2013}]{2013ApJ...768...15P}
{Pullen} A.~R.,  {Chang} T.-C.,  {Dor{\'e}} O.,   {Lidz} A.,  2013, \mn@doi
  [\apj] {10.1088/0004-637X/768/1/15}, \href
  {http://adsabs.harvard.edu/abs/2013ApJ...768...15P} {768, 15}

\bibitem[\protect\citeauthoryear{{Pullen}, {Serra}, {Chang}, {Dor{\'e}}  \&
  {Ho}}{{Pullen} et~al.}{2018}]{2018MNRAS.478.1911P}
{Pullen} A.~R.,  {Serra} P.,  {Chang} T.-C.,  {Dor{\'e}} O.,   {Ho} S.,  2018,
  \mn@doi [\mnras] {10.1093/mnras/sty1243}, \href
  {https://ui.adsabs.harvard.edu/abs/2018MNRAS.478.1911P} {478, 1911}

\bibitem[\protect\citeauthoryear{{Reddy}, {Steidel}, {Erb}, {Shapley}  \&
  {Pettini}}{{Reddy} et~al.}{2006}]{2006ApJ...653.1004R}
{Reddy} N.~A.,  {Steidel} C.~C.,  {Erb} D.~K.,  {Shapley} A.~E.,   {Pettini}
  M.,  2006, \mn@doi [\apj] {10.1086/508851}, \href
  {https://ui.adsabs.harvard.edu/abs/2006ApJ...653.1004R} {653, 1004}

\bibitem[\protect\citeauthoryear{Reid et~al.,}{Reid et~al.}{2016}]{reid2016}
Reid B.,  et~al., 2016, Monthly Notices of the Royal Astronomical Society, 455,
  1553

\bibitem[\protect\citeauthoryear{{Righi}, {Hern{\'a}ndez-Monteagudo}  \&
  {Sunyaev}}{{Righi} et~al.}{2008}]{2008A&A...489..489R}
{Righi} M.,  {Hern{\'a}ndez-Monteagudo} C.,   {Sunyaev} R.~A.,  2008, \mn@doi
  [\aap] {10.1051/0004-6361:200810199}, \href
  {http://adsabs.harvard.edu/abs/2008A%26A...489..489R} {489, 489}

\bibitem[\protect\citeauthoryear{{Schaan} \& {White}}{{Schaan} \&
  {White}}{2021}]{2021JCAP...05..068S}
{Schaan} E.,  {White} M.,  2021, \mn@doi [\jcap]
  {10.1088/1475-7516/2021/05/068}, \href
  {https://ui.adsabs.harvard.edu/abs/2021JCAP...05..068S} {2021, 068}

\bibitem[\protect\citeauthoryear{{Sheth} \& {Tormen}}{{Sheth} \&
  {Tormen}}{2002}]{2002MNRAS.329...61S}
{Sheth} R.~K.,  {Tormen} G.,  2002, \mn@doi [\mnras]
  {10.1046/j.1365-8711.2002.04950.x}, \href
  {https://ui.adsabs.harvard.edu/abs/2002MNRAS.329...61S} {329, 61}

\bibitem[\protect\citeauthoryear{{Silva}, {Santos}, {Cooray}  \&
  {Gong}}{{Silva} et~al.}{2015}]{2015ApJ...806..209S}
{Silva} M.,  {Santos} M.~G.,  {Cooray} A.,   {Gong} Y.,  2015, \mn@doi [\apj]
  {10.1088/0004-637X/806/2/209}, \href
  {http://adsabs.harvard.edu/abs/2015ApJ...806..209S} {806, 209}

\bibitem[\protect\citeauthoryear{{Singal} et~al.,}{{Singal}
  et~al.}{2011}]{ARCADE2_Instrument_Singal_2011}
{Singal} J.,  et~al., 2011, \mn@doi [\apj] {10.1088/0004-637X/730/2/138}, \href
  {https://ui.adsabs.harvard.edu/abs/2011ApJ...730..138S} {730, 138}

\bibitem[\protect\citeauthoryear{{Somerville} \& {Primack}}{{Somerville} \&
  {Primack}}{1999}]{1999MNRAS.310.1087S}
{Somerville} R.~S.,  {Primack} J.~R.,  1999, \mn@doi [\mnras]
  {10.1046/j.1365-8711.1999.03032.x}, \href
  {https://ui.adsabs.harvard.edu/abs/1999MNRAS.310.1087S} {310, 1087}

\bibitem[\protect\citeauthoryear{{Somerville}, {Popping}  \&
  {Trager}}{{Somerville} et~al.}{2015}]{2015MNRAS.453.4337S}
{Somerville} R.~S.,  {Popping} G.,   {Trager} S.~C.,  2015, \mn@doi [\mnras]
  {10.1093/mnras/stv1877}, \href
  {https://ui.adsabs.harvard.edu/abs/2015MNRAS.453.4337S} {453, 4337}

\bibitem[\protect\citeauthoryear{{Switzer} et~al.,}{{Switzer}
  et~al.}{2013}]{2013MNRAS.434L..46S}
{Switzer} E.~R.,  et~al., 2013, \mn@doi [\mnras] {10.1093/mnrasl/slt074}, \href
  {https://ui.adsabs.harvard.edu/abs/2013MNRAS.434L..46S} {434, L46}

\bibitem[\protect\citeauthoryear{{Switzer}, {Chang}, {Masui}, {Pen}  \&
  {Voytek}}{{Switzer} et~al.}{2015}]{2015ApJ...815...51S}
{Switzer} E.~R.,  {Chang} T.~C.,  {Masui} K.~W.,  {Pen} U.~L.,   {Voytek}
  T.~C.,  2015, \mn@doi [\apj] {10.1088/0004-637X/815/1/51}, \href
  {https://ui.adsabs.harvard.edu/abs/2015ApJ...815...51S} {815, 51}

\bibitem[\protect\citeauthoryear{{Switzer}, {Anderson}, {Pullen}  \&
  {Yang}}{{Switzer} et~al.}{2019}]{2019ApJ...872...82S}
{Switzer} E.~R.,  {Anderson} C.~J.,  {Pullen} A.~R.,   {Yang} S.,  2019,
  \mn@doi [\apj] {10.3847/1538-4357/aaf9ab}, \href
  {https://ui.adsabs.harvard.edu/abs/2019ApJ...872...82S} {872, 82}

\bibitem[\protect\citeauthoryear{{Switzer} et~al.,}{{Switzer}
  et~al.}{2021}]{2021JATIS...7d4004S}
{Switzer} E.~R.,  et~al., 2021, \mn@doi [Journal of Astronomical Telescopes,
  Instruments, and Systems] {10.1117/1.JATIS.7.4.044004}, \href
  {https://ui.adsabs.harvard.edu/abs/2021JATIS...7d4004S} {7, 044004}

\bibitem[\protect\citeauthoryear{{Thorne}, {Dunkley}, {Alonso}  \&
  {N{\ae}ss}}{{Thorne} et~al.}{2017}]{2017MNRAS.469.2821T}
{Thorne} B.,  {Dunkley} J.,  {Alonso} D.,   {N{\ae}ss} S.,  2017, \mn@doi
  [\mnras] {10.1093/mnras/stx949}, \href
  {https://ui.adsabs.harvard.edu/abs/2017MNRAS.469.2821T} {469, 2821}

\bibitem[\protect\citeauthoryear{{Visbal} \& {Loeb}}{{Visbal} \&
  {Loeb}}{2010}]{2010JCAP...11..016V}
{Visbal} E.,  {Loeb} A.,  2010, \mn@doi [\jcap]
  {10.1088/1475-7516/2010/11/016}, \href
  {https://ui.adsabs.harvard.edu/abs/2010JCAP...11..016V} {2010, 016}

\bibitem[\protect\citeauthoryear{{Visbal}, {Trac}  \& {Loeb}}{{Visbal}
  et~al.}{2011}]{2011JCAP...08..010V}
{Visbal} E.,  {Trac} H.,   {Loeb} A.,  2011, \mn@doi [\jcap]
  {10.1088/1475-7516/2011/08/010}, \href
  {https://ui.adsabs.harvard.edu/abs/2011JCAP...08..010V} {2011, 010}

\bibitem[\protect\citeauthoryear{{Volpert} et~al.,}{{Volpert}
  et~al.}{2022}]{2022arXiv220802786V}
{Volpert} C.~G.,  et~al., 2022, arXiv e-prints, \href
  {https://ui.adsabs.harvard.edu/abs/2022arXiv220802786V} {p. arXiv:2208.02786}

\bibitem[\protect\citeauthoryear{{Walter} et~al.,}{{Walter}
  et~al.}{2016}]{2016ApJ...833...67W}
{Walter} F.,  et~al., 2016, \mn@doi [\apj] {10.3847/1538-4357/833/1/67}, \href
  {https://ui.adsabs.harvard.edu/abs/2016ApJ...833...67W} {833, 67}

\bibitem[\protect\citeauthoryear{{Walter} et~al.,}{{Walter}
  et~al.}{2020}]{2020ApJ...902..111W}
{Walter} F.,  et~al., 2020, \mn@doi [\apj] {10.3847/1538-4357/abb82e}, \href
  {https://ui.adsabs.harvard.edu/abs/2020ApJ...902..111W} {902, 111}

\bibitem[\protect\citeauthoryear{{Wang} et~al.,}{{Wang}
  et~al.}{2022}]{2022ApJ...928....1W}
{Wang} Y.,  et~al., 2022, \mn@doi [\apj] {10.3847/1538-4357/ac4973}, \href
  {https://ui.adsabs.harvard.edu/abs/2022ApJ...928....1W} {928, 1}

\bibitem[\protect\citeauthoryear{{Wolz}, {Blake}  \& {Wyithe}}{{Wolz}
  et~al.}{2017}]{2017MNRAS.470.3220W}
{Wolz} L.,  {Blake} C.,   {Wyithe} J.~S.~B.,  2017, \mn@doi [\mnras]
  {10.1093/mnras/stx1388}, \href
  {https://ui.adsabs.harvard.edu/abs/2017MNRAS.470.3220W} {470, 3220}

\bibitem[\protect\citeauthoryear{{Wolz} et~al.,}{{Wolz}
  et~al.}{2022}]{2022MNRAS.510.3495W}
{Wolz} L.,  et~al., 2022, \mn@doi [\mnras] {10.1093/mnras/stab3621}, \href
  {https://ui.adsabs.harvard.edu/abs/2022MNRAS.510.3495W} {510, 3495}

\bibitem[\protect\citeauthoryear{{Yang}, {Pullen}  \& {Switzer}}{{Yang}
  et~al.}{2019}]{2019MNRAS.489L..53Y}
{Yang} S.,  {Pullen} A.~R.,   {Switzer} E.~R.,  2019, \mn@doi [\mnras]
  {10.1093/mnrasl/slz126}, \href
  {https://ui.adsabs.harvard.edu/abs/2019MNRAS.489L..53Y} {489, L53}

\bibitem[\protect\citeauthoryear{{Yang}, {Popping}, {Somerville}, {Pullen},
  {Breysse}  \& {Maniyar}}{{Yang} et~al.}{2022}]{2022ApJ...929..140Y}
{Yang} S.,  {Popping} G.,  {Somerville} R.~S.,  {Pullen} A.~R.,  {Breysse}
  P.~C.,   {Maniyar} A.~S.,  2022, \mn@doi [\apj] {10.3847/1538-4357/ac5d57},
  \href {https://ui.adsabs.harvard.edu/abs/2022ApJ...929..140Y} {929, 140}

\bibitem[\protect\citeauthoryear{{Yue}, {Ferrara}, {Pallottini}, {Gallerani}
  \& {Vallini}}{{Yue} et~al.}{2015}]{2015MNRAS.450.3829Y}
{Yue} B.,  {Ferrara} A.,  {Pallottini} A.,  {Gallerani} S.,   {Vallini} L.,
  2015, \mn@doi [\mnras] {10.1093/mnras/stv933}, \href
  {http://adsabs.harvard.edu/abs/2015MNRAS.450.3829Y} {450, 3829}

\makeatother
\end{thebibliography}
\bsp	% typesetting comment
\label{lastpage}
\end{document}